\documentclass[12pt]{article}
\pdfoutput=1
\usepackage{jheppub}
\usepackage{ucs}
\usepackage[utf8x]{inputenc}
\usepackage{amsmath}
\usepackage{amsfonts}
\usepackage{amssymb}
\usepackage{amsthm}
\usepackage{graphicx}
\usepackage{subfigure}
\usepackage[american]{babel}
\usepackage{fontenc}
\usepackage{graphicx}
\usepackage{mathtools}
\usepackage{float}
\usepackage{bm}  %
\usepackage{ulem}  %
\usepackage[sans]{dsfont}  %

\title{
Zero-Range Effective Field Theory \\
for Resonant Wino  Dark Matter\\  I. Framework}
\author{Eric Braaten,}
\author{Evan Johnson,}
\author{and Hong Zhang}

\affiliation{Department of Physics,
         The Ohio State University, Columbus, OH\ 43210, USA}

\emailAdd{braaten.1@osu.edu}
\emailAdd{johnson.6036@osu.edu}
\emailAdd{zhang.5676@osu.edu}

\abstract{
The most dramatic ``Sommerfeld enhancements'' of neutral-wino-pair annihilation occur when the wino mass is near a critical value where there is a zero-energy S-wave resonance at the neutral-wino-pair threshold. Near such a critical mass, low-energy winos can be described by a zero-range effective field theory in which the winos interact nonperturbatively through a contact interaction. The effective field theory is controlled by a renormalization-group fixed point at which the neutral and charged winos are degenerate in mass and their scattering length is infinite. The parameters of the zero-range effective field theory can be determined by matching wino-wino scattering amplitudes calculated  by solving the Schr\"odinger equation for  winos interacting through a potential due to the exchange of weak gauge bosons. If the wino mass is larger than the critical value, the resonance is a wino-pair bound state. The power of the zero-range effective field theory is illustrated by calculating the rate for formation of the bound state in the collision of two neutral winos through the emission of two soft photons.
}

\smallskip

\begin{document}
\maketitle
\flushbottom

\section{Introduction}
\label{sec:Introduction}

A weakly interacting massive particle ({\it wimp}) is one of the best motivated candidates for a dark-matter particle that provides  most of the mass of the universe. A stable particle with weak interactions and whose mass is roughly at the electroweak scale is naturally produced in the early universe with a relic abundance comparable to the observed mass density of dark matter \cite{Kolb:1990vq,Steigman:2012nb}. The annihilation of a wimp and its antiparticle provides an indirect detection signal for dark matter. In the present universe, wimps that make up the dark matter are highly nonrelativistic; for example, the typical local wimp velocity $v$ in the Milky Way galaxy is about $10^{-3}$. If the wimp mass $M$ is in the TeV range, the wimp annihilation rate is complicated by a nonperturbative effect first pointed out by Hisano et al.\ \cite{Hisano:2002fk}. Weak interactions between low-energy wimps are nonperturbative in the same sense as Coulomb interactions between low-energy charged particles: the exchange of gauge bosons must be summed to all orders in the gauge coupling constant. The resummation can  significantly increase the annihilation rate. In particular, there can be critical values of the wimp mass  near which the annihilation rate is enhanced by orders of magnitude  \cite{Hisano:2003ec,Hisano:2004ds}.

The increase of the annihilation rate of wimps from the nonperturbative exchange of electroweak gauge bosons or other mediators is commonly referred to as {\it Sommerfeld enhancement}. This phrase was popularized in ref.~\cite{ArkaniHamed:2008qn}, which considered the nonperturbative effect of the exchange of mediators from a dark sector. The nonperturbative effect of the Coulomb rescattering of charged particles was  derived by Sommerfeld around 1920 \cite{Sommerfeld:1921}. The cross section is multiplied by a {\it Sommerfeld factor} that depends on the velocity  of the charged particles and cancels the factor of velocity from the phase space, so the cross section has a nonzero limit as the energy approaches the threshold. The phrase ``Sommerfeld enhancement'' applied to the orders-of-magnitude enhancement  of a wimp annihilation rate near a critical value of the wimp mass is somewhat of a misnomer. The physical origin of the orders-of-magnitude enhancement is actually much simpler than that of the Sommerfeld factor. Above each critical value of the wimp mass, there is an additional bound state of two dark-matter particles. The ``Sommerfeld enhancement'' comes from the resonance associated with that bound state. Like the Sommerfeld factor, the resonance can be produced by the exchange of gauge bosons. However it could also be produced by any potential between wimps that is sufficiently attractive to produce a bound state. A resonance that produces a dramatic Sommerfeld enhancement of the wimp-pair annihilation rate also has dramatic effects on the self-interactions of wimps. In particular, it produces dramatic velocity dependence in low-energy wimp-wimp scattering cross sections.

If the Sommerfeld enhancement of the wimp annihilation rate is produced by a resonance near threshold with orbital-angular-momentum quantum number $L$, the rate has an angular momentum suppression factor of $v^{2L}$. A resonance in an S-wave channel ($L=0$) can generally produce the most dramatic enhancement over the broadest range of $M$. There is also a qualitative difference between a near-threshold resonance in an S-wave channel and in a channel with higher orbital angular momentum. The S-wave resonance generates dynamically a length scale that is much larger than the  range of the interactions. This length scale is the absolute value of the S-wave scattering length $a$, which can be orders of magnitude larger than the range. If there are no pair-annihilation channels,  the scattering length can even be infinitely large. A bound state typically has a size of order the range of the interactions. The exception is when $M$ is just above the critical mass for an S-wave resonance, in which case the size of the S-wave bound state closest to threshold is of order $a$.

Bound states of dark-matter particles could have significant effects on the phenomenology of dark matter. The formation of bound states followed by the annihilation of their constituents provides new channels for the indirect detection of dark matter. The bound states may also modify the signals for direct detection of dark matter. Dark-matter bound states could also affect the cosmological history and thermal relic abundance of dark matter. In most investigations of dark-matter bound states, the bound states are produced by the exchange of a new light mediator \cite{MarchRussell:2008tu,Wise:2014jva,vonHarling:2014kha,Petraki:2015hla,An:2015pva,Bi:2016gca,An:2016gad,Kouvaris:2016ltf,Nozzoli:2016coi,Cirelli:2016rnw,Mitridate:2017izz}. The weak interaction of the Standard Model can also produce bound states of wimps \cite{Shepherd:2009sa,Asadi:2016ybp}. The effects of a bound state on dark matter can be particularly dramatic if the bound state is near the threshold.

Dramatic velocity dependence in the low-energy scattering of dark-matter particles with each other can significantly affect the small-scale structure of the universe. The standard model of cosmology ($\Lambda$CDM), with cold dark matter and a cosmological constant, is very successful in describing the mass-energy distribution of the universe on large scales down to the size of galaxies like the Milky Way. Dark-matter-only simulations of dark-matter halos predict a cusp at the center of a galaxy, while observed rotation curves of dwarf galaxies favor a  profile with a constant density in the center \cite{KuziodeNaray:2007qi,Oh:2008ww}. Dark-matter-only simulations also predict larger satellite galaxies of the Milky Way and other galaxies than those that are observed \cite{BoylanKolchin:2011dk,Ferrero:2011au}. Self-interacting dark matter models have been introduced  to explain the observations that are in tension with predictions of collisionless cold dark matter in the $\Lambda$CDM model \cite{Spergel:1999mh}. (See ref.~\cite{Tulin:2017ara} for a recent review.) In these models, dark-matter particles have velocity-dependent scattering at low energy that does not affect the large-scale structure that is described so successfully by the $\Lambda$CDM model, but does affect the small-scale structure. Self-interacting dark matter models typically invoke a dark sector consisting of new particles and forces that can generate a rich phenomenology but can be difficult to constrain. The mass scale of the dark-matter particle that is required to have the desired effect on small-scale structure is roughly 100~MeV \cite{Tulin:2017ara}. A simple way to produce the dramatic velocity dependence of the cross section at low energy is to have an S-wave resonance close to the threshold.

In a fundamental quantum field theory, wimps interact through the exchange of electroweak gauge bosons to which they couple through local gauge interactions. The Sommerfeld enhancement of the wimp annihilation rate can be calculated by summing an infinite set of diagrams in that quantum field theory. The enhancement can be calculated more simply using a nonrelativistic effective field theory (NREFT) in which the wimps have instantaneous interactions at a distance through a potential generated by the exchange of the electroweak gauge bosons. In NREFT, the Sommerfeld  enhancement of the wimp annihilation rate can be calculated by the numerical solution of a Schr\"odinger equation \cite{Hisano:2002fk,Hisano:2003ec,Hisano:2004ds}. A thorough development of NREFT for nearly degenerate neutralinos and charginos in the MSSM has been presented in refs.~\cite{Beneke:2012tg,Hellmann:2013jxa} and applied to the Sommerfeld enhancement of pair annihilation in ref.~\cite{Beneke:2014gja}. One limitation of the results on pair annihilation is that they do not apply in the case of a narrow S-wave resonance near the threshold. NREFT can also be used to calculate numerically other few-body reaction rates of nonrelativistic wimps, such as  scattering cross sections. If the wimp mass is about 1~TeV, the relative velocity must be at most about 0.1. The velocity dependence of the low-energy cross section can affect the relic abundance of dark matter \cite{Hisano:2006nn,Cirelli:2007xd}. NREFT has recently been used to calculate the capture rates of two neutral winos into wino-pair bound states through the radiation of a photon \cite{Asadi:2016ybp}.

In the case of an S-wave resonance near threshold, low-energy wimps can be described more simply using a nonrelativistic effective field theory with zero-range interactions (ZREFT). ZREFT exploits the large length scale that is generated dynamically  by an S-wave resonance. 
If there are no wimp-pair annihilation channels, the S-wave scattering length $a$ diverges at critical values of the wimp mass $M$. ZREFT is applicable if $M$ is close enough to a critical value that $|a|$  is  large compared  to the range $1/m_W$ of the weak interactions.  ZREFT can be used to obtain analytic results for wimp-wimp scattering cross sections, provided the relative momentum of the wimps is less than $m_W$. If the wimp mass is above the critical value, the S-wave resonance is a bound state below the neutral-wimp-pair threshold with a size of order $|a|$. ZREFT can be used to simplify calculations of few-body reaction rates involving this bound state.  
If there are wimp-pair annihilation channels, ZREFT can be used to obtain analytic results for inclusive wimp-pair annihilation rates.  In particular,  it provides analytic results for Sommerfeld enhancement factors.

There have been several previous applications of zero-range effective field theories to dark matter with resonant S-wave self-interactions \cite{Braaten:2013tza,Laha:2013gva,Laha:2015yoa}. Braaten and Hammer pointed out that the elastic scattering cross section of the dark-matter particles, their total annihilation cross section, and the binding energy and width of a dark matter bound state are all determined by the complex S-wave scattering length \cite{Braaten:2013tza}. Laha and Braaten studied the nuclear recoil energy spectrum in dark-matter direct detection experiments due to both elastic scattering and breakup scattering of an incident dark-matter bound state \cite{Laha:2013gva}. Laha extended that analysis to the angular recoil spectrum in directional detection experiments \cite{Laha:2015yoa}.

In this paper, we develop the ZREFT for wimps that consist of the neutral dark-matter particle $w^0$ and charged wimps $w^+$ and $w^-$ that are nearly degenerate in mass with $w^0$. We refer to these wimps as {\it winos}, because the fundamental theory describing them could be the minimal supersymmetric standard model (MSSM) in a region of parameter space where the neutral wino is the lightest supersymmetric particle. We take the neutral-wino mass $M$ close enough to a critical mass for an S-wave resonance at the neutral-wino-pair threshold that the neutral-wino scattering length $a_0$ is large compared to the range $1/m_W$ of the weak interactions. We take the wino mass splitting $\delta$ small  enough that $\sqrt{2M\delta}$ is smaller  than $m_W$. The transition between a pair of neutral winos and a pair of charged winos is then within the domain of  validity of ZREFT. If $M$ is above the critical mass, there is an S-wave wino-pair bound state whose constituents are a superposition of $w^0w^0$ and $w^+ w^-$. We ignore the effects of wimp-pair annihilation in this paper.

The minimal wino-dark-matter scenario assumes all the dark matter consists of pure neutral winos. For wino mass below 3 TeV, this scenario is almost ruled out by the absence of signals in direct detection experiments \cite{Cohen:2013ama,Fan:2013faa,Beneke:2016jpw}. Even if the minimal wino-dark-matter scenario is completely ruled out, the development of ZREFT for wino dark matter is still useful for the wino component of dark matter in non-minimal scenarios. The ZREFT for winos can be used as a simple model for self-interacting dark matter in which the dark-matter particle is a member of an $SU(2)$ triplet in a dark sector. The ZREFT for winos can also be generalized to other wimps with small mass splittings from the dark matter particle, such as those in Higgsino dark matter.

This paper is organized as follows. We begin in section~\ref{sec:QFT} by describing the Lorentz-invariant quantum field theory that provides a fundamental description of winos and their electroweak interactions. In section~\ref{sec:NREFT}, we describe the nonrelativistic effective field theory NREFT for low-energy winos in which they interact through a potential generated by the exchange of weak gauge bosons and in which charged winos also have electromagnetic interactions. We use the Schr\"odinger equation to numerically calculate 2-body observables for winos, including cross sections for $w^0w^0$ and $w^+ w^-$ and the binding energy of a wino pair bound state. In section~\ref{sec:ZREFT}, we introduce the zero-range effective field theory ZREFT for low-energy winos in which they interact through zero-range self-interactions and through couplings to the electromagnetic field. In the absence of electromagnetic interactions, ZREFT is a systematically improvable effective field theory that can be defined by deformations of a renormalization-group fixed point with  scale-invariant interactions. In section~\ref{sec:ZREFTLO}, we use ZREFT at leading order (LO) and without electromagnetism to analytically calculate two-body observables for low-energy winos in terms of the scattering length and one adjustable parameter. By comparing with results from NREFT with $\alpha=0$, we show that ZREFT at LO gives surprisingly accurate predictions for most observables. In section~\ref{sec:ZREFTNLO}, we use ZREFT at next-to-leading order (NLO) to analytically calculate low-energy wino cross sections in terms of the scattering length and three adjustable parameters. By comparing with results from NREFT, we show that ZREFT at NLO systematically improves upon the predictions of ZREFT at LO. In section~\ref{sec:DoubleRadiative},  we illustrate the power of ZREFT by using it  to calculate the rate for the double-radiative formation of a wino-pair bound state through the emission of two soft photons. Our results are summarized in section~\ref{sec:Conclusion}.

For many observables in NREFT, including the double-radiative formation of the wimp-pair bound state, the electromagnetic interactions can be treated as a perturbation. However Coulomb interactions between low-energy charged winos have dramatic effects on wino-wino scattering in the threshold region, so Coulomb exchange must be treated to all orders. In this paper, all the calculation in ZREFT are carried out with Coulomb interactions and the effects of wino-pair annihilation  omitted. In a companion paper, the effects of Coulomb interactions on two-body observables for low-energy winos are calculated analytically in ZREFT  \cite{BJZ-Coulomb}. In another companion paper, the effects of wino-pair annihilation into electroweak gauge bosons are taken into account through the analytic continuation of the interaction parameters of  ZREFT  \cite{BJZ-Annihilation}.

\section{Fundamental Theory}
\label{sec:QFT}

We assume the dark-matter particle is the neutral member of an $SU(2)$ triplet of Majorana fermions with zero hypercharge. The Lorentz-invariant quantum field theory that provides a fundamental  description of these fermions could simply be an extension of the Standard Model with this additional $SU(2)$ multiplet and with a symmetry that forbids the decay of the fermion into Standard Model particles. The fundamental theory could also be the Minimal Supersymmetric Standard Model (MSSM) in a region of parameter space where the lightest supersymmetric particle is a wino-like neutralino. In either case, we refer to  the particles in the $SU(2)$ multiplet as {\it winos}. We denote the neutral wino by $w^0$ and the charged winos by $w^+$ and $w^-$.

The relic density of the neutral wino is compatible with the observed mass density of dark matter if the neutral wino mass $M$  is roughly at the electroweak scale \cite{Kolb:1990vq}. We are particularly interested in a mass $M$ at the TeV scale so that effects from the exchange of electroweak gauge bosons must be summed to all orders. For the neutral wino to be stable, the charged wino must have a larger mass $M+\delta$. In the MSSM, the mass splitting $\delta$ arises from radiative corrections. The splitting from one-loop radiative corrections is determined by $M$ and Standard Model parameters only  \cite{Cheng:1998hc,Feng:1999fu,Gherghetta:1999sw}. As $M$ ranges from 1~TeV to 10~TeV, the one-loop splitting $\delta$ remains very close to 174~MeV. The two-loop radiative corrections decrease $\delta$ by a few MeV  \cite{Ibe:2012sx}. We  take $\delta= 170$~MeV to be the preferred mass splitting, but we also consider the effect of decreasing $\delta$ to zero.

The winos can be represented by a triplet $\chi_i$ of 4-component Majorana spinor fields, where the neutral-wino field is $\chi_3$ and the charged-wino fields are linear combinations of $\chi_1$ and $\chi_2$. The most important interactions of the winos are those with the electroweak gauge bosons: the photon, the $W^\pm$, and the $Z^0$. The Lagrangian for the winos is
\begin{equation}
\mathcal{L}_{\rm wino} = 
\sum_i \left( \tfrac{i}{2} \chi_i^T C \gamma^\mu D_\mu \chi_i  
 - \tfrac{1}{2} M \chi_i^T C \chi_i   \right),
\label{eq:Lwino}
\end{equation}
where $D_\mu$ is the $SU(2)$ gauge-covariant derivative and $C$ is a charge conjugation matrix. The mass $M$ of the winos is an adjustable parameter. The splitting $\delta$ between the masses of $w^\pm$ and $w^0$ arises from electroweak radiative corrections. The relevant Standard Model parameters are the mass $m_W = 80.4$~GeV of the $W^\pm$, the mass $m_Z = 91.2$~GeV of the $Z^0$, the $SU(2)$ coupling constant $\alpha_2= 1/29.5$, the electromagnetic coupling constant $\alpha = 1/137.04$, and the weak mixing angle, which is given by $\sin^2 \theta_w = 0.231$.

\begin{figure}[t]
\centering
\includegraphics[width=0.98\linewidth]{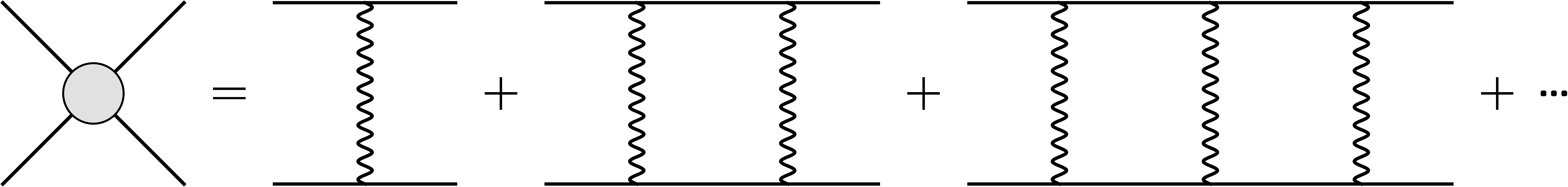}
\caption{Feynman diagrams in the fundamental theory for wino-wino scattering
through the exchange of electroweak gauge bosons.
The solid lines are neutral winos or charged winos, and the wavy lines are electroweak gauge bosons.
If the winos are nonrelativistic,
ladder diagrams from the exchange of electroweak gauge bosons must be summed to all orders.
}
\label{fig:FundamentalSum}
\end{figure}

Hisano, Matsumoto, and Nojiri pointed out that if the mass of the wino is large enough that $\alpha_2 M$ is of order $m_W$ or larger, loop diagrams in which electroweak gauge bosons are exchanged between a pair of nonrelativistic winos are not suppressed  \cite{Hisano:2002fk}. The electroweak interactions between a pair of nonrelativistic winos must therefore be treated nonperturbatively by summing ladder  diagrams from the exchange of electroweak bosons between the winos to all orders. For the elastic scattering of winos, the first few diagrams in the sum are shown in Fig.~\ref{fig:FundamentalSum}. In the corresponding diagrams for neutral-wino-pair annihilation into two electroweak gauge bosons,  the last interaction is the annihilation of a pair of charged winos.

A particular dramatic consequence of the resummation of ladder diagrams from the exchange of electroweak gauge bosons is the existence of a zero-energy resonance at the neutral-wino-pair threshold $2M$ at a sequence of critical values of $M$ \cite{Hisano:2003ec,Hisano:2004ds}. Near these resonances,  the annihilation rate of a wino pair into a pair of electroweak gauge bosons is increased by orders of magnitude \cite{Hisano:2003ec}. Two-body observables for winos can be calculated in the fundamental theory by summing ladder diagrams to all orders. However the calculations can be greatly facilitated by using a nonrelativistic effective field theory for the winos, which will be discussed in the next section.

\section{Nonrelativistic  effective field theory}
\label{sec:NREFT}

Low-energy winos can be described by a nonrelativistic effective field theory in which they interact through potentials that arise from the exchange of weak gauge bosons and in which charged winos also have electromagnetic interactions. We call this effective field theory {\it NREFT}. In this section, we use NREFT to calculate cross sections for nonrelativistic wino-wino scattering and the  binding energy for a wino-pair bound state. We study the dependence of the wino cross sections on the wino mass $M$, the electromagnetic coupling constant $\alpha$, and the wino mass splitting $\delta$.

\subsection{Lagrangian}

In refs.~\cite{Hisano:2003ec,Hisano:2004ds}, Hisano, Matsumoto, and Nojiri used a nonrelativistic effective field theory for winos to calculate the  nonperturbative effect of the exchange of electroweak gauge bosons between winos. The nonrelativistic wino fields are 2-component spinor fields: $\zeta$ which annihilates a neutral wino $w^0$, $\eta$ which annihilates a charged wino $w^-$, and $\xi$ which creates a charged wino $w^+$. The kinetic terms for winos in the Lagrangian density are
\begin{equation}
\mathcal{L}_{\rm kinetic} = \zeta^\dagger\left(i\partial_0+\frac{\bm{\nabla}^2}{2M}\right) \zeta 
+ \eta^\dagger\left(i\partial_0+\frac{\bm{\nabla}^2}{2M} -\delta\right) \eta
+ \xi^\dagger\left(i\partial_0-\frac{\bm{\nabla}^2}{2M} +\delta\right) \xi.
\label{eq:kineticLHMN}
\end{equation}
The mass splitting $\delta$ has been taken into account through the rest energy of the charged winos. The interaction terms in the Hamiltonian are instantaneous interactions at a distance through a potential produced by the exchange of electroweak gauge bosons:
\begin{eqnarray}
H_{\rm potential} &=& - \frac12 \int\!\! d^3x \int\!\! d^3y
\bigg( \left[ \frac{\alpha}{|\bm{x}-\bm{y}|}  
+  \frac{\alpha_2 \cos^2 \theta_w}{|\bm{x}-\bm{y}|} e^{-m_Z|\bm{x}-\bm{y}|} \right]
\eta^\dagger(\bm{x}) \xi(\bm{y})\,  \xi^\dagger(\bm{y}) \eta(\bm{x})~~
\nonumber\\
&&+ \frac{\alpha_2}{|\bm{x}-\bm{y}|}  e^{-m_W|\bm{x}-\bm{y}|} 
\left[ \zeta^\dagger(\bm{x}) \zeta^c(\bm{y})\,  \xi^\dagger(\bm{y}) \eta(\bm{x})
+ \zeta^{c\dagger}(\bm{x}) \zeta(\bm{y})\,  \eta^\dagger(\bm{y}) \xi(\bm{x})\right] \bigg),
\label{eq:LintHMN}
\end{eqnarray}
where $\zeta^c = -i \sigma_2 \xi^*$ and $\sigma_2$ is a Pauli matrix. The Coulomb potential $\alpha/|\bm{x}-\bm{y}|$ from photon exchange has infinite range. The potentials from the exchange of $W^\pm$ and $Z^0$ have ranges of order $1/m_W$.

Since low-energy photons can be radiated from the charged winos, the electromagnetic field should be included explicitly in the effective field theory. This can be accomplished by replacing the derivatives acting on the charged-wino fields $\eta$ and $\xi$ in eq.~\eqref{eq:kineticLHMN} by electromagnetic covariant derivatives, omitting the Coulomb potential $\alpha/|\bm{x}-\bm{y}|$ in the interaction term in eq.~\eqref{eq:LintHMN}, and adding the kinetic term $-\frac14 F_{\mu\nu} F^{\mu\nu}$  for the electromagnetic field. We refer to the resulting nonrelativistic effective field theory as NREFT.

Ladder diagrams from the exchange of electroweak gauge bosons between a pair of wimps can be summed to all orders in NREFT by solving a Schr\"odinger equation. The coupled-channel radial Schr\"odinger equation for S-wave scattering in the spin-singlet channel is
\begin{equation}
\left[ -\frac{1}{M} \begin{pmatrix} 1~  & ~0 \\ 0~ & ~1 \end{pmatrix}\left( \frac{d\ }{dr} \right)^2
+ 2\delta \begin{pmatrix} 0~  & ~0 \\ 0~ & ~1 \end{pmatrix}
+\bm{V}(r) \right] r \binom{R_0(r)}{R_1(r)} = E\,  r \binom{R_0(r)}{R_1(r)},
\label{eq:radialSchrEq}
\end{equation}
where $R_0(r)$ and $R_1(r)$ are the radial wavefunctions for a pair of neutral winos and a pair of charged winos, respectively. The $2 \times 2$ matrix of potentials is
\begin{equation}
\bm{V}(r) = -  
\alpha_2 \begin{pmatrix}                0        & \sqrt{2}\, e^{-m_Wr}/r \\ 
                         \sqrt{2}\, e^{-m_Wr}/r  &  c_w^2\, e^{-m_Zr}/r   \end{pmatrix}
-  \alpha \begin{pmatrix} 0~  & ~0 \\ 0~ & ~1/r \end{pmatrix},
\label{eq:V-matrix}
\end{equation}
where $c_w = \cos \theta_w$. There is a continuum of positive energy eigenvalues $E$ that correspond to S-wave scattering states. There may also be discrete negative eigenvalues that correspond to bound states.

\subsection{Wino-wino scattering}
\label{sec:CrossSections}

For center-of-mass energy $E>0$, the coupled-channel radial Schr\"odinger equation in eq.~\eqref{eq:radialSchrEq} can be solved for the radial wavefunctions $R_0(r)$ and $R_1(r)$. For energy $E > 2 \delta$ above the charged-wino-pair threshold, the asymptotic solutions for $R_0(r)$ and $R_1(r)$ as $r \to \infty$ determine a unitary and symmetric $2 \times 2$ S-matrix $\bm{S}(E)$. The $2\times2$ T-matrix $\bm{T}(E)$ is defined by
\begin{equation}
\bm{S}(E) =
\mathds{1} + i \, \bm{T}(E),
\label{eq:S-high}
\end{equation}
where $\mathds{1}$ is the $2\times 2$ unit matrix. The T-matrix satisfies the unitarity equation
\begin{equation}
2 \, \textrm{Im}\, \bm{T}(E) = \bm{T} ^\dagger(E)\, \bm{T}(E) .
\label{eq:T-unitarity}
\end{equation}
This implies that the imaginary parts of the components of the inverse of $\bm{T}$ are particularly simple:
\begin{equation}
2 \, \textrm{Im}\,\bm{T}^{-1}(E) = - \mathds{1}.
\label{eq:T-unitarity2}
\end{equation}
For energy in the range $0 < E < 2 \delta$ below the charged-wino-pair threshold, the only nonzero entry in the T-matrix is $T_{00}$. Unitarity implies that it can be expressed in the form
\begin{equation}
T_{00}(E) =\frac{2}{\cot \delta_0(E) -i},
\label{eq:T00}
\end{equation}
where $\delta_0(E)$ is the S-wave phase shift for neutral winos, which is real in this energy range. 

We denote the contribution to the cross section for elastic scattering from channel $i$ to channel $j$ at energy $E$ from scattering in the S-wave spin-singlet channel by $\sigma_{i \to j}(E)$. The expressions for these cross sections in terms of the T-matrix elements $T_{ji}$ are
\begin{subequations}
\begin{eqnarray}
\sigma_{0 \to j}(E) &=& \frac{2\pi}{M^2 v_0(E)^2} \big| T_{j0}(E) \big|^2,
\label{eq:sig0j-T}
\\
\sigma_{1\to j}(E) &=&   \frac{\pi}{M^2 v_1(E)^2}\big| T_{j1}(E) \big|^2,
\label{eq:sig1j-T}
\end{eqnarray}
\label{eq:sigij-T}%
\end{subequations}
where $v_0(E)$ and $v_1(E)$ are the wino velocities  in the center-of-mass frame for a neutral-wino pair and a charged-wino pair with total energy $E$:
\begin{subequations}
\label{eq:v0,1-E}
\begin{eqnarray}
v_0(E) &=&  \sqrt{E/M},
\label{eq:v0-E}
 \\
v_1(E) &=&  \sqrt{(E-2 \delta)/M}.
\label{eq:v1-E}
\end{eqnarray}
\end{subequations}
The cross sections in eqs.~\eqref{eq:sigij-T} have been averaged over initial spins and summed over final spins.

\begin{figure}[t]
\centering
\includegraphics[width=0.8\linewidth]{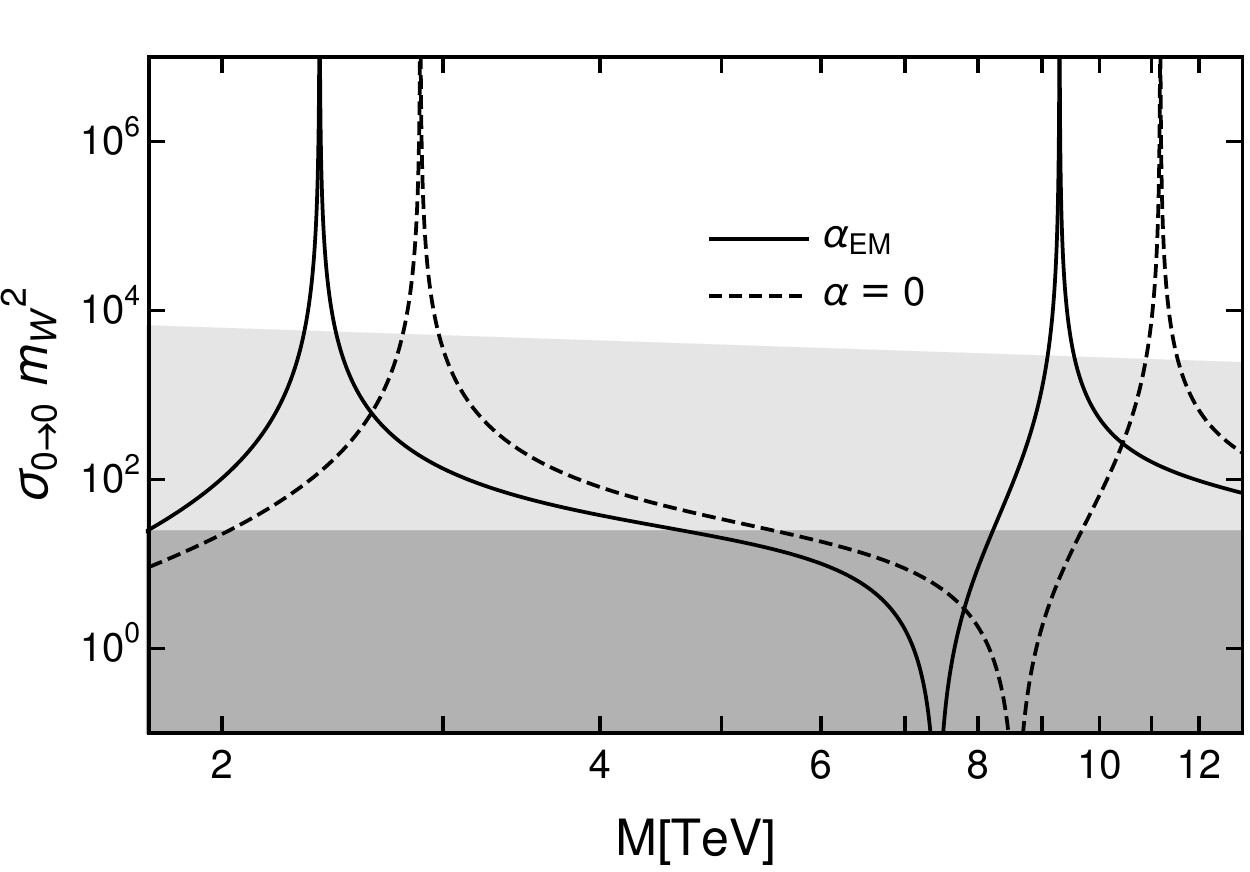}
\caption{The neutral-wino elastic cross section $\sigma_{0 \to 0}$ at zero energy as a function of the wino mass $M$. The cross section is shown for $\alpha=1/137$  (solid curve) and for $\alpha = 0$ (dashed curve).  If $\sigma_{0 \to 0}$ is above the  darker shaded region ($\sigma_{0 \to 0} < 8 \pi/m_W^2$), a zero-range effective field theory for winos is applicable. If $\sigma_{0 \to 0}$ is above the lighter shaded region ($\sigma_{0 \to 0} < 4 \pi/M \delta$), a zero-range effective field theory for neutral winos only is applicable.}
\label{fig:sigma00vsMcode}
\end{figure}

The neutral-wino elastic cross section $\sigma_{0 \to 0}(E=0)$ at zero energy for $\delta=170$~MeV is shown as a function of the wino mass $M$ in figure~\ref{fig:sigma00vsMcode}. The cross section diverges at critical values of $M$. The first critical mass is $M_*=2.39$~TeV and the second is 9.23~TeV. The divergence indicates that there is a zero-energy resonance at the neutral-wino-pair threshold. The {\it S-wave unitarity bounds} for neutral winos, which are identical spin-$\tfrac12$ particles, 
and  for charged winos, which are distinguishable spin-$\tfrac12$ particles, are
\begin{subequations}
\label{sigma-unitarity}
\begin{eqnarray}
\sigma_{0\to 0}(E)  &\le& \frac{8 \pi}{ME},
\label{sigma-unitarity0}
 \\
\sigma_{1\to 1}(E)  &\le& \frac{4 \pi}{M(E- 2 \delta)}.
\label{sigma-unitarity1}
\end{eqnarray}
\end{subequations}
At a critical mass where there is an S-wave resonance at the neutral-wino-pair threshold, 
the neutral-wino elastic cross section $\sigma_{0\to 0}(E)$
saturates the unitarity bound in the limit $E\to 0$. We therefore refer  to such a mass as a {\it unitarity mass} or simply as {\it unitarity}.

The neutral-wino elastic cross section at zero energy depends sensitively on the strength $\alpha$  of the Coulomb potential. The Coulomb potential can be  turned off by setting $\alpha = 0$ in the  potential matrix in eq.~\eqref{eq:V-matrix}. The resulting cross section for neutral winos with zero energy is compared to the cross section at the physical value $\alpha= 1/137$ in figure~\ref{fig:sigma00vsMcode}. If the Coulomb potential is turned off by setting $\alpha = 0$, the shape of the curve is almost the same, but the first two unitarity masses are shifted  upward by about 20\% to 2.88~TeV and 11.18~TeV. For $\alpha$ in the range between 0 and 1/137, the unitarity masses are accurately parameterized by expressions linear in $\alpha$.

\begin{figure}[t]
\centering
\includegraphics[width=0.8\linewidth]{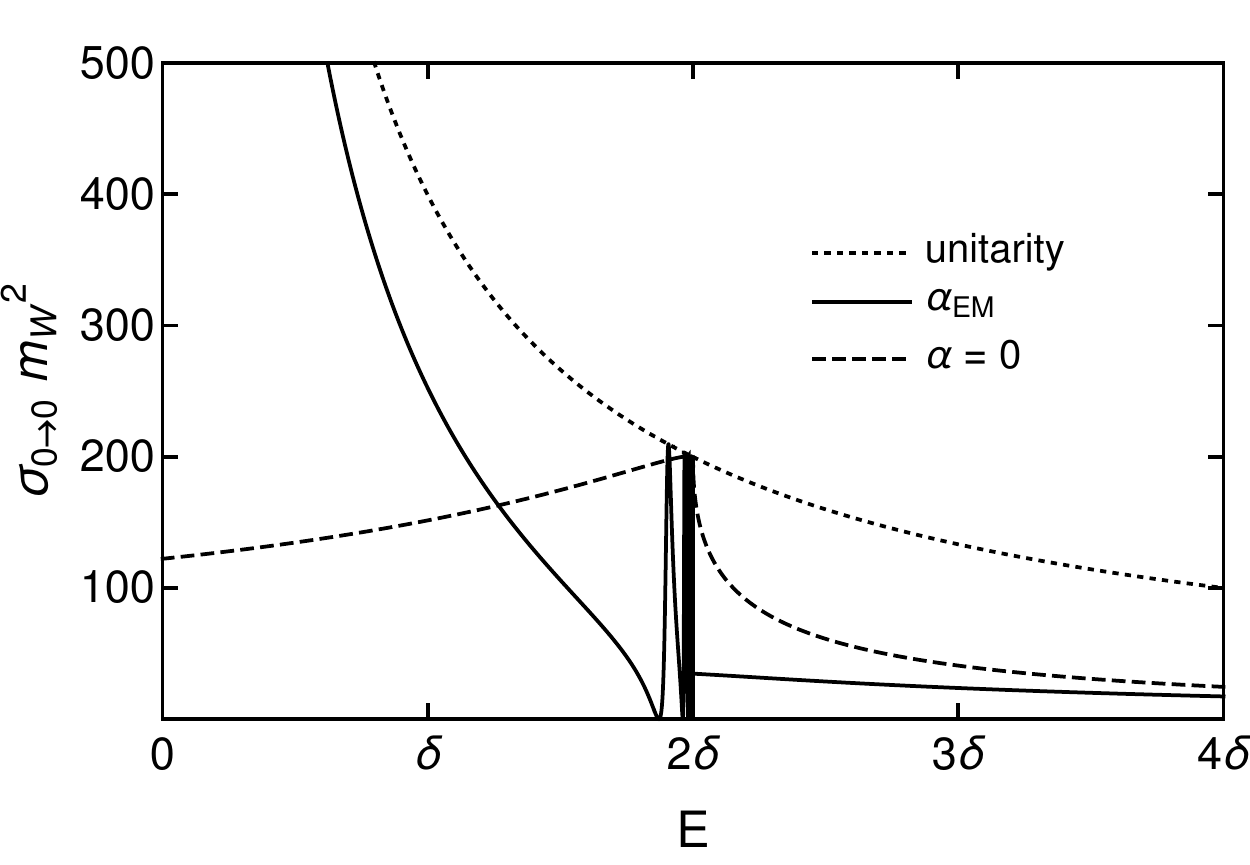}
\caption{The neutral-wino elastic cross section $\sigma_{0 \to 0}$  as a function of the energy $E$. The cross section for $M=2.39$~TeV is shown for $\alpha = 1/137$ (solid curve) and for $\alpha = 0$ (dashed curve). The S-wave unitarity bound is shown as a dotted curve.
}
\label{fig:sigma00-NREFT}
\end{figure}

The neutral-wino elastic cross section $\sigma_{0 \to 0}(E)$ has the most dramatic energy dependence at a unitarity mass, such as $M_*=2.39$~TeV for which the cross section is shown in figure~\ref{fig:sigma00-NREFT}. As $E$ approaches 0, the cross section approaches the unitarity bound in eq.~\eqref{sigma-unitarity0} from below, saturating the bound in the limit. Just below the charged-wino-pair threshold $2 \delta$, the cross section with the Coulomb potential has a sequence of narrow resonances whose peaks saturate the unitarity bound. The resonances can be interpreted as bound states in the Coulomb potential for  the charged-wino pair $w^+w^-$. If the Coulomb potential is turned off by setting $\alpha = 0$ while keeping $M$ fixed at 2.39~TeV, the cross section is finite at $E=0$ and the resonances disappear, as illustrated in figure~\ref{fig:sigma00-NREFT}. As $E$ increases from 0, the cross section increases monotonically until the threshold $2 \delta$, where it has a kink, and it then decreases as $E$ increases further.

Neutral winos with energies well below the charged-wino-pair threshold $2\delta$ have short-range interactions, because the Coulomb interaction enters only through virtual charged winos. The short-range interactions guarantee that $v_0(E) \cot \delta_0(E)$, where $\delta_0(E)$ is the S-wave phase shift defined in eq.~\eqref{eq:T00}, has an expansion in integer powers of $E$. This implies that  $v_0(E)/T_{00}(E)$ can be expanded in powers of the  relative momentum $p = \sqrt{ME}$:
\begin{equation}
\frac{2 Mv_0(E)}{T_{00}(E)} =
-\frac{1}{a_0}  -ip+ \frac12 r_0\,p^2 + \frac18 s_0\,p^4 + {\cal O}(p^6).
\label{eq:T00NRinv}
\end{equation}
The only odd power of $p$ is the imaginary term $-ip$.  The coefficients of the even powers of $p$ are real.
The leading term in the expansion defines the  {\it neutral-wino S-wave scattering length} $a_0$. It diverges at a unitarity mass where the zero-energy cross section $\sigma_{0 \to 0}(E=0)$ is infinite. The coefficients of $p^2$ and $p^4$ define the {\it effective range}  $r_0$ and a {\it shape parameter} $s_0$.

\begin{figure}[t]
\centering
\includegraphics[width=0.8\linewidth]{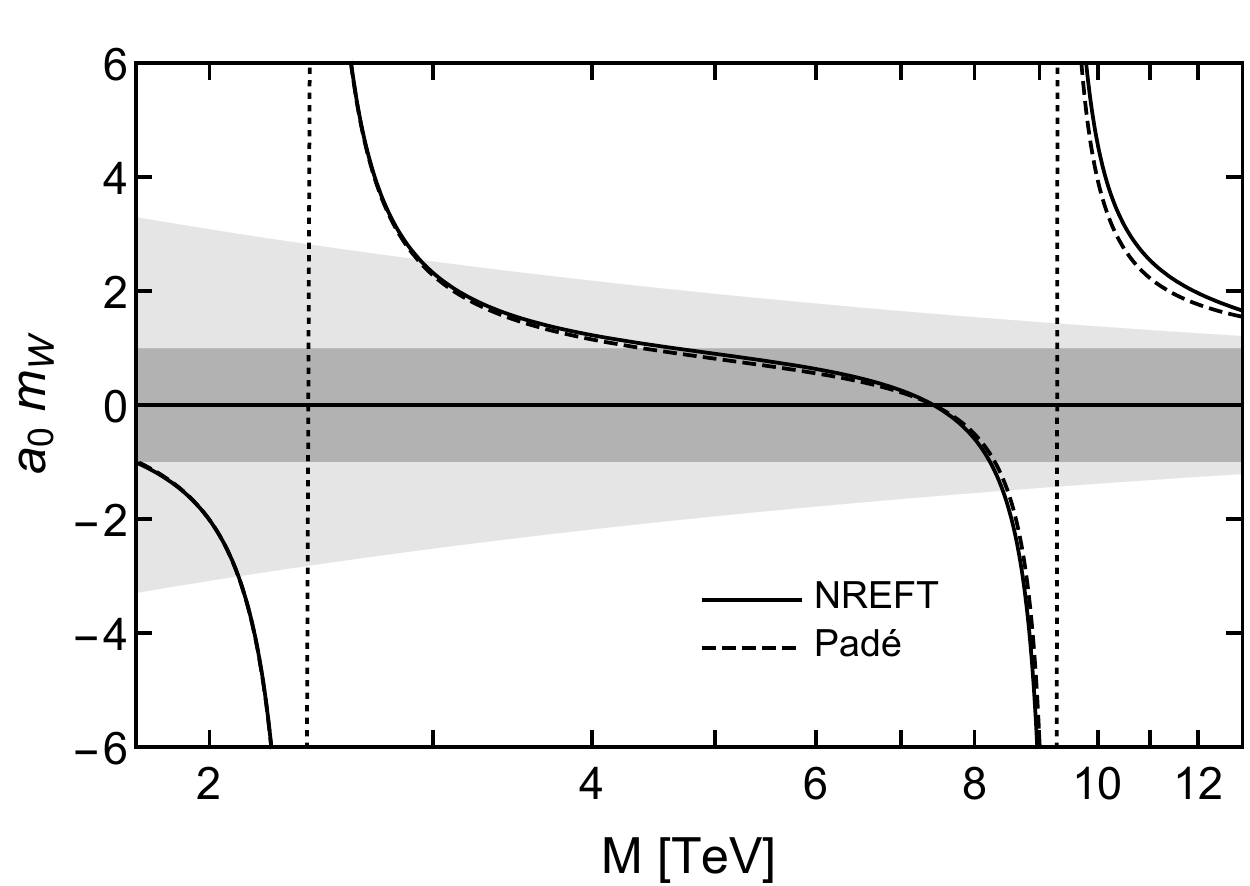}
\caption{The neutral-wino scattering length $a_0$ as a function of the wino mass $M$ (solid curve). The dashed curve is the Pad\'e approximant  given in eq.~\eqref{eq:a0Pade}. The vertical dotted lines indicate the first and second unitarity masses $M_* = 2.39$~TeV and 9.23~TeV.  If $a_0$ is outside the  darker shaded region ($|a_0| <1/m_W$), a zero-range effective field theory for winos is applicable. If $a_0$ is outside the lighter shaded region ($|a_0| <1/\sqrt{2 M \delta}$), a zero-range effective field theory for neutral winos only is applicable.
}
\label{fig:Real_a_vsM}
\end{figure}

The coefficients in the expansion in eq.~\eqref{eq:T00NRinv} are functions of $M$ and $\delta$ that can be determined numerically by solving the Schr\"odinger equation. The scattering length $a_0$ for $\delta = 170$~MeV and $\alpha=1/137$ is shown as a function of the mass $M$ in figure~\ref{fig:Real_a_vsM}. The dependence of $a_0$ on $M$ can be fit surprisingly well by a Pad\'e approximant in $M$ of order [2,2] whose poles match the first and second resonances at $M_*=2.39$~TeV and $M_*' = 9.23$~TeV and whose zeros match the first and second zero crossings at $M_0 = 0.0027$~TeV and $M_0' = 7.39$~TeV. The only adjustable parameter is an overall prefactor. We can improve the fit near the resonance at $M_*$ significantly by fitting $M_0$ as well as the prefactor. The resulting fit is
\begin{equation}
\label{eq:a0Pade}
a_0(M) = \frac{0.905}{m_W} \, \frac{(M-M_0)(M-M_0')}{(M-M_*)(M-M_*')},
\end{equation}
where $M_0 = 0.775$~TeV. The difference between the Pad\'e approximant and  $a_0(M)$ in figure~\ref{fig:Real_a_vsM} is less than 5\% if $M$ is in the range $|M - M_*| < 1.0$~TeV. If we use the correct value $M_0 = 0.0027$~TeV for the first zero crossing, the prefactor for the best fit is 0.593, but the difference between the Pad\'e approximant and  $a_0(M)$ is less than  5\% only if $M$ is in the much narrower range $|M - M_*| < 0.2$~TeV.

For NREFT with the Coulomb potential turned off by setting $\alpha = 0$, the scattering length $a_0(M)$ can be accurately approximated by an expression like that in eq.~\eqref{eq:a0Pade} but with different parameters. The first and second resonances are at $M_* = 2.88$~TeV and $M_*' = 11.18$~TeV. The first and second zero crossings are at $M_0 = 0.058$~TeV and $M_0' = 8.59$~TeV. We can improve the fit near the resonance at $M_*$ by fitting $M_0$ as well as the prefactor. The best fit is obtained with $M_0 = 0.87$~TeV and a  prefactor of $0.995/m_W$. The relatively small difference between the parameters for $\alpha = 1/137$ and the parameters for $\alpha = 0$ suggests that the difference may be perturbative in $\alpha$.

If $M$ is near the unitarity mass $M_*$, the coefficients of the positive powers of $p^2$ in the range expansion in eq.~\eqref{eq:T00NRinv}
can be expanded  in powers of $M - M_*$ or, alternatively, in powers of the inverse scattering length $\gamma_0=1/a_0$. The expansions in $\gamma_0$ are accurate only in a narrow range of $M - M_*$, because the coefficients $r_0$ and $s_0$ diverge at the zero-crossings of $a_0(M)$. For $\delta = 170$~MeV and $\alpha=1/137$, the effective range, its derivative with respect to the inverse scattering length, and the shape parameter at the unitarity mass $M_* = 2.39$~TeV are
\begin{subequations}
\begin{eqnarray}
r_0(M_*) &=& -1.653/\Delta_*,
\label{eq:r0*EM}
\\
\frac{dr_0}{d\gamma_0}(M_*) &=& 0.806/\Delta_*^2,
\label{eq:dr0*EM}
\\
s_0(M_*) &=& - 2.653/\Delta_*^3,
\label{eq:s0*EM}
\end{eqnarray}
\label{eq:r0,dr0,s0*EM}%
\end{subequations}
where  $\Delta_*  = \sqrt{2 M_* \delta} = 28.5$~GeV. The absolute values of the coefficients are order 1, indicating that $\Delta_*$ is an appropriate momentum scale. If the Coulomb potential is turned off by setting $\alpha = 0$, the right sides of eqs.~\eqref{eq:r0*EM}, \eqref{eq:dr0*EM}, and \eqref{eq:s0*EM} are changed to $-1.224/\Delta$, $ 0.284/\Delta^2$, and $- 1.878/\Delta^3,$ where $\Delta  = \sqrt{2 M \delta} = 28.5$~GeV. Thus the coefficients in the range expansion are somewhat sensitive to $\alpha$.

\begin{figure}[t]
\centering
\includegraphics[width=0.48\linewidth]{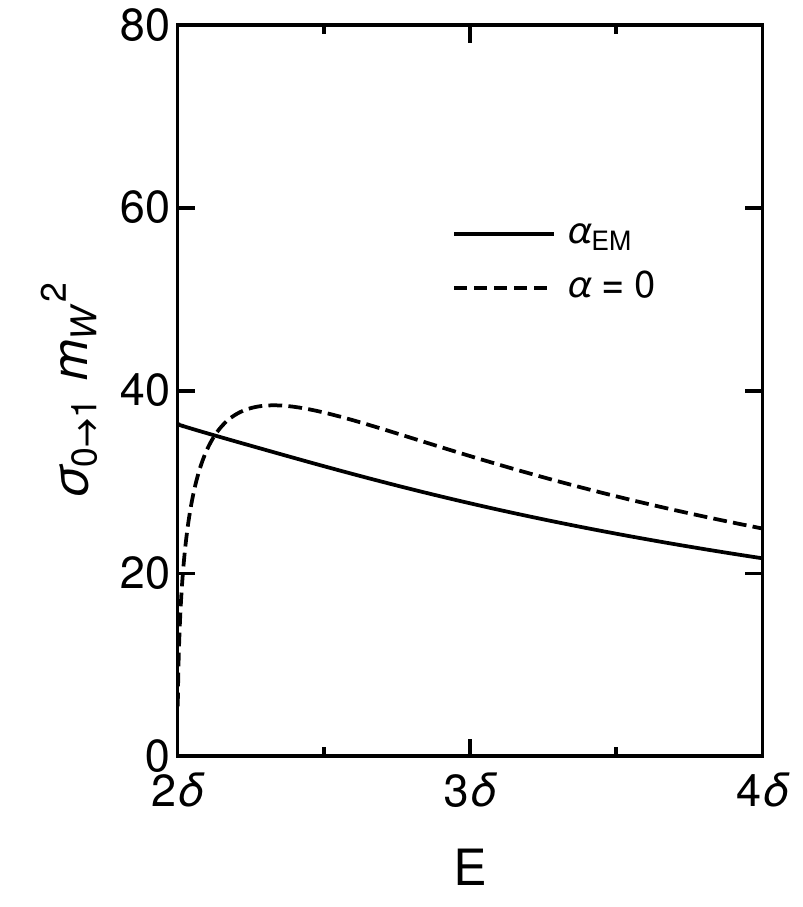}
~
\includegraphics[width=0.48\linewidth]{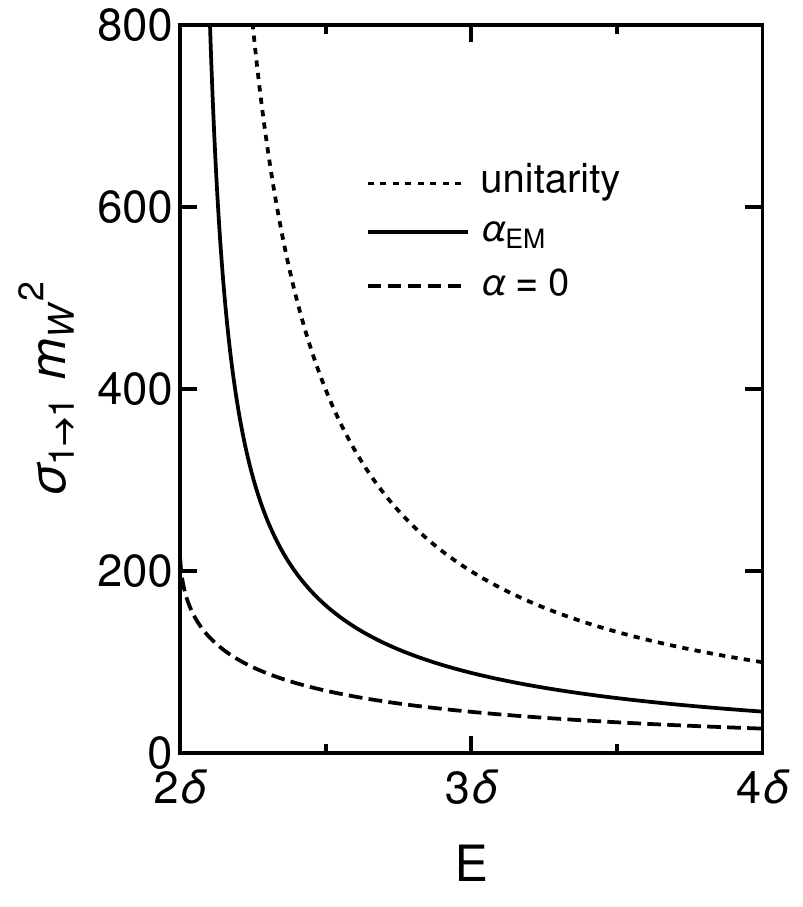}
\caption{The  neutral-to-charged transition cross section $\sigma_{0 \to 1}$  (left panel) and the charged-wino elastic cross section $\sigma_{1 \to 1}$ (right panel) as functions of the energy $E$. The cross sections for $M=2.39$~TeV are shown for $\alpha = 1/137$ (solid curves)  and for $\alpha = 0$ (dashed curves). In the right panel, the dotted curve is the S-wave unitarity bound in eq.~\eqref{sigma-unitarity1}.
}
\label{fig:sigma01,11-NREFT}
\end{figure}

The energy dependence of the neutral-to-charged transition cross section $\sigma_{0 \to 1}(E)$ at the unitarity mass $M_*=2.39$~TeV is illustrated in the left panel of figure~\ref{fig:sigma01,11-NREFT}. As $E$ decreases to the threshold $2 \delta$, the cross section increases mononotically to a finite maximum. The cross section with $\alpha =0$ is also shown in figure~\ref{fig:sigma01,11-NREFT}. As $E$ decreases to $2 \delta$, the cross section with $\alpha =0$ increases  to a maximum near $E=2.33\,\delta$, and it then decreases to zero because of a phase space factor proportional to $v_1(E)$. The nonzero cross section at the threshold for $\alpha=1/137$ is due to a Sommerfeld factor for Coulomb rescattering of the final-state $w^+$ and $w^-$.

The energy dependence of the charged-wino elastic cross section $\sigma_{1 \to 1}(E)$ at the unitarity mass $M_*=2.39$~TeV is illustrated in the right panel of figure~\ref{fig:sigma01,11-NREFT}. As $E$ decreases to the threshold $2 \delta$, the cross section increases to infinity. The cross section with $\alpha =0$ is also shown in figure~\ref{fig:sigma01,11-NREFT}. As $E$ decreases to  $2 \delta$, the cross section increases mononotically to a finite maximum. The difference between the cross sections for $\alpha=1/137$ and $\alpha = 0$ is due to a Sommerfeld factor for Coulomb rescattering of  $w^+$ and $w^-$.

\subsection{Wino-pair bound state}
\label{sec:boundstate}

\begin{figure}[t]
\centering
\includegraphics[width=0.8\linewidth]{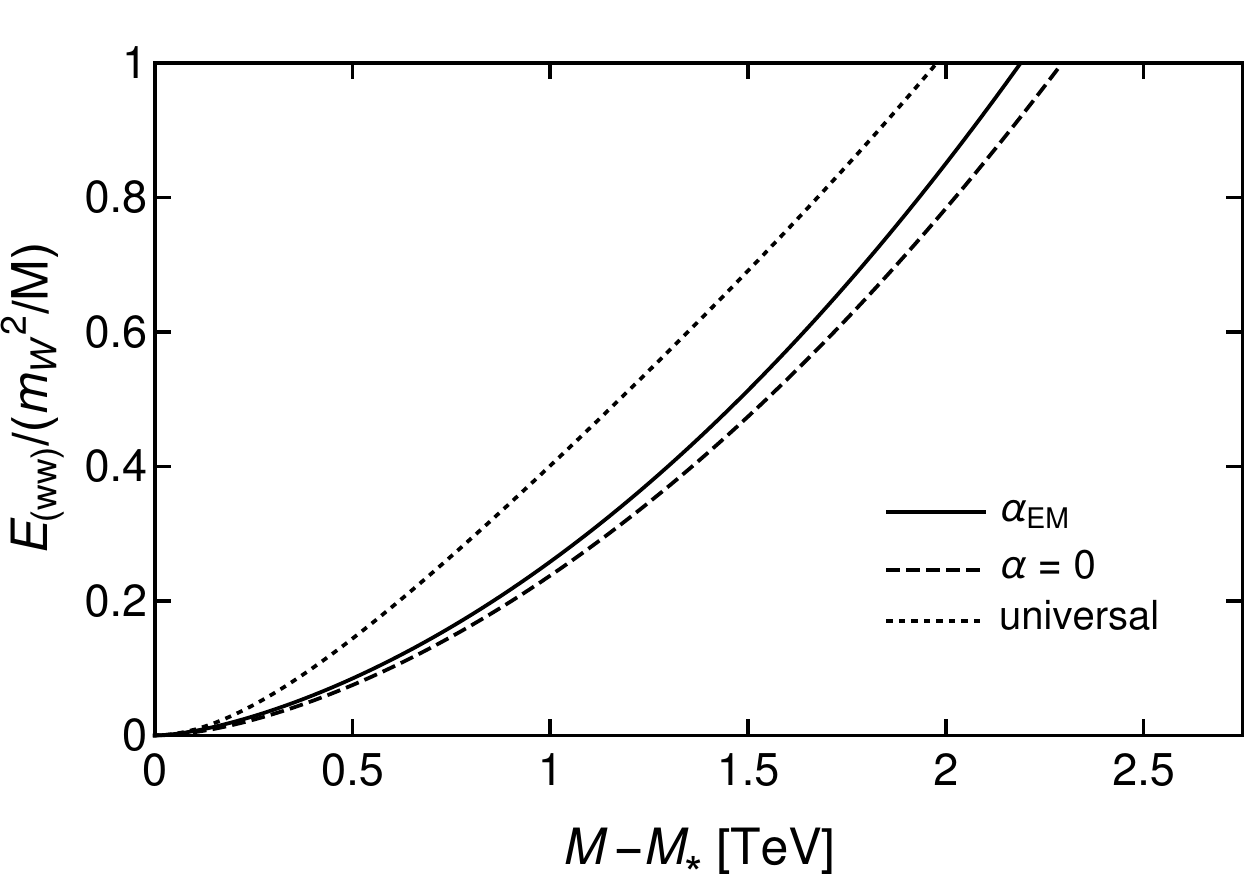}
\caption{The binding energy $E_{(ww)}$  of the wino-pair bound state as a function of the difference between the wino mass $M$ and the unitarity mass $M_*$. The binding energy is shown for $\alpha = 1/137$ (solid curve), for which $M_*=2.39$~TeV, and for $\alpha = 0$ (dashed curve), for which $M_*=2.88$~TeV. The universal approximation in eq.~\eqref{eq:Eww-largea}, with the Pad\'e approximant for $a_0(M)$ in eq.~\eqref{eq:a0Pade}, is  shown as a dotted curve. 
}
\label{fig:bindingenergy}
\end{figure}

If the neutral wino mass $M$ is larger than the unitarity mass where the neutral-wino scattering length $a_0(M)$ diverges, the S-wave resonance is a bound state below the neutral-wino-pair threshold. The bound state is a superposition of a neutral-wino pair and a charged-wino pair, and we denote it by $(ww)$. The coupled-channel radial Schr\"odinger equation in eq.~\eqref{eq:radialSchrEq} has a negative eigenvalue $- E_{(ww)}$, where $E_{(ww)}$ is the binding energy. In figure~\ref{fig:bindingenergy}, the binding energy for $\delta=170$~MeV is shown as a function of $M-M_*$, where $M_*=2.39$~TeV is the unitarity mass. The binding energy goes to zero as $M$ approaches the unitarity mass from above. The Coulomb potential between the charged winos can be turned off by setting $\alpha=0$. This shifts the unitarity mass $M_*$ to 2.88~TeV. As illustrated in figure~\ref{fig:bindingenergy}, the effect of the Coulomb potential on the binding energy is a small effect if $E_{(ww)}$ is expressed as a function of $M-M_*$.

\subsection{Universal approximations}
\label{sec:Univesal}

Particles with short-range interactions that produce an S-wave resonance sufficiently close to their scattering threshold have universal low-energy behavior that is completely determined by their S-wave scattering length \cite{Braaten:2004rn}. The scattering length must be large compared to the range $\ell$ set by the interactions. The universal behavior appears at energies that are small compared to $1/M\ell^2$. For neutral winos, the appropriate range $\ell$ is the maximum of the range $1/m_W$ of the weak interactions and the length scale $(2 M \delta)^{-1/2}$ associated with transitions to a charged-wino pair. The corresponding momentum scale is $\Delta = (2 M \delta)^{1/2}$. For $\delta = 170$~MeV and the unitarity mass $M_*=2.39$~TeV, the transition momentum scale is $\Delta_* = 28.5$~GeV. The neutral winos have universal behavior when $|a_0(M)| \gg 1/\Delta$, and the universal region of the energy $E$  is $|E| \ll 2\delta$. In figure~\ref{fig:Real_a_vsM}, the universal region of $M$ is when $a_0$ is well outside the lighter shaded region. 

\begin{figure}[t]
\centering
\includegraphics[width=0.8\linewidth]{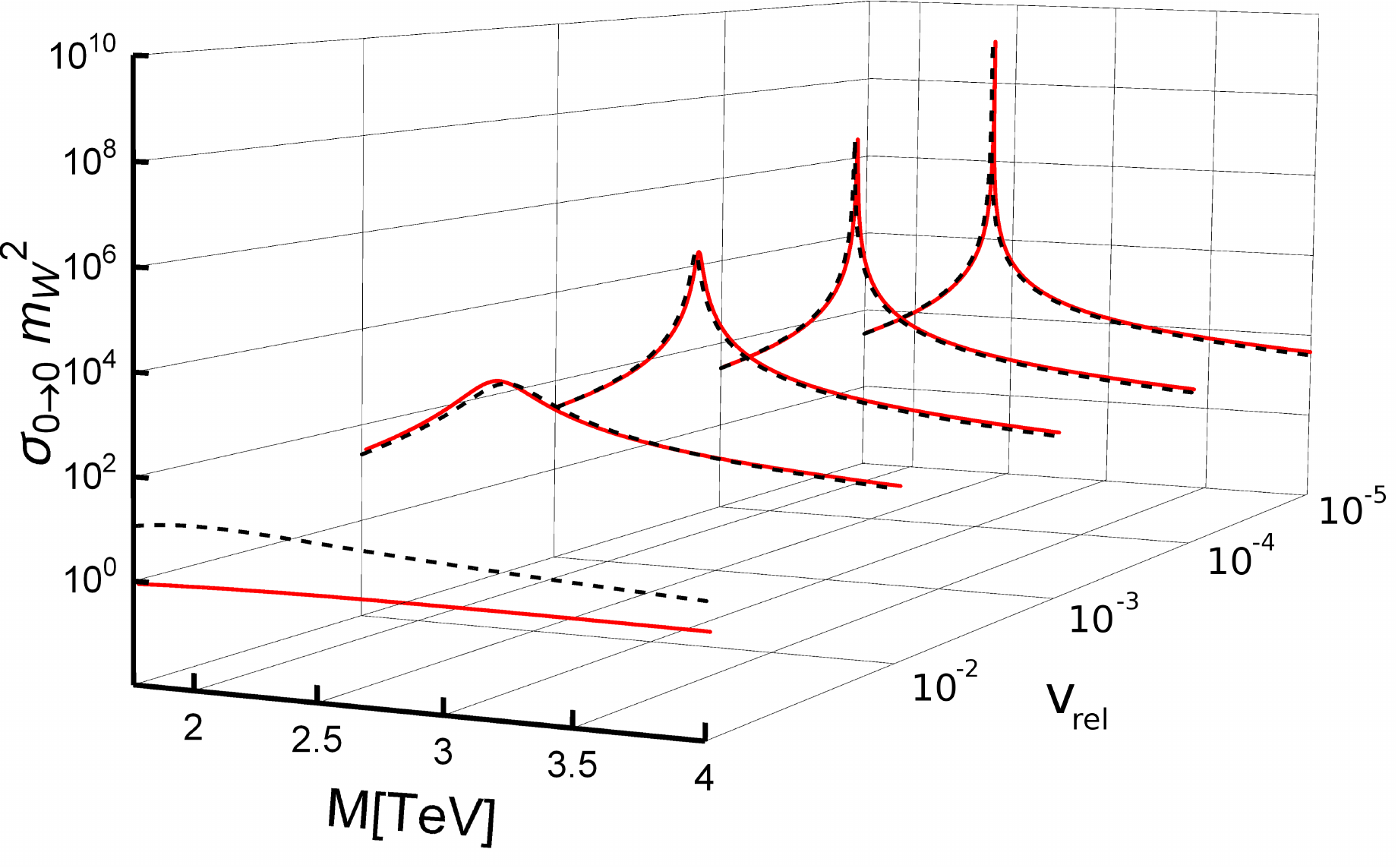}
\caption{The neutral-wino elastic cross section $\sigma_{0 \to 0}$ at fixed relative velocity $v_\text{rel}=2v_0$ as a function of the mass $M$.  The relative velocities $v_\text{rel}$  are $10^{-1}$, $10^{-2}$, $10^{-3}$, $10^{-4}$, and $10^{-5}$, decreasing into the page. As $v_\text{rel}$ decreases, the universal cross section in eq.~\eqref{eq:sigma00-largea} (dashed curves) provides an increasingly accurate approximation to the cross section from solving the Schr\"odinger equation (solid curves) for $M$ near the resonance.
}
\label{fig:sigma010-largea}
\end{figure}

In the universal region  of energy $E \ll 2\delta$, the T-matrix element for neutral-wino elastic scattering in  eq.~\eqref{eq:T00} can be approximated as
\begin{equation}
T_{00}(E) = \frac{-2 \sqrt{ME}}{1/a_0 + i \sqrt{ME}}.
\label{eq:T00-largea}
\end{equation}
The cross section in this region is completely determined by the single parameter $a_0$:
\begin{equation}
\sigma_{0 \to 0}(E) = \frac{8\pi}{1/a_0^2 +  ME}.
\label{eq:sigma00-largea}
\end{equation}
In figure~\ref{fig:sigma010-largea},  this universal approximation for $\sigma_{0 \to 0}(E)$ is compared to the cross section calculated numerically by solving the Schr\"odinger equation for several relative  velocities $v_\text{rel}$. The universal approximation is not valid for $v_\text{rel} = 10^{-1}$, but it provides an increasingly accurate approximation to the cross section near the resonance as $v_\text{rel}$ decreases.

If the scattering length is large compared to the range set by the interactions and also positive, the S-wave bound state close to threshold has universal properties that are determined only by the scattering length. The binding energy $E_{(w w)}$ can be determined from the pole in the energy $E$ in the expression for the T-matrix element in eq.~\eqref{eq:T00-largea}:
\begin{equation}
E_{(w w)} = 1/(M a_0^2).
\label{eq:Eww-largea}
\end{equation}
In figure~\ref{fig:bindingenergy}, this universal approximation for $E_{(w w)}$\, with the Pad\'e approximant for $a_0(M)$ in eq.~\eqref{eq:a0Pade}, is compared to the binding energy calculated numerically by solving the Schr\"odinger equation.

In the universal limit where $a_0(M)$ is much larger than $1/\Delta$, the wavefunction  of the wino-pair bound state $(ww)$ has a remarkable structure. The spacial extent of the wavefunction of the charged-wino pair is roughly the range $1/m_W$. The spacial extent of the wavefunction of the neutral-wino pair is roughly $a_0(M)$, which approaches $\infty$ as $M$ approaches $M_*$ from above. That the size of the bound state can be orders of magnitude larger than the range is a particularly remarkable universal aspect of nonrelativistic particles with short-range interactions and an S-wave resonance near threshold \cite{Braaten:2004rn}.

There have been several previous applications of the universal approximations to dark matter with resonant S-wave self-interactions \cite{Braaten:2013tza,Laha:2013gva,Laha:2015yoa}. Although these papers considered explicitly only a single scattering channel, the results of ref.~\cite{Braaten:2013tza,Laha:2013gva,Laha:2015yoa} are model independent. They apply to a multi-channel problem in which there is a small mass splitting $\delta$ between wimps provided the wimp mass $M$ is close enough to a critical value for an S-wave resonance at threshold that $|a|$ is large compared to  $1/\sqrt{2M\delta}$  as well as $1/m_W$ and provided the relative momentum of the wimps is small compared to  $\sqrt{2M\delta}$ as well as $m_W$. The additional conditions are more stringent if $\delta < m_W^2/(2M)$. If the wimp mass is about 1~TeV and if the mass splitting $\delta$ is about 100~MeV, the relative velocity must be much less than about 0.01. The more stringent restrictions on $a$ and on the relative momentum can be removed by developing a multi-channel ZREFT that includes explicitly all the wimps with small mass splittings of order $\delta$.

\subsection{Scattering without Coulomb potential at unitarity}
\label{sec:noCoulomb}

Low-energy winos with resonant S-wave interactions can be described by NREFT, but they can be described more simply by an effective field theory for nonrelativistic winos with zero-range  interactions and electromagnetic interactions. This effective field theory, which we call ZREFT, is introduced in section~\ref{sec:ZREFT}. The behavior of low-energy winos with resonant S-wave interactions is simpler if electromagnetic interactions are turned off by setting $\alpha = 0$, because then all interactions have a short range. Results given below from NREFT with $\alpha = 0$ at unitarity will be used in section~\ref{sec:ZREFTNLO} to verify that ZREFT is a systematically improvable effective field theory. There is a critical point in parameter space where S-wave interactions have the remarkable property of scale invariance at low energies. Results given below from NREFT  near this critical point, which is $\alpha = 0$ at unitarity in the limit $\delta \to 0$, will be explained in section~\ref{sec:ZREFTLO}  by a renormalization group fixed point for ZREFT.

As shown in figure~\ref{fig:sigma00vsMcode}, the first unitarity mass where the neutral-wino elastic cross section $\sigma_{0 \to 0}$ at zero energy diverges  is  $M_* = 2.88$~TeV for $\delta = 170$~MeV and $\alpha = 0$. The effective range, its derivative with respect to  the inverse scattering length $\gamma_0=1/a_0$, and the shape parameter at  unitarity are
\begin{subequations}
\begin{eqnarray}
r_0(M_*) &=& -0.693/ \Delta_*,
\label{eq:r0*}
\\
\frac{dr_0}{d\gamma_0}(M_*) &=& 0.496/\Delta_*^2,
\label{eq:dr0*}
\\
s_0(M_*) &=& - 0.989/\Delta_*^3,
\label{eq:s0*}
\end{eqnarray}
\label{eq:r0,dr0,s0*}%
\end{subequations}
where $\Delta_*  = \sqrt{2M_*\delta}=31.3$~GeV for $\alpha = 0$. These results will be used in section~\ref{sec:ZREFTNLO} to fit the parameters of ZREFT at NLO.

For $\alpha = 0$, the short-range interactions of both neutral and charged winos constrains the behavior of the T-matrix elements $T_{ij}(E)$ as the energy $E$ approaches the charged-wino-pair threshold from above. Nonzero limits as  $E \to 2 \delta$ can be obtained by pre-multiplying and post-multiplying $\bm{T}$ by  the inverse of the square root of the diagonal $2 \times 2$ matrix $\bm{v}(E)$ whose diagonal entries are the velocities defined in eq.~\eqref{eq:v0,1-E}:
\begin{equation}
\bm{v}(E) = 
\begin{pmatrix} v_0(E) & 0 \\ 0 & v_1(E) \end{pmatrix}.
\label{eq:vmatrix}
\end{equation}
The limits of the T-matrix elements as $E$ approaches $2 \delta$ from above define complex scattering lengths $a_{0 \to 0}$, $a_{0 \to 1}$, and $a_{1 \to 1}$:
\begin{equation}
\bm{v}(E)^{-1/2}\, \bm{T}(E)\,  \bm{v}(E)^{-1/2} \longrightarrow - 2M 
\begin{pmatrix} a_{0 \to 0} & a_{0 \to 1}
\\ a_{0 \to 1} & a_{1 \to 1} \end{pmatrix}.
\label{eq:aijdef}
\end{equation}
The imaginary parts of the diagonal entries $a_{0 \to 0}$ and $a_{1 \to 1}$ are required by unitarity to be negative. The scattering lengths  $a_{i \to j}(M)$ can be determined numerically as a function of $M$ by solving the Schr\"odinger equation. For $\delta = 170$~MeV and $\alpha = 0$, the complex scattering lengths at the unitarity mass $M_*=2.88$~TeV are determined to be
\begin{subequations}
\begin{eqnarray}
a_{0 \to 0}(M_*) &=& ( 0.483 - 0.629 \, i)/ \Delta_*,
\label{eq:a00NR*}
\\
a_{0 \to 1}(M_*) &=& ( 0.424 - 0.553 \, i)/ \Delta_*,
\label{eq:a01NR*}
\\
a_{1 \to 1}(M_*) &=& ( 0.982 - 0.486 \, i)/ \Delta_*,
\label{eq:a11NR*}
\end{eqnarray}
\label{eq:aijNR*}%
\end{subequations}
where $\Delta_* =  31.3$~GeV. These results will be used in section~\ref{sec:ZREFTNLO} to test the accuracy of predictions of ZREFT at NLO.

The interactions of winos in the threshold region involve multiple length scales, including
\begin{itemize}
\item
the range $1/m_W$ of the weak  interactions,
\item
the length scale $1/\sqrt{2M\delta}$ associated with neutral-to-charged transitions,
\item
the absolute value $|a_0|$ of the neutral-wino scattering length,
\item
the Bohr radius  $1/(\alpha M)$ for the charged winos.
\end{itemize}
Some of the length scales can be eliminated by taking appropriate limits. The Bohr radius can be eliminated by setting $\alpha = 0$, which turns off the Coulomb potential. The neutral-wino scattering length  can be eliminated by tuning the wino mass $M$ to the unitarity value $M_*(\delta)$ where $a_0$ diverges. The length scale associated with neutral-to-charged transitions can be eliminated by taking the limit $\delta \to 0$. If all these limits are taken simultaneously, the range $1/m_W$ is the only remaining length scale for S-wave interactions. But S-wave interactions have a well-behaved zero-range limit as  $1/m_W \to 0$ if the strength $\alpha_2$ of the weak interaction potential is tuned to keep the scattering length $a_0$ fixed. Therefore if we set $\alpha=0$, tune $M$  to $M_*(\delta)$, and then take the limit $\delta \to 0$, S-wave interactions must become scale invariant in the low-energy limit. Scale-invariant interactions can be described by a zero-range effective field theory that is a renormalization-group fixed point.

The  unitarity mass  where the neutral-wino cross section at zero energy diverges depends on the mass splitting $\delta$ and on $\alpha$.  If we set $\alpha=0$ and decrease $\delta$ from 170~MeV to 0, the first unitarity mass $M_*(\delta)$ decreases smoothly from $M_*=2.88$~TeV to $M_{*0}=2.22$~TeV, approaching its limiting value as $\sqrt{\delta}$. It can be parametrized as
\begin{equation}
M_*(\delta) = 
M_{*0}  \left[ 1 + 0.787 \frac{\sqrt{2 M_{*0} \delta}}{m_W} 
+  0.216 \left(\frac{\sqrt{2 M_{*0} \delta}}{m_W}\right)^2 \right].
\label{eq:M*-delta}
\end{equation}
We have expressed it as an expansion in the  momentum scale $\sqrt{2 M_{*0} \delta}$ divided by the natural momentum scale $m_W$ set by the range of the potential.

We first consider low-energy neutral-wino elastic scattering with $\alpha = 0$. By solving the Schr\"odinger equation, the coefficients in the range expansion for the reciprocal of the T-matrix element $T_{00}(E)$  in eq.~\eqref{eq:T00NRinv}, such as the effective range $r_0$ and the shape parameter $s_0$, are found to be smooth functions of $\sqrt{\delta}$ as $\delta \to 0$. The coefficients in the range expansion can be made dimensionless by multiplying them by appropriate powers of $\Delta(\delta) = \sqrt{2 M \delta}$, which is the relative momentum for neutral winos at the charged-wino-pair threshold. At the unitarity mass $M_*(\delta)$, the momentum scale  is
\begin{equation}
\Delta_*(\delta) = 
\big[2M_*(\delta)\, \delta \big]^{1/2}.
\label{eq:Delta*}
\end{equation}
If the mass is kept at the unitarity value $M_*(\delta)$ as $\delta$ changes, the coefficients in the range expansion diverge in the limit $\delta \to 0$. At unitarity, each dimensionless coefficient in the range expansion can be fit accurately  over the range of $\delta$ from 0 to 170~MeV by a quadratic polynomial in $\Delta_*(\delta) /m_W$ with coefficients of order 1. At unitarity, the limiting behaviors of the effective range $r_0$ and the shape parameter $s_0$    as $\delta \to 0$ are found to be
\begin{subequations}
\begin{eqnarray}
 r_0\big(M_*(\delta),\delta \big) &\longrightarrow& - 1.552/ \Delta_*(\delta) ,
\label{eq:r0-delta}
\\
s_0\big(M_*(\delta),\delta \big) &\longrightarrow& - 1.552/ \Delta_*(\delta)^3.
\label{eq:s0-delta}
\end{eqnarray}
\label{eq:r0dr0s0-delta}%
\end{subequations}
The numerical coefficients seem to have the same value $-t^2$, where $t=1.246$. This qualitative feature will be explained in section~\ref{sec:ZREFTLO} in terms of a renormalization-group fixed point of ZREFT.

We next consider scattering with $\alpha = 0$ at the energy $E=2 \delta$  of the charged-wino-pair threshold. By solving the Schr\"odinger equation, the complex scattering lengths  $a_{i \to j}(M,\delta)$ defined in eqs.~\eqref{eq:aijdef} are found to be smooth functions of $\sqrt{\delta}$ as $\delta \to 0$. The complex scattering lengths can be made dimensionless by multiplying them by $\Delta(\delta) = \sqrt{2 M \delta}$. If the mass is kept at the unitarity value $M_*(\delta)$ for unitarity as $\delta$ changes, the complex scattering lengths diverge in the limit $\delta \to 0$. At unitarity, each complex scattering length can be fit accurately over the range of $\delta$ from 0 to 170~MeV by a quadratic  polynomial in $\Delta_*(\delta) /m_W$ with coefficients  of order 1. At unitarity, the limiting behaviors of the complex scattering lengths as $\delta \to 0$ are found to be
\begin{subequations}
\begin{eqnarray}
a_{0 \to 0}\big(M_*(\delta),\delta \big) &\longrightarrow& 
\big( 0.455  -0.291\, i \big)/\Delta_*(\delta) ,
\label{eq:Rea00-delta}
\\
a_{0 \to 1}\big(M_*(\delta),\delta \big) &\longrightarrow& 
\big(0.569  - 0.366\, i \big)/\Delta_*(\delta) ,
\label{eq:Rea01-delta}
\\
a_{1 \to 1}\big(M_*(\delta),\delta \big) &\longrightarrow & 
\big( 0.706   -0.455\, i \big)/ \Delta_*(\delta).
\label{eq:Rea11-delta}
\end{eqnarray}
\label{eq:Reaij-delta}%
\end{subequations}
As $\delta \to 0$, the ratios $\mathrm{Re}(a_{1 \to 1})/\mathrm{Re}(a_{0 \to 1})$ and $\mathrm{Re}(a_{0 \to 1})/\mathrm{Re}(a_{0 \to 0})$ of the real parts and the ratios $\mathrm{Im}(a_{1 \to 1})/\mathrm{Im}(a_{0 \to 1})$ and $\mathrm{Im}(a_{0 \to 1})/\mathrm{Im}(a_{0 \to 0})$ of the imaginary parts seem to approach the same limit  $t \approx 1.25$. The ratio $\mathrm{Im}(a_{1 \to 1})/\mathrm{Re}(a_{0 \to 0})$  seems to approach the limit $-1$. These ratios will be explained in section~\ref{sec:ZREFTLO} in terms of a renormalization-group fixed point of ZREFT.

\section{Zero-range effective field theory}
\label{sec:ZREFT}

If the wino mass $M$ is tuned to near a zero-energy resonance in neutral-wino-pair scattering, low-energy winos can be described  by a nonrelativistic effective field theory in which the winos interact through zero-range self-interactions and through their couplings to the electromagnetic field. We call this effective field theory {\it ZREFT}. If electromagnetic couplings are turned off by setting $\alpha = 0$, ZREFT can be defined as a systematically improvable effective field theory obtained through deformations of  a renormalization-group fixed point. In this section, we identify the scale-invariant theory associated with the  renormalization-group fixed point. We present a parametrization of the wino-pair  transition amplitudes that incorporates the deformations of the fixed point that can be used to systematically improve the accuracy of the effective field theory.

\subsection{Zero-range model}
\label{sec:ZRM}

\begin{figure}[t]
\centering
\includegraphics[width=0.25\linewidth]{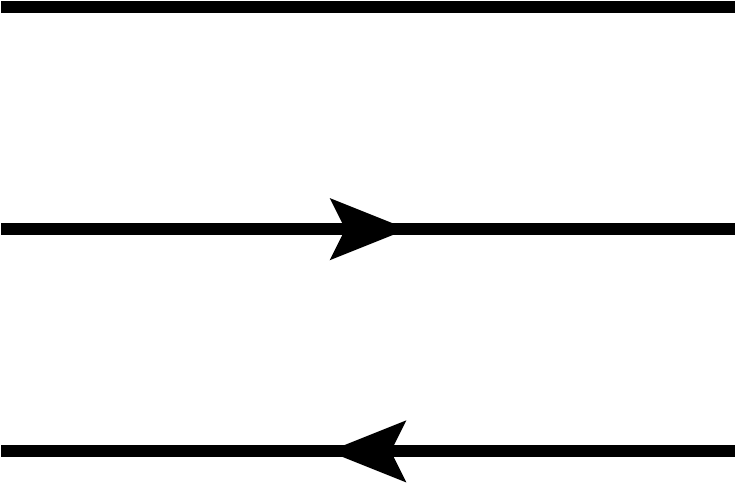}
\caption{The propagators for the neutral-wino field $w_0$ and the charged-wino fields $w_+$ and $w_-$. The Feynman rules are $i\delta^{ab}/[E - p^2/(2M) + i \epsilon]$  for the neutral-wino propagator and $i\delta^{ab}/[E - p^2/(2M) - \delta + i \epsilon]$  for the charged-wino propagators, where $a$ and $b$ are Pauli spinor indices.}
\label{fig:Propagators}
\end{figure}

The range of the weak interactions between two winos is $1/m_W$. Near a unitarity mass $M_*$, where there is an S-wave resonance at the neutral-wino-pair threshold, the neutral-wino scattering length $a_0$ is a dynamically generated length scale that can be much larger than $1/m_W$. For winos with energies small compared to $m_W^2/M$, the effects of the exchange of weak bosons can be mimicked by zero-range interactions. The effective field theory that describes these zero-range interactions must be nonperturbative in order to dynamically generate the length scale $a_0$.

\begin{figure}[t]
\centering
\includegraphics[width=0.4\linewidth]{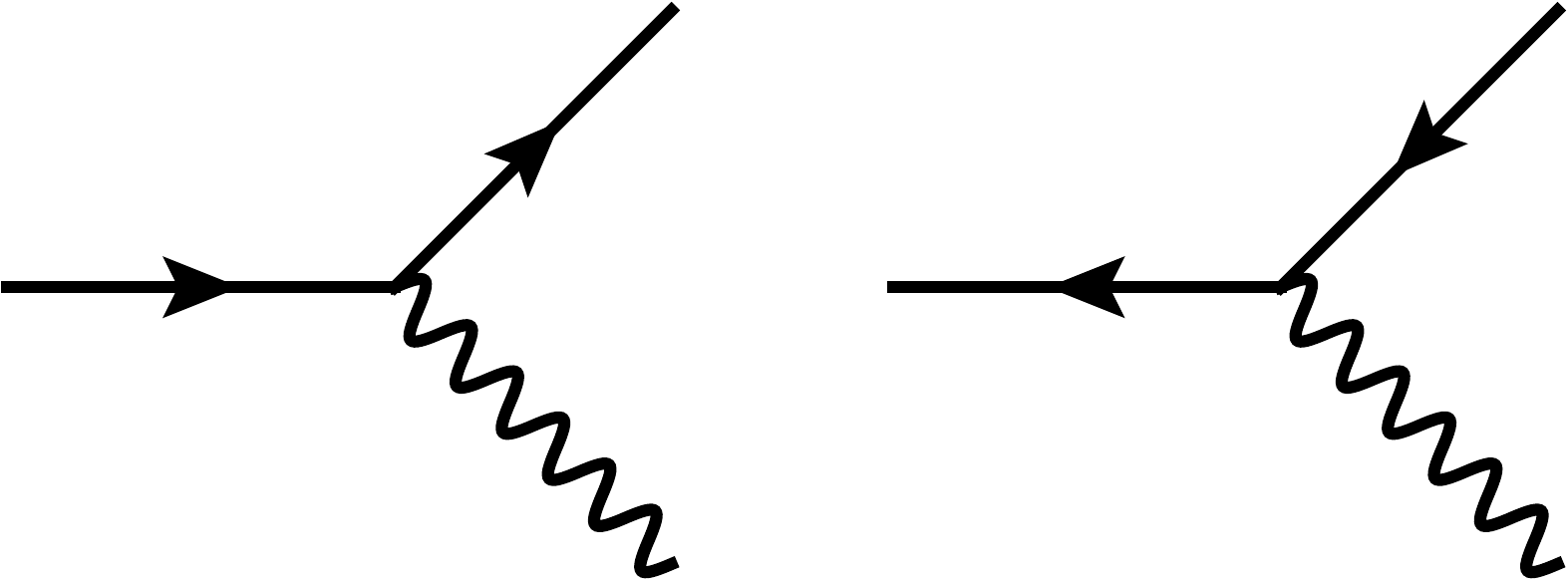}~
\includegraphics[width=0.4\linewidth]{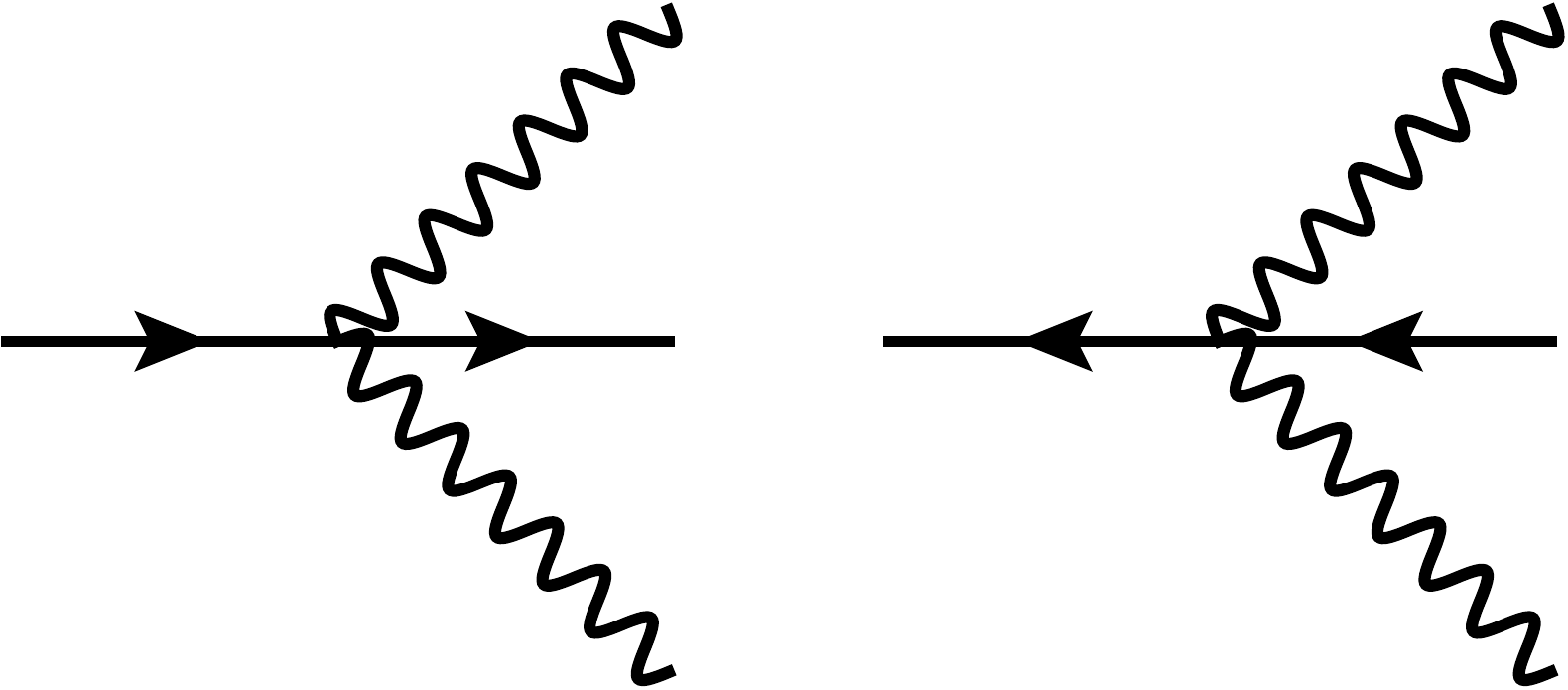}
\caption{The interaction vertices for emission of one or two photons from positively and negatively charged wino lines.}
\label{fig:EMvertices}
\end{figure}

The winos can be described by nonrelativistic two-component spinor fields. We denote the fields that annihilate $w^0$, $w^+$, and $w^-$ by $w_0$, $w_+$, and $w_-$, respectively. They can be identified with the fields $\zeta$, $\xi^\dagger$, and $\eta$ in NREFT, respectively. The kinetic terms  in the Lagrangian for ZREFT are
\begin{equation}
\mathcal{L}_{\rm kinetic} = w_0^\dagger\left(i\partial_0+\frac{\bm{\nabla}^2}{2M}\right) w_0  
+ \sum_\pm w_\pm^\dagger\left(iD_0+\frac{\bm{D}^2}{2M}-\delta\right) w_\pm +\frac12 \left(\bm{E}^2 - \bm{B}^2 \right).  
\label{eq:kineticL}
\end{equation}
The electromagnetic covariant derivatives are
\begin{eqnarray}
D_0 w_\pm = (\partial_0 \pm ieA_0)w_\pm,
\qquad
\bm{D} w_\pm = (\bm{\nabla} \mp ie\bm{A})w_\pm.
\end{eqnarray}
The neutral and charged winos have the same kinetic mass $M$, and the mass splitting $\delta$ is taken into account through the rest energy of the charged winos. The equal kinetic masses ensures Galilean invariance in the absence of electromagnetism. The propagators for the wino fields $w_0$, $w_+$, and $w_-$ are represented by solid lines with no arrow, a forward arrow, and a backward arrow, respectively, as illustrated in figure~\ref{fig:Propagators}. The charged winos have electromagnetic couplings to the photon field with the vertices illustrated in figure~\ref{fig:EMvertices}.

\begin{figure}[t]
\centering
\includegraphics[width=0.7\linewidth]{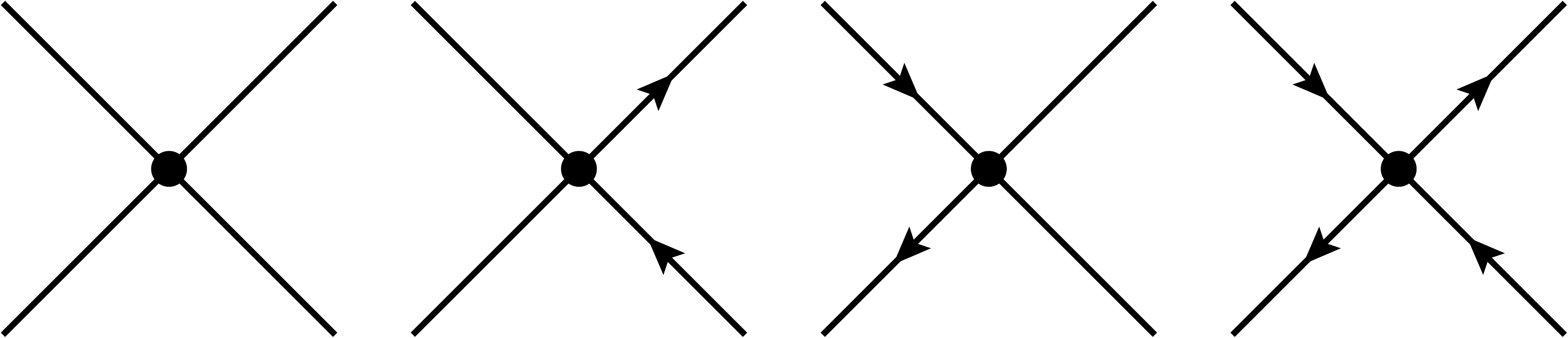}
\caption{The zero-range  interaction vertices for $w_0 w_0 \rightarrow w_0 w_0$, $w_0 w_0 \rightarrow w_+ w_-$, $w_+ w_- \rightarrow w_0 w_0$, and $w_+ w_- \rightarrow w_+ w_-$. The corresponding Feynman rules are $-i(\delta^{ac}\delta^{bd} - \delta^{ad} \delta^{bc})/2$ multiplied by $\lambda_{00}$, $\lambda_{01}$, $\lambda_{01}$, and $\lambda_{11}$, respectively, where $a$ and $b$ are Pauli spinor indices for the incoming lines while $c$ and $d$ are Pauli spinor indices for the outgoing lines.
}
\label{fig:LOcontacts}
\end{figure}

Since the neutral wino is a Majorana fermion, a pair of neutral winos can have an S-wave resonance at threshold only in the spin-singlet channel. That channel is coupled to the spin-singlet channel for charged winos. The Lagrangian for zero-range interactions in the spin-singlet channels can be expressed as
\begin{eqnarray}
\mathcal{L}_{\rm zero-range} &=& 
-\tfrac{1}{4} \lambda_{00} ( w_0^{c\dagger} w_0^{d\dagger} )
\tfrac12 ( \delta^{ac}\delta^{bd}- \delta^{ad}\delta^{bc}) ( w_0^a w_0^b )
\nonumber\\
&& 
-\tfrac{1}{2} \lambda_{01} (  w_+^{c\dagger} w_-^{d\dagger} )
\tfrac12 ( \delta^{ac}\delta^{bd}- \delta^{ad}\delta^{bc}) ( w_0^a w_0^b )
\nonumber\\
&& 
-\tfrac{1}{2} \lambda_{01} (  w_0^{c\dagger} w_0^{d\dagger} )
\tfrac12 ( \delta^{ac}\delta^{bd}- \delta^{ad}\delta^{bc}) ( w_+^a w_-^b )
\nonumber\\
&& 
- \lambda_{11} ( w_+^{c\dagger} w_-^{d\dagger} )
\tfrac12 ( \delta^{ac}\delta^{bd}- \delta^{ad}\delta^{bc})
( w_+^a w_-^b ),
\label{eq:ZRint-symmetric}
\end{eqnarray}
where $\lambda_{00}$, $\lambda_{01}$, and $\lambda_{11}$ are bare coupling constants. The factor $\frac12( \delta^{ac}\delta^{bd}- \delta^{ad}\delta^{bc})$ is the projector onto the spin-singlet channel. The Lagrangian can be written in a more compact form with implicit spinor indices:
\begin{eqnarray}
\mathcal{L}_{\rm zero-range} = 
\tfrac{1}{4}\lambda_{00}  ({w_0}^\dagger w_0)^2  
+\tfrac{1}{2} \lambda_{01} 
 \big[  (w_+^\dagger w_0) (w_-^\dagger w_0)
 + ( w_0^\dagger  w_+) (w_0^\dagger w_-) \big]  
 \nonumber\\
+ \tfrac12 \lambda_{11} 
 \big[  (w_+^\dagger w_+) (w_-^\dagger w_-)
 + (w_+^\dagger  w_-) (w_-^\dagger w_+) \big].
\label{eq:ZRint}
\end{eqnarray}
The zero-range  interaction vertices are illustrated in figure~\ref{fig:LOcontacts}.

\begin{figure}[t]
\centering
\includegraphics[width=0.98\linewidth]{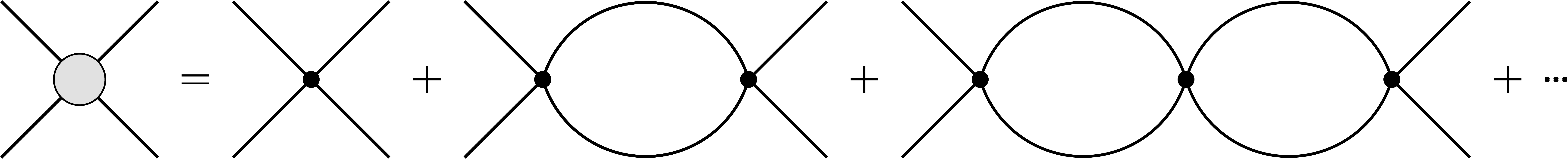}
\caption{Feynman diagrams for wino-wino scattering in the zero-range model without electromagnetism. 
Bubble diagrams must be summed to all orders.
The solid lines are winos, which can be either neutral or charged.}
\label{fig:ZREFTSum}
\end{figure}

We define the {\it zero-range model} by declaring its Lagrangian to be the sum of the kinetic terms in eqs.~\eqref{eq:kineticL} and the zero-range interaction terms in eq.~\eqref{eq:ZRint}. The zero-range interactions must be treated  nonperturbatively by summing bubble diagrams involving the vertices in figure~\ref{fig:LOcontacts} to all orders. For wino-wino scattering, the first few diagrams in the sum are shown in figure~\ref{fig:ZREFTSum}. The T-matrix elements for low-energy wino-wino scattering can be obtained from the amplitudes $\mathcal{A}_{ij}(E)$ for transitions among the two coupled channels. The Lippmann-Schwinger equations for the zero-range model are solved nonperturbatively in Appendix~\ref{app:LSeq}. The solution involves functions $\kappa_0(E)$ and $\kappa_1(E)$ of the complex energy $E$ with branch points at the neutral-wino-pair threshold and the charged-wino-pair threshold, respectively:
\begin{subequations}
\begin{eqnarray}
\kappa_0(E) &=& \sqrt{-ME-i\varepsilon},
\label{eq:kappa0}
\\
\kappa_1(E) &=& \sqrt{-M(E-2\delta)-i\varepsilon},
\label{eq:kappa1}
\end{eqnarray}
\label{eq:kappa01}%
\end{subequations}
The solution is expressed most simply by giving the inverse of the $2\times2$ matrix $\mathcal{A}_{ij}(E)$ of transition amplitudes 
in eq.~\eqref{eq:Ainverse2}, which can be expressed as
\begin{equation}
\label{eq:Ainverse0}
\bm{\mathcal{A}}^{-1}(E) = \frac{1}{8\pi} \bm{M}^{1/2}
\Big[ - \bm{\gamma} + \bm{\kappa}(E) \Big] \bm{M}^{1/2},
\end{equation}
where  $\bm{\gamma}$ is a symmetric matrix of renormalized parameters,
\begin{equation}
\label{eq:gammamatrix}
\bm{\gamma}= 
\begin{pmatrix} \gamma_{00}   & \gamma_{01} \\ 
 \gamma_{01} & \gamma_{11} 
\end{pmatrix},
\end{equation}
 $\bm{\kappa}(E)$ is the diagonal matrix whose diagonal entries are the functions of $E$ defined in eqs.~\eqref{eq:kappa01},
\begin{equation}
\bm{\kappa}(E) =
\begin{pmatrix} \kappa_0(E)  &          0       \\ 
                0            & \kappa_1(E)
\end{pmatrix} ,
\label{eq:kappamatrix}
\end{equation}
and $\bm{M}$ is the $2\times2$ matrix
\begin{equation}
\label{eq:Mmatrix}
\bm{M}=
\begin{pmatrix}  M &   0   \\ 
                0  & 2M   
\end{pmatrix}.
\end{equation}
The different diagonal entries of $\bm{M}$ take into account that the neutral channel consists of a pair of identical Majorana fermions while the charged channel consists of two distinguishable Majorana fermions. The constraints on the T-matrix elements from S-wave unitarity can be derived from the identity in eq.~\eqref{A-unitarity} for real energy $E$, 
which can be expressed as
\begin{equation}
{\rm Im}\bm{\mathcal{A}}(E+ i \epsilon) = 
\frac{1}{8 \pi} \bm{\mathcal{A}}(E) \bm{M}^{1/2} \Big[ 
{\rm Im} \bm{\gamma} - {\rm Im}\bm{\kappa}(E)  \Big]  \bm{M}^{1/2} \bm{\mathcal{A}}^*(E).
\label{eq:A-unitarity}
\end{equation}
S-wave unitarity requires that the first term inside the square brackets is zero, which implies that the parameters $\gamma_{00}$, $\gamma_{01}$, and $\gamma_{11}$ are real valued.

Range corrections can be incorporated into the zero-range model by adding terms to the Lagrangian with more and  more gradients acting on the fields. Alternatively, for S-wave interactions, range corrections can be incorporated by adding terms to the Lagrangian with more and  more time derivatives acting on the fields. To be specific, the three bare coupling constants $\lambda_{ij}$ in eq.~\eqref{eq:ZRint} are replaced by coupling operators $\lambda_{ij}(i \partial/\partial t)$ that can be expanded in powers of the time derivative operator $i \partial/\partial t$. This defines a theory with infinitely many parameters. One choice for the infinitely many parameters is the bare parameters defined by the coefficients $\lambda_{ij}^{(n)}(0)$ in the Taylor series expansion of the coupling operators. In the solution in eq.~\eqref{eq:Ainverse0} to the Lippmann-Schwinger equation, the effect of replacing the bare coupling constants by the coupling operators $\lambda_{ij}(i \partial/\partial t)$ is to replace the three renormalized parameters $\gamma_{ij}$ by functions $\gamma_{ij}(p^2)$ of the energy that can be expanded in powers of $p^2 = ME$. An alternative choice for the infinitely many parameters are the renormalized parameters defined by the coefficients $\gamma_{ij}^{(n)}(0)$ in the Taylor series expansion of those functions.

\subsection{Renormalization-group fixed points}

The zero-range model with transition amplitudes given by the matrix in eq.~\eqref{eq:Ainverse0} has two coupled scattering channels with different energy thresholds. This model is analogous to the leading order (LO) approximation to the {\it pion-less effective field theory} ($\pi \!\!\!\slash$EFT) that has been widely used in nuclear physics to describe low-energy nucleons \cite{Kaplan:1998tg,Kaplan:1998we}. In $\pi \!\!\!\slash$EFT at LO,  nucleon pairs have two decoupled S-wave scattering channels (the spin-singlet isospin-triplet channel and the spin-triplet isospin-singlet channel) with the same energy threshold. Zero-range models that have two coupled scattering channels with different energy thresholds have been applied previously to nucleon-nucleus interactions \cite{Cohen:2004kf}, to ultracold atoms \cite{Braaten:2007nq}, and to charm meson pairs near the $X(3872)$ resonance \cite{Braaten:2007ft}.

The zero-range model with coupling constants replaced by coupling functions of $\partial/\partial t$, which includes range corrections to all orders, has infinitely many parameters. An {\it effective field theory} can be  defined as a sequence of models with an increasing finite number of parameters that take into account range corrections with systematically improving accuracy. An effective field theory can be defined most rigorously by identifying a renormalization-group fixed point. Systematically improving accuracy is then ensured by adding to the Lagrangian operators  with increasingly higher scaling dimensions. The finite number of parameters can be chosen to be the coefficients of those operators that have been included. Coefficients in the Lagrangian are not necessarily the most convenient parameters. Other parameters, such as the coefficients $\gamma_{ij}^{(n)}(0)$ in the low-energy expansion of  the entries of  $\bm{\mathcal{A}}^{-1}(E)$, can be used if we can identify a power counting that determines the improvement in the accuracy of a model that can be obtained by including each of the parameters.  To define the power counting, we introduce the generic momentum scale $Q$ described by the effective theory. We take the energy $E$ and the mass splitting $\delta$ to be of order $Q^2/M$. We also introduce a momentum scale $\Lambda$ that can be regarded as the ultraviolet cutoff of the effective field theory. The physical interpretation of $\Lambda$ is  the smallest momentum scale beyond the domain of applicability of the effective field theory. In the case of winos, $\Lambda$ is the inverse range $m_W$ of the weak interactions. The power-counting scheme identifies how each parameter scales as a power of $Q$ and $\Lambda$. The transition amplitudes $\mathcal{A}_{ij}(E)$ can be expanded in powers of $Q/\Lambda$. The systematically improving accuracy of the effective field theory is ensured by including parameters whose leading contributions to $\mathcal{A}_{ij}(E)$ scale with increasingly higher powers of $Q/\Lambda$.

Lensky and Birse have carried out a careful renormalization-group (RG) analysis of the two-particle sector for the field theory with two coupled scattering channels with zero-range interactions \cite{Lensky:2011he}. They identified three distinct RG fixed points. The first RG fixed point is the {\it noninteracting fixed point}, which is the scale-invariant theory in which the transition amplitudes $\mathcal{A}_{ij}(E)$ are all zero. The power counting for the parameters $\gamma_{ij}^{(n)}(0)$ is that they scale as $\Lambda$  raised to the power required by dimensional analysis:  $\gamma_{ij}^{(n)}(0) \sim \Lambda^{1-2n}$. The expansion of the transition amplitudes $\mathcal{A}_{ij}(E)$ in powers of $Q/\Lambda$ can also be obtained by renormalized perturbation theory in the parameters $\lambda_{ij}^{(n)}(0)$, and then by an expansion in powers of $E$ and $\delta$. At leading order in the power counting, the inverse of the matrix of transition amplitudes is given by eq.~\eqref{eq:Ainverse0} with the substitution $\bm{\kappa}(E) \to 0$.

The second RG fixed point is a theory in which the spin-singlet scattering amplitudes saturate the S-wave unitarity bounds in eq.~\eqref{sigma-unitarity}. We will refer to this fixed point as the {\it two-channel-unitarity fixed point}. At this fixed point, the two scattering channels have the same threshold ($\delta = 0$) and the matrix of scattering amplitudes $\mathcal{T}_{ij}(E)$ is obtained by setting $\bm{\gamma}=0$  in eq.~\eqref{eq:Ainverse0}:
\begin{equation}
\label{eq:Tfp2}
\bm{\mathcal{T}}_*(E) = \frac{8\pi i}{\sqrt{ME}} \, \bm{M}^{-1/2} 
\begin{pmatrix} 1~  & ~0\\ 0~ & ~1 \end{pmatrix} \bm{M}^{-1/2}.
\end{equation}
The cross sections in both channels saturate the S-wave unitarity bound. The cross sections have the scaling behavior $1/E$. The power-law dependence on $E$ implies that the interactions  are scale invariant. Away from the fixed point, the power counting for the parameters is that $\gamma_{ij}^{(0)}(0) \sim Q$ while $\gamma_{ij}^{(n)}(0)$  for $n \ge 1$ scales as $\Lambda$ raised to the power required by dimensional analysis:  $\gamma_{ij}^{(n)}(0) \sim \Lambda^{1-2n}$. At leading order in the power counting, the inverse of the matrix of transition amplitudes is given by eq.~\eqref{eq:Ainverse0}. The zero-range models in refs.~\cite{Cohen:2004kf,Braaten:2007nq,Braaten:2007ft}, which have two coupled scattering channels with different energy thresholds, can be interpreted as LO approximations to effective field theories near the two-channel-unitarity fixed point.

In ref.~\cite{Lensky:2011he}, Lensky and Birse pointed out that there is a third RG fixed point. We will refer to it as the {\it single-channel-unitarity fixed point}. At this fixed point, the two scattering channels have the same threshold  ($\delta = 0$) and the  matrix of  scattering amplitudes $\mathcal{T}_{ij}(E)$ is
\begin{equation}
\label{eq:Tfp3}
\bm{\mathcal{T}}_*(E) = \frac{8\pi i}{\sqrt{ME}} \, \bm{M}^{-1/2} 
\begin{pmatrix} \cos^2\phi  & \cos\phi  \sin\phi \\ 
 \cos\phi  \sin\phi & \sin^2\phi \end{pmatrix}
 \bm{M}^{-1/2}.
\end{equation}
There is nontrivial scattering in a single channel that is a linear combination of the neutral channel $w^0w^0$ and the charged channel $w^+ w^-$  with mixing angle $\phi$. In that channel, the cross section saturates the S-wave unitarity bound. There is no scattering in the orthogonal channel. The single-channel-unitarity fixed point is the most natural one for describing a system with a single fine tuning, such as the tuning of the wino mass $M$  to a unitarity value where there is an S-wave resonance at the threshold.

\subsection{Power counting}

In order to define a power counting for the effective field theory associated with the single-channel-unitarity fixed point, we give an explicit parametrization of the transition amplitudes $\mathcal{A}_{ij}(E)$. We introduce two 2-dimensional unit vectors that depend on the mixing angle $\phi$:
\begin{equation}
\label{eq:u,v-def}
\bm{u}(\phi) = \binom{\cos\phi}{\sin\phi}, \qquad \bm{v}(\phi) = \binom{-\sin\phi}{~~\cos\phi}.
\end{equation}
We use these vectors to define two projection matrices and another symmetric matrix:
\begin{subequations}
\begin{eqnarray}
\bm{\mathcal{P}}_u(\phi) &=& \bm{u}(\phi)\, \bm{u}(\phi)^T
= \begin{pmatrix} \cos^2\phi  & \cos\phi  \sin\phi \\  \cos\phi  \sin\phi & \sin^2\phi \end{pmatrix}, 
\label{eq:Pu}
\\ 
\bm{\mathcal{P}}_v(\phi) &=& \bm{v}(\phi)\, \bm{v}(\phi)^T
= \begin{pmatrix} \sin^2\phi  & - \cos\phi  \sin\phi \\  - \cos\phi  \sin\phi & \cos^2\phi \end{pmatrix},
\label{eq:Pv}
\\ 
\bm{\mathcal{P}}_m(\phi) &=& \bm{u}(\phi)\, \bm{v}(\phi)^T + \bm{v}(\phi)\, \bm{u}(\phi)^T
= \begin{pmatrix} -\sin(2\phi)  & \cos(2\phi) \\  \cos(2\phi) & \sin(2\phi) \end{pmatrix}.
\label{eq:Pm}
\end{eqnarray}
\label{eq:Pu,Pv,Pm}%
\end{subequations}
The superscript $T$ on $\bm{u}$ or $\bm{v}$ indicates the transpose of the column vector. In the T-matrix at the critical point in eq.~\eqref{eq:Tfp3}, the matrix sandwiched between the factors of $\bm{M}^{-1/2}$ is $\bm{\mathcal{P}}_u(\phi)$. The three matrices defined in eqs.~\eqref{eq:Pu,Pv,Pm} form a basis for $2 \times2$ symmetric matrices. This set of matrices is closed under differentiation:
\begin{subequations}
\begin{eqnarray}
\bm{\mathcal{P}}_u'(\phi) &=& \bm{\mathcal{P}}_m(\phi), 
\label{eq:dPu}
\\ 
\bm{\mathcal{P}}_v'(\phi) &=& -\bm{\mathcal{P}}_m(\phi), 
\label{eq:dPv}
\\ 
\bm{\mathcal{P}}_m'(\phi) &=& -2\, \bm{\mathcal{P}}_u(\phi) +2\, \bm{\mathcal{P}}_v(\phi).
\label{eq:dPm}
\end{eqnarray}
\label{eq:dPu,dPv,dPm}%
\end{subequations}

In ref.~\cite{Lensky:2011he}, Lensky and Birse diagonalized the RG flow near the unitarity fixed point whose T-matrix $\bm{\mathcal{T}}_*(E)$ is given  in eq.~\eqref{eq:Tfp3}, identifying all the scaling perturbations and their scaling dimensions. The scaling perturbations to $\bm{M}^{-1/2} \bm{\mathcal{T}}^{-1}(E) \bm{M}^{-1/2}$ have the form $(p^2)^i (\Delta^2)^j$, where $p = \sqrt{ME}$,  $\Delta = \sqrt{2 M \delta}$, and $i$ and $j$ are nonnegative integers, multiplied by either $\bm{\mathcal{P}}_u(\phi)$ or $\bm{\mathcal{P}}_v(\phi)$ or $\bm{\mathcal{P}}_m(\phi)$. The scaling dimensions are $-1+2i+2j$ in the $\bm{\mathcal{P}}_u$ channel, $1+2i+2j$ in the $\bm{\mathcal{P}}_v$ channel, and $2i+2j$ in the $\bm{\mathcal{P}}_m$ channel. The scaling perturbations provide a basis for the vector space of perturbations near the fixed point. They can be used to provide a complete parametrization of the T-matrix:
\begin{eqnarray}
\label{eq:TZREFT-RG}
\bm{\mathcal{T}}^{-1}(E) = \frac{1}{8\pi} \bm{M}^{1/2} 
\left[ \bigg( \sum_{i,j=0}^\infty c^{(u)}_{ij} (p^2)^i (\Delta^2)^j  \bigg) \bm{\mathcal{P}}_u(\phi)
 + \bigg( \sum_{i,j=0}^\infty c^{(v)}_{ij} (p^2)^i (\Delta^2)^j  \bigg) \bm{\mathcal{P}}_v(\phi) \right.
 \nonumber
 \\
\left.
 + \bigg( \sum_{i,j=0}^\infty c^{(m)}_{ij} (p^2)^i (\Delta^2)^j  \bigg) \bm{\mathcal{P}}_m(\phi) 
 + \bm{\kappa}(E) \right]
\bm{M}^{1/2}.~~~
\end{eqnarray}
Unitarity constrains the coefficients of the expansions in powers of $p^2$ and $\Delta^2$ to be real. At the fixed point, there is a single relevant operator with scaling dimension $-1$. It corresponds to the  parameter $c^{(u)}_{00}$ in the coefficient of $\bm{\mathcal{P}}_u(\phi)$ in eq.~\eqref{eq:TZREFT-RG}. Since the operator is relevant, the parameter $c^{(u)}_{00}$ must be treated nonperturbatively. There is a single marginal operator with scaling dimension 0. It corresponds to the  parameter $c^{(m)}_{00}$ in the coefficient of $\bm{\mathcal{P}}_m(\phi)$ in eq.~\eqref{eq:TZREFT-RG}. Because of the identity in eq.~\eqref{eq:dPu}, an infinitesimal change in this parameter can be compensated by an infinitesimal change in the mixing angle $\phi$ in the fixed point  T-matrix $\bm{\mathcal{T}}_*(E)$ in eq.~\eqref{eq:Tfp3}. Thus the parameter $c^{(m)}_{00}$ can be absorbed into the mixing angle $\phi$. All the other operators are irrelevant operators with scaling dimensions 1 or higher.  The corresponding parameters can be treated perturbatively. The sums in eq.~\eqref{eq:TZREFT-RG} can be truncated to include only terms with scaling dimensions below some maximum. This truncation defines a field theory with a finite number of parameters. By increasing the maximum scaling dimension, we obtain a systematically improvable sequence of field theories. They define an effective field theory that we refer to  as {\it zero-range effective field theory} or {\it ZREFT}.

In ref.~\cite{Lensky:2011he}, Lensky and Birse defined a simpler power counting for ZREFT by summing the expansions in powers of $\Delta^2$ in eq.~\eqref{eq:TZREFT-RG} into  $\Delta$-dependent parameters:
\begin{eqnarray}
\label{eq:TZREFT-LB}
\bm{\mathcal{T}}^{-1}(E) = \frac{1}{8\pi} \bm{M}^{1/2} 
\left[ \bigg( -\gamma_u + \sum_{i=1}^\infty c^{(u)}_i(p^2)^i  \bigg) \bm{\mathcal{P}}_u(\phi)
 + \bigg( \sum_{i=0}^\infty c^{(v)}_i (p^2)^i \bigg) \bm{\mathcal{P}}_v(\phi) \right.
 \nonumber \\
\left.
 + \bigg( \sum_{i=1}^\infty c^{(m)}_i (p^2)^i  \bigg) \bm{\mathcal{P}}_m(\phi) 
 + \bm{\kappa}(E) \right]
\bm{M}^{1/2}.
\end{eqnarray}
The terms $c^{(u)}_{0j}  (\Delta^2)^j$ in eq.~\eqref{eq:TZREFT-RG} have been summed into the $\Delta$-dependent parameter  $-\gamma_u$. The terms $c^{(m)}_{0j}  (\Delta^2)^j$ in eq.~\eqref{eq:TZREFT-RG} have been absorbed up into the $\Delta$-dependent  mixing angle $\phi$. The power counting for the  $\Delta$-dependent  parameters $c^{(u)}_i$, $c^{(v)}_i$, and $c^{(m)}_i$ is the same as for the leading terms in their expansions in $\Delta^2$, which are the parameters $c^{(u)}_{i0}$, $c^{(v)}_{i0}$, and $c^{(m)}_{i0}$, respectively. The power counting for the  $\Delta$-dependent  parameter $\gamma_u$ is the same as for  $c^{(u)}_{00}$. Because the parameter  $c^{(u)}_{00}$ is the coefficient of a relevant operator, $\gamma_u$ must be treated nonperturbatively. A simple way to do this is to eliminate $\gamma_u$ in favor of the inverse scattering length $1/a_0$ for neutral winos.

The T-matrix elements $\mathcal{T}_{ij}(E)$ are the entries $\mathcal{A}_{ij}(E)$ of the matrix of transition amplitudes  evaluated at a positive energy $E$. The general parametrization of the T-matrix in eq.~\eqref{eq:TZREFT-LB} also provides a general parametrization of the transition amplitude. We choose to express the parameterization in the form
\begin{eqnarray}
\label{eq:AinverseZREFT}
\bm{\mathcal{A}}^{-1}(E) = \frac{1}{8\pi} \bm{M}^{1/2} 
\Big[ \big(- \gamma_u + \tfrac12 r_u p^2 + \ldots \big) \bm{\mathcal{P}}_u(\phi)
 + \big(-1/a_v + \ldots \big) \bm{\mathcal{P}}_v(\phi)
 \nonumber
 \\
 + \big(\tfrac12 r_m p^2 + \ldots \big) \bm{\mathcal{P}}_m(\phi) + \bm{\kappa}(E) \Big]
\bm{M}^{1/2},
\end{eqnarray}
where $p^2 = ME$ and $\bm{\kappa}(E)$ is the diagonal matrix in eq.~\eqref{eq:kappamatrix}. The mixing angle $\phi$ and the coefficients of the expansions in powers of $p^2$, such as $\gamma_u$, $r_u$, $1/a_v$, and $r_m$, should all be regarded as functions of $M$ and $\delta$ with expansions in powers of $\Delta^2$. Successive truncations of the expansions in $p^2$ of the coefficients of $\bm{\mathcal{P}}_u$, $\bm{\mathcal{P}}_v$, and $\bm{\mathcal{P}}_m$ in eq.~\eqref{eq:AinverseZREFT} define successive improvements of ZREFT. The parameter $\gamma_u$, which is associated with a relevant operator, is the only expansion parameter in ZREFT at leading order (LO). The other parameters are $M$, $\delta$, and the mixing angle $\phi$. The expansion parameters $r_u$ and $1/a_v$ are associated with irrelevant operators with scaling dimensions 1. They are the only additional parameters in ZREFT at next-to-leading order (NLO). The expansion parameter $r_m$ is the only one associated with an operator with scaling dimension 2. It is therefore the only additional parameter in ZREFT at next-to-next-to-leading order  (NNLO).

The power-counting rules of Lensky and Birse can be verified by truncating the expansions in the matrix $\bm{\mathcal{A}}^{-1}(E)$  in eq.~\eqref{eq:AinverseZREFT}, inverting the matrix to get $\bm{\mathcal{A}}(E)$, and then expanding $\bm{\mathcal{A}}(E)$ in powers of $Q/\Lambda$. The functions  $\kappa_0(E)$ and $\kappa_1(E)$ are order $Q$ and $p^2=ME$ is order $Q^2$. The parameter $\gamma_u$ is naturally order $\Lambda$, but the fine-tuning to the critical region makes it order $Q$. The inverse scattering length $1/a_0$ is a momentum scale that can be arbitrarily small. Large cancellations associated with this small momentum scale can be avoided by eliminating $\gamma_u$ in favor of $1/a_0$. All the other coefficients in the expansions in powers of $p^2$  in eq.~\eqref{eq:AinverseZREFT} scale as $\Lambda$ raised to a negative power. The expansion of $\bm{\mathcal{A}}(E)$ in powers of $Q/\Lambda$ reveals that the parameters $a_v$ and $r_u$ first enter at first order in $Q/\Lambda$, confirming that they are NLO parameters. The first-order terms are linear in $a_v$ and $r_u$, so we conclude that $a_v$ and $r_u$ are both order $1/\Lambda$. The expansion of $\bm{\mathcal{A}}(E)$ in powers of $Q/\Lambda$ reveals that the parameter $r_m$ first enters at second order in $Q/\Lambda$ through terms proportional to $a_v r_m$, confirming that $r_m$ is an NNLO parameter. Since $a_v$ is order $1/\Lambda$, we conclude that $r_m$ is also order $1/\Lambda$. The power counting rules can be summarized very simply by stating that all the expansion parameters in eq.~\eqref{eq:AinverseZREFT} except $\gamma_u$ scale as $\Lambda$ raised to the power expected from dimensional analysis. 

The parameters of ZREFT at LO are the mass $M$, the splitting $\delta$, and the mixing angle $\phi$, and $\gamma_u$. At NLO, there are two additional scattering parameters: $r_u$ and $a_v$. At NNLO, the only additional scattering parameter is $r_m$. The region of validity of ZREFT is momenta smaller than $m_W = 80.4$~GeV. The region of validity includes the transition momentum scale $\Delta = \sqrt{2M\delta}$ set by the wino mass splitting. Near the unitarity mass $M_* = 2.88$~TeV for $\alpha=0$, our preferred mass splitting $\delta = 170$~MeV implies a momentum scale $\Delta_* \approx 31$~GeV. Since this is approximately $0.4\,m_W$, accurate results at the momentum scale $\Delta$ may require ZREFT beyond LO.

ZREFT can be extended to an effective field theory for winos and photons. In ZREFT at LO, the only electromagnetic coupling is that of the charged winos through the covariant derivatives acting on the charged wino fields in eq.~\eqref{eq:kineticL}. In ZREFT beyond LO, gauge invariance requires some  of the terms proportional to powers of $p^2=ME$ in the inverse matrix of transition amplitudes in eq.~\eqref{eq:AinverseZREFT} to be accompanied by additional interaction  terms proportional to powers of $A_0$. There may also be additional interaction terms involving the gauge invariant electromagnetic field strength $F_{\mu\nu}$. 

\section{ZREFT at LO}
\label{sec:ZREFTLO}

In this section, we consider ZREFT without electromagnetism at LO. We determine the single adjustable parameter by matching to NREFT with $\alpha = 0$. We show that ZREFT at LO gives surprisingly accurate predictions for most two-body observables for winos in the threshold region.

\subsection{Transition amplitude}
\label{sec:AmpZREFT}

The interaction parameters of ZREFT at LO are the mixing angle $\phi$ and the scattering parameter $\gamma_u$. 
The matrix of transition amplitudes can be obtained by setting all the higher-order coefficients in  eq.~\eqref{eq:AinverseZREFT} except $a_v$ to 0, inverting that matrix, and then taking the limit $a_v \to 0$:
\begin{equation}
\label{eq:AmatrixLOLu}
\bm{\mathcal{A}}(E) = 
\frac{8\pi}{L_u(E)}  \, 
\bm{M}^{-1/2} \, \bm{\mathcal{P}}_u(\phi)\,  \bm{M}^{-1/2},
\end{equation}
where $\bm{\mathcal{P}}_u(\phi)$ is the projection matrix defined in eq.~\eqref{eq:Pu} and $\bm{M}$ is the diagonal matrix in eq.~\eqref{eq:Mmatrix}. The denominator in eq.~\eqref{eq:AmatrixLOLu} is
\begin{equation}
L_u(E)=-\gamma_u +\cos^2\phi\,  \kappa_0(E)  +\sin^2\phi \, \kappa_1(E),
\label{eq:Ku}
\end{equation}
where $\kappa_0(E)$ and $\kappa_1(E)$ are given in eqs.~\eqref{eq:kappa01}. If we consider $E>0$ and set $\delta = 0$ and $\gamma_u=0$, we recover the T-matrix at the fixed point in eq.~\eqref{eq:Tfp3}.

The neutral-wino scattering length $a_0$ can be obtained by evaluating the transition amplitude $\mathcal{A}_{00}(E)$ at the neutral-wino-pair threshold:
\begin{equation}
\mathcal{A}_{00}(E=0) = - 8\pi a_0 /M .
\label{eq:T00-a0}
\end{equation}
The inverse neutral-wino scattering length $\gamma_0 \equiv 1/a_0$ is
\begin{equation}
\gamma_0 = (1 +  t_\phi^2)\gamma_u - t_\phi^2\,  \Delta ,
\label{eq:a0-gammauLO}
\end{equation}
where $t_\phi = \tan \phi$ and $\Delta = \sqrt{2 M \delta}$. This equation can be solved for $\gamma_u$ as a function of $\gamma_0$:
\begin{equation}
\gamma_u = \frac{t_\phi^2\,  \Delta + \gamma_0}{1 +  t_\phi^2}.
\label{eq:gammau-a0LO}
\end{equation}
Large cancellations in the denominator $L_u(E)$ in eq.~\eqref{eq:Ku} can be avoided by eliminating $\gamma_u$ in favor of $\gamma_0$. The resulting expression for the matrix of transition amplitudes is
\begin{equation}
\label{eq:AmatrixLOL0}
\bm{\mathcal{A}}(E) = 
 \frac{8\pi}{L_0(E)}  \bm{M}^{-1/2}   
\begin{pmatrix}    ~1~     & t_\phi\\  t_\phi & t_\phi^2 \end{pmatrix}  \bm{M}^{-1/2} .
\end{equation}
The denominator  is
\begin{equation}
L_0(E)=-\gamma_0 +t_\phi^2 \big[ \kappa_1(E) -\Delta \big] + \kappa_0(E),
\label{eq:K0-E}
\end{equation}
where $\kappa_0(E)$ and $\kappa_1(E)$ are given in eqs.~\eqref{eq:kappa01} and $\Delta = \sqrt{2 M \delta}$. 

\subsection{Wino-wino scattering}
\label{sec:CrossSectionLO}

The cross section for elastic scattering from channel $i$ to channel $j$ at energy $E$, averaged over initial spins and summed over final spins, is denoted by  $\sigma_{i \to j}(E)$.  The expressions for these cross sections in terms of the T-matrix elements ${\cal T}_{ij}(E)$ for states with the standard normalizations of a nonrelativistic field theory are
\begin{subequations}
\begin{eqnarray}
\sigma_{i \to 0}(E) &=&\frac{M^2}{8\pi}
\big| {\cal T}_{i0}(E) \big|^2 \frac{v_0(E)}{v_i(E)},
\label{eq:sig0E-calT}
\\
\sigma_{i\to 1}(E) &=&\frac{M^2}{4\pi}
\big| {\cal T}_{i1}(E) \big|^2 \frac{v_1(E)}{v_i(E)},
\label{eq:sig1E-calT}
\end{eqnarray}
\label{eq:sigE-calT}%
\end{subequations}
where $v_i(E)$ and $v_j(E)$ are the velocities of the incoming and outgoing winos, which are given in eqs.~\eqref{eq:v0,1-E}. The extra factor of $1/2$ in the cross sections  $\sigma_{i \to 0}$ in eq.~\eqref{eq:sig0E-calT} for producing a neutral-wino pair compensates for  overcounting by integrating over the entire phase space of the two identical particles. For the neutral-wino elastic cross section $\sigma_{0 \to 0}$, the energy threshold is $E=0$. For the other three cross sections $\sigma_{1 \to 0}$, $\sigma_{0 \to 1}$,  and $\sigma_{1 \to 1}$, the energy threshold is $E=2\delta$.

The T-matrix elements ${\cal T}_{ij}(E)$ in the spin-singlet channel are obtained by evaluating the transition amplitudes $\mathcal{A}_{ij}(E)$ on the appropriate energy shell. For a neutral-wino pair $w^0 w^0$ with relative momentum $p$, the energy shell is $E = p^2/M$. For a charged-wino pair $w^+ w^-$ with relative momentum $p$, the energy shell is $E = 2 \delta + p^2/M$. The transition amplitudes $\mathcal{A}_{ij}(E)$ for ZREFT at LO are given by the $2\times 2$ matrix in eq.~\eqref{eq:AmatrixLOL0}.

For center-of-mass energy in the range $0 \leq E < 2\delta$ below the charged-wino-pair threshold, only the neutral-wino-pair channel is open. The LO  T-matrix element for  $w^0w^0 \to w^0w^0$ is
\begin{equation}
{\cal T}_{00}(E) = \frac{8\pi/M}{L_0(E)}.
\label{eq:T00LOloE}
\end{equation}
where $L_0(E)$ is given in eq.~\eqref{eq:K0-E}.
If $\gamma_0 = 0$, the neutral-wino elastic cross section $\sigma_{0 \to 0}(E)$ saturates the unitarity bound in 
eq.~\eqref{sigma-unitarity0} in the limit $E \to 0$. For this reason, we refer to the critical value $\gamma_0 = 0$ as {\it unitarity}.
For energy in the range $ E > 2\delta$ above the charged-wino-pair threshold, the $w^0\, w^0$ and $w^+ w^-$ channels are both open. The T-matrix element  in ZREFT at LO for  $w^0w^0 \to w^0w^0$ is given in eq.~\eqref{eq:T00LOloE}.
The T-matrix elements  in ZREFT at LO for  $w^0w^0 \to w^+w^-$ and $w^+w^- \to w^+w^-$
are given by the 01 and 11 entries of the matrix in eq.~\eqref{eq:AmatrixLOL0}:
\begin{subequations}
\begin{eqnarray}
{\cal T}_{01}(E) &=& \frac{(4\sqrt2\, \pi/M) t_\phi}{L_0(E)},
\label{eq:T01LO}
\\
{\cal T}_{11}(E) &=& \frac{(4 \pi/M) t_\phi^2}{L_0(E)}.
\label{eq:T11LO}
\end{eqnarray}
\label{eq:T01,11LO}%
\end{subequations}

The reciprocal of the T-matrix element ${\cal T}_{00}(E)$ for neutral-wino elastic scattering can be expanded in powers of the relative momentum $p = \sqrt{ME}$:
\begin{equation}
\frac{8\pi/M}{{\cal T}_{00}(E)} =
- \gamma_0  -ip - \frac{t_\phi^2}{2 \Delta}  p^2
 - \frac{t_\phi^2}{8 \Delta^3} p^4  + {\cal O}(p^6).
\label{eq:T00LOinv}%
\end{equation}
The only odd power of $p$ in the expansion is the imaginary term $-ip$. The real part has an expansion in even-integer powers of $p$. The leading term $-\gamma_0$ vanishes at the unitarity mass $M_*$. The effective range  $r_0$ and the shape parameter $s_0$ can be determined by expanding the expression for $1/{\cal T}_{00}(E)$ from eq.~\eqref{eq:T00LOloE} in powers of $p$ and comparing to eq.~\eqref{eq:T00NRinv}:
\begin{subequations}
\begin{eqnarray}
r_0 = - t_\phi^2/\Delta,
\label{eq:r0LO}
\\
s_0 = - t_\phi^2/\Delta^3,
\label{eq:s0LO}
\end{eqnarray}
\label{eq:r0s0LO}%
\end{subequations}
where $\Delta = \sqrt{2 M \delta}$. The predictions for these coefficients are independent of $\gamma_0$.

The T-matrix elements at the threshold $E = 2 \delta$ define complex scattering lengths $a_{i \to j}$:
\begin{eqnarray}
\label{eq:T-aijdef}
\bm{\mathcal T}(E=2 \delta) = - 8\pi\, \bm{M}^{-1/2}\, 
\begin{pmatrix}
a_{0 \to 0} & a_{0 \to 1} \\
a_{0 \to 1} & a_{1 \to 1} \end{pmatrix}
\bm{M}^{-1/2}.
\end{eqnarray}
At LO, these complex scattering lengths are predicted to be
\begin{subequations}
\label{eq:aijLO}
\begin{eqnarray}
a_{0 \to 0} &=&
\frac{1}{t_\phi^2\, \Delta + \gamma_0   + i \Delta},
\label{eq:a00LO}
\\
a_{0 \to 1} &=&
\frac{t_\phi}{t_\phi^2\, \Delta + \gamma_0  + i \Delta},
\label{eq:a01LO}
\\
a_{1 \to 1} &=&
\frac{t_\phi^2}{t_\phi^2\, \Delta + \gamma_0 + i \Delta}.
\label{eq:a11LO}
\end{eqnarray}
\end{subequations}
In the high-energy limit, the matrix of amplitudes in eq.~\eqref{eq:AmatrixLOL0} approaches the fixed-point T-matrix  in eq.~\eqref{eq:Tfp3}.

The T-matrix elements in eqs.~\eqref{eq:T00LOloE} and \eqref{eq:T01,11LO} are for ZREFT at LO with $\alpha = 0$. As illustrated by the NREFT cross sections in figs.~\ref{fig:sigma00-NREFT} and \ref{fig:sigma01,11-NREFT}, the Coulomb potential between charged winos can have a dramatic affect on the T-matrix elements in the region of the charged-wino-pair threshold. If $\alpha$ is not 0, the effects of the Coulomb potential between charged winos must be taken into account in the T-matrix elements of ZREFT. The resummation of Coulomb exchange to all orders in $\alpha$ is calculated analytically in a companion paper \cite{BJZ-Coulomb}.

\subsection{Matching with NREFT}
\label{sec:MatchingLO}

The scattering parameters of ZREFT can be determined by matching  T-matrix elements in ZREFT with low-energy T-matrix elements in NREFT for values of the  wino mass $M$ and the wino mass splitting $\delta$ that are close enough to the RG fixed point, which is $\delta = 0$, $\alpha = 0$ and  $M = M_*(\delta = 0)$, that the neutral-wino scattering length $a_0$ is large compared to the range $1/m_W$. The matching can be carried out with Coulomb resummation to all orders in both ZREFT and NREFT. If we choose matching quantities that are perturbative in $\alpha$, the matching can also be carried out without Coulomb resummation in both ZREFT and NREFT. We choose to carry out the matching with $\alpha = 0$ and at the unitarity mass $M_* = 2.88$~TeV for $\delta = 170$~MeV, where the neutral-wino elastic cross section has the most dramatic energy dependence. The resulting parameters can be used as estimates of the parameters of ZREFT with $\alpha=1/137$. The parameters are determined accurately with Coulomb resummation in a companion paper \cite{BJZ-Coulomb}.

The dimensionless T-matrix elements $T_{ij}(E)$ for elastic wino scattering in NREFT without Coulomb resummation can be calculated numerically by solving the coupled-channel Schr\"odinger equation in eq.~\eqref{eq:radialSchrEq} with $\alpha=0$. The T-matrix elements ${\cal T}_{ij}(E)$ for elastic wino scattering in ZREFT at LO are given analytically  in eqs.~\eqref{eq:T00LOloE} and \eqref{eq:T01,11LO}. The relation between the T-matrix in NREFT and the T-matrix in ZREFT can be deduced by comparing the unitarity equation for $\bm{T}(E)$ in eq.~\eqref{eq:T-unitarity} with the unitarity equation for $\bm{\mathcal{T}}(E)$, which can be deduced from the equation for the imaginary part of $\bm{\mathcal{A}}(E)$ in eq.~\eqref{eq:A-unitarity}:
\begin{equation}
{\rm Im}\bm{\mathcal{T}}(E) = 
-\frac{1}{8 \pi} \bm{\mathcal{T}}(E) \bm{M}^{1/2}\,
{\rm Im}\bm{\kappa}(E)\,  \bm{M}^{1/2} \,
 \bm{\mathcal{T}}^{\, *}(E).
\label{eq:T-unitarityZR}
\end{equation}
For $E>2 \delta$, the relation between the T-matrices is
\begin{equation}
\frac{1}{2M} \, \bm{v}(E)^{-1/2} \, \bm{T}(E)  \, \bm{v}(E)^{-1/2}= 
\frac{1}{8\pi} \, \bm{M}^{1/2} \, \bm{\mathcal{T}}(E) \, \bm{M}^{1/2},
\label{eq:TNR-TZR}
\end{equation}
where $\bm{v}(E)$ is the diagonal matrix of velocities in eq.~\eqref{eq:vmatrix} and $\bm{M}$ is the diagonal matrix of masses in eq.~\eqref{eq:Mmatrix}. For $0<E<2 \delta$, the relation between the T-matrix elements for neutral-wino scattering is
\begin{equation}
\frac{1}{2Mv_0(E)} T_{00}(E) = \frac{M}{8 \pi}\, {\cal T}_{00}(E).
\label{eq:TNR-TZR00}
\end{equation}

The scattering parameters of ZREFT at LO are $\phi$ and $\gamma_u$. We have eliminated $\gamma_u$ in favor of the neutral-wino scattering length $a_0$. An accurate parametrization of $a_0(M)$ for NREFT with $\delta = 170$~MeV and $M$ near the unitarity mass $M_*$ is given in eq.~\eqref{eq:a0Pade}. The angle $\phi$ can be determined by matching some other physical quantity in ZREFT and in NREFT. There are many possible choices for the matching quantity that determines $\phi$. If $\delta$ is fixed, it is better to use a value of $M$ close to the critical value  $M_*(\delta)$ and to match a T-matrix element at an energy $E$ close to 0. The expansion of the reciprocal of the T-matrix element ${\cal T}_{00}(E)$ for neutral-wino elastic scattering in powers of the relative momentum $p = \sqrt{ME}$ is given in eq.~\eqref{eq:T00LOinv}. The corresponding expansion in powers of $p$  in NREFT is given in eq.~\eqref{eq:T00NRinv}. The lowest-energy quantity that can be used for matching is the effective range. In ZREFT at LO, the prediction for the effective range is given in eq.~\eqref{eq:r0LO}. We choose to determine the angle $\phi$ at LO by matching the effective range  for $\delta =170$~MeV at the unitarity mass $M_* = 2.88$~TeV:
\begin{equation}
t_\phi^2 =  -\sqrt{2 M_* \delta} \, r_0(M_*).
\label{eq:matchr0}
\end{equation}
 The numerical result for the effective range at unitarity in NREFT with $\alpha =0$ is given in eq.~\eqref{eq:r0*}. Thus our matching condition gives $\tan\phi = 0.832$. The angle $\phi$  at LO is determined to be $\phi = 0.694$, which corresponds to about $40^\circ$. ZREFT at LO can be applied for $M$ near $M_*$ by replacing $\gamma_0 = 1/a_0$ by a Pad\'e approximant for the inverse scattering length. The Pad\'e approximant is obtained from eq.~\eqref{eq:a0Pade}  by replacing the parameters by those for $\alpha = 0$ given in the subsequent paragraph.

\begin{figure}[t]
\centering
\includegraphics[width=0.8\linewidth]{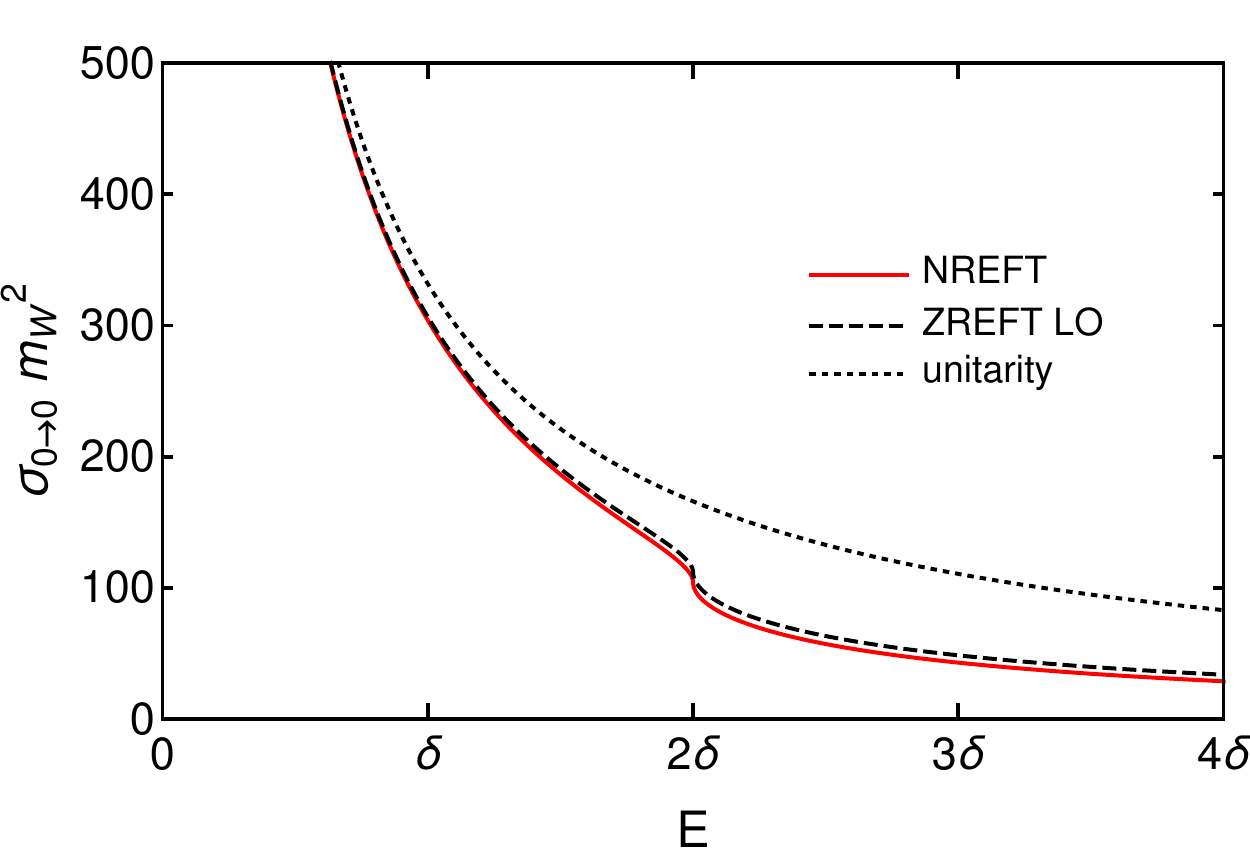}
\caption{The neutral-wino elastic cross section $\sigma_{0 \to 0}$ as a function of the energy $E$. The cross section for $\alpha=0$ at the unitarity mass $M_*=2.88$~TeV is shown for NREFT (solid line)  and for ZREFT at LO (dashed line). The S-wave unitarity bound is shown as a dotted line.
}
\label{fig:sigma00vsE-LO}
\end{figure}

In figure~\ref{fig:sigma00vsE-LO}, we compare the cross section for neutral-wino elastic scattering from NREFT with $\alpha = 0$ with the prediction of ZREFT at LO at the unitarity mass $M_*=2.88$~TeV. In the limit $E \to 0$, both cross sections saturate the unitarity bound. The angle $\phi = 0.694$ was tuned so that the next-to-leading terms in the low-energy expansions also agree. Somewhat surprisingly,  the ZREFT cross section continues to track the NREFT cross section quite closely out to the charged-wino-pair threshold at $2 \delta$ and beyond. In particular, the cross sections agree very well near the threshold, where the NREFT cross section decreases sharply. In ZREFT, this sharp change comes from the $\kappa_1(E)$ term in the denominator of eq.~\eqref{eq:AmatrixLOL0}, which switches from pure real to pure imaginary as the energy $E$ crosses the  threshold.

\begin{figure}[t]
\centering
\includegraphics[width=0.48\linewidth]{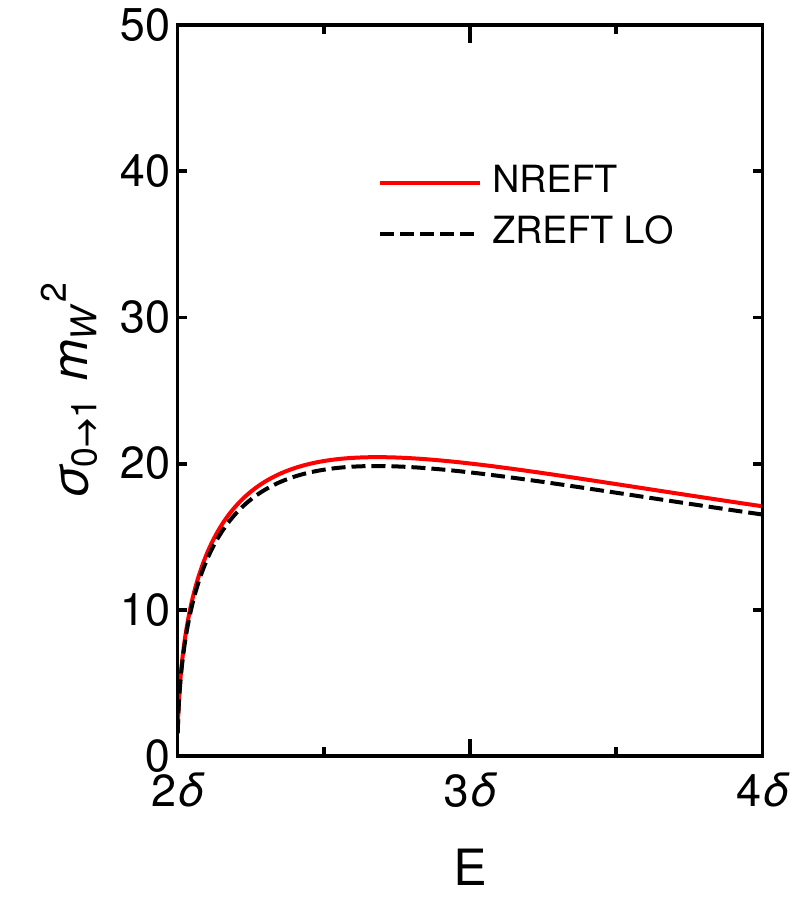}
~
\includegraphics[width=0.48\linewidth]{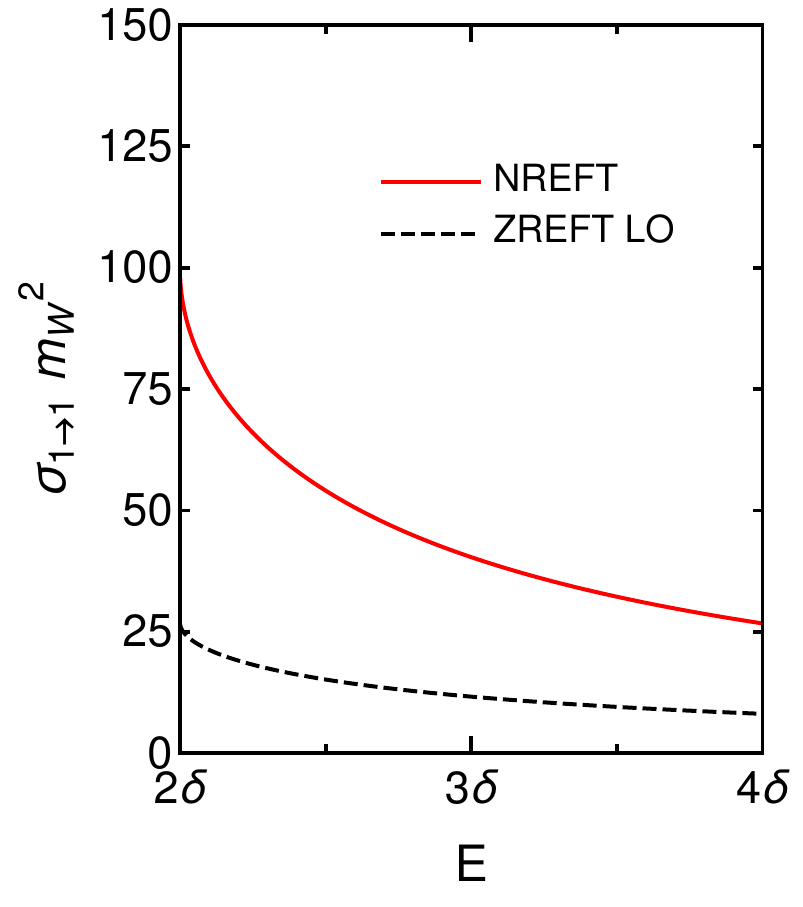}
\caption{The neutral-to-charged transition cross section $\sigma_{0 \to 1}$  (left panel) and the charged-wino elastic cross section $\sigma_{1 \to 1}$  (right panel) as functions of the energy $E$. The cross sections for $\alpha =0$ at the unitarity mass $M_*=2.88$~TeV are shown for NREFT (solid line)  and for ZREFT at LO (dashed line).  
}
\label{fig:sigma01,11vsE-LO}
\end{figure}

In the left panel of figure~\ref{fig:sigma01,11vsE-LO}, we compare the cross sections for the neutral-to-charged transition from NREFT with $\alpha = 0$ and from ZREFT at LO  at the unitarity mass  $M_*=2.88$~TeV. The cross sections agree very well over the range from $2 \delta$ to  $4 \delta$. In the right panel of figure~\ref{fig:sigma01,11vsE-LO}, we compare the cross sections for charged-wino elastic scattering from NREFT with $\alpha = 0$ and from ZREFT at LO at the unitarity mass $M_*=2.88$~TeV. The cross sections have similar shapes, but they differ significantly in magnitude. The NREFT cross section is larger by a factor that changes slowly from 3.7 at $2 \delta$ to 3.3 at $4 \delta$. Apparently the mass splitting $\delta$ is too large for ZREFT at LO to give a good approximation for the T-matrix element in this channel.

The predictions of ZREFT  at LO for the scattering lengths $a_{i \to j}$ at the charged-wino-pair threshold are given in eqs.~\eqref{eq:aijLO}. The predictions at unitarity are obtained by setting  $\Delta = \Delta_* = 31.3$~GeV, $\gamma_0 =0$, and $t_\phi=0.832$. The predictions for the real and imaginary parts of $a_{0 \to 0}(M_*)$, $a_{0 \to 1}(M_*)$, and $a_{1 \to 1}(M_*)$ all differ from the results from NREFT in eqs.~\eqref{eq:aijNR*} by less than 8\% with one glaring exception. The prediction for the real part of the charged-wino scattering length $a_{1 \to 1}(M_*)$ is smaller  than the result from NREFT in eq.~\eqref{eq:a11NR*}  by 67\%. It is the difference between the prediction for Re$(a_{1 \to 1}(M_*))$  and the result from NREFT that is primarily  responsible for the large discrepancy in the cross section for charged-wino scattering at threshold that can be seen in the right panel of Figure~\ref{fig:sigma01,11vsE-LO}. We have verified that the error in Re$(a_{1 \to 1}(M_*))$ decreases to 0 as $\delta$ decreases to 0. Apparently the mass splitting $\delta$ is too large for ZREFT at LO to give a good approximation for the real part of $a_{1 \to 1}(M_*)$.

\subsection{Wino-pair bound state}
\label{sec:BoundStateLO}

\begin{figure}[t]
\centering
\includegraphics[width=0.5\linewidth]{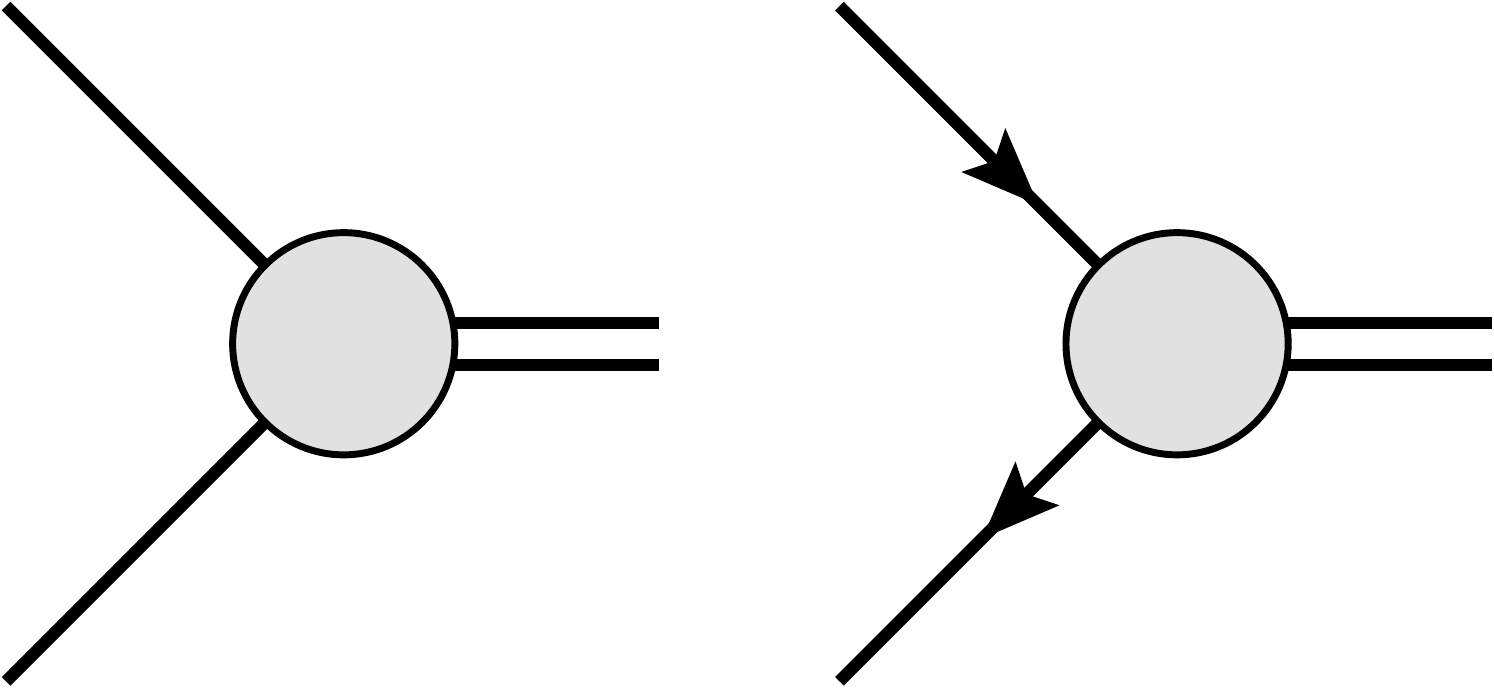}
\caption{The amplitudes for the transition from a pair of neutral winos or a pair of charged winos 
to the wino-pair bound state.
The bound state $(ww)$ is represented by an external solid double line. 
The Feynman rules for these amplitudes are 
$-i \mathcal{Z}_0^{1/2}$ and $-i \mathcal{Z}_1^{1/2}$, respectively,
where $\mathcal{Z}_0$ and $\mathcal{Z}_1$ are the residue factors given in eqs.~\eqref{eq:Z01}.}
\label{fig:BoundState}
\end{figure}

\begin{figure}[t]
\centering
\includegraphics[width=0.8\linewidth]{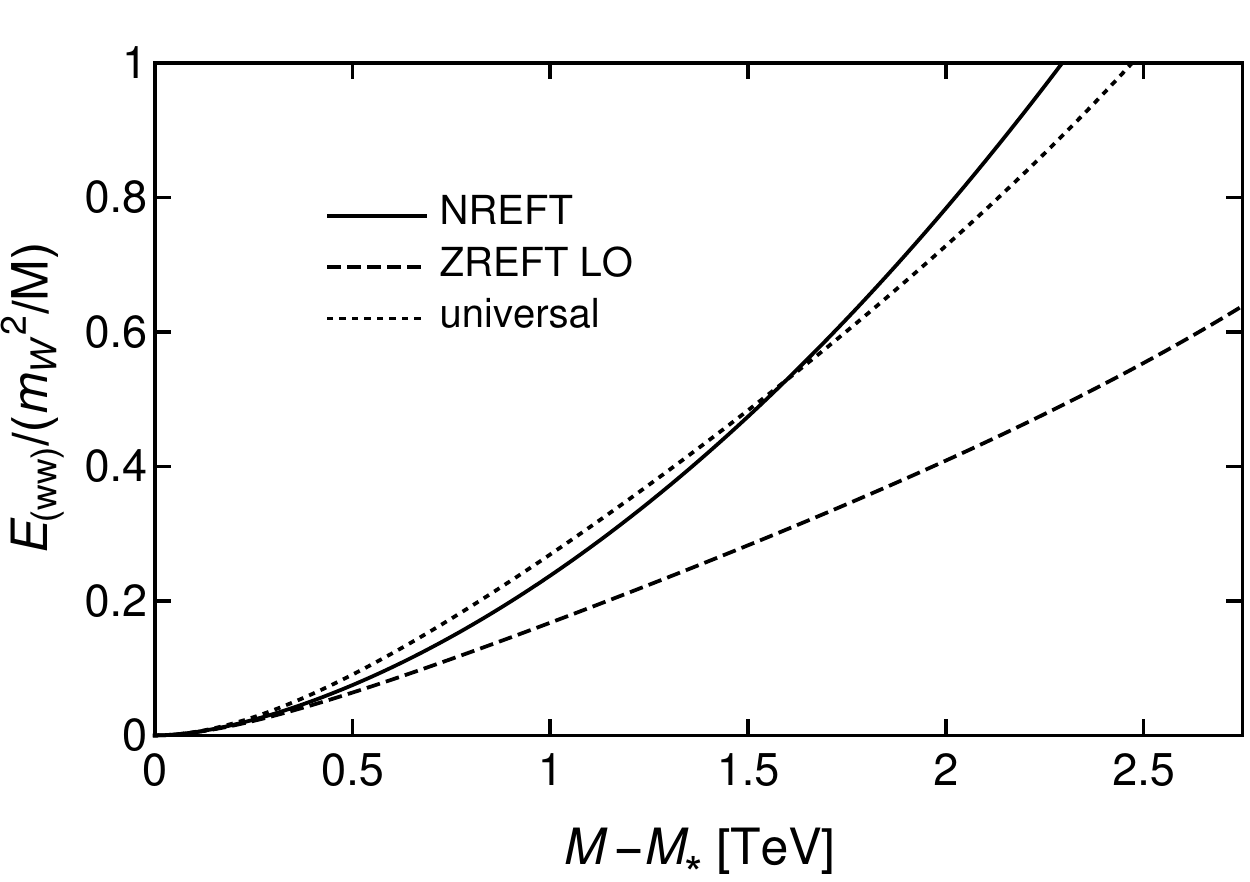}
\caption{The binding energy of the wino-pair bound state as a function of the wino mass $M$. The binding energy for $\alpha = 0$ is shown for NREFT  (solid line) and  for ZREFT at LO (dashed line). The universal approximation in eq.~\eqref{eq:Eww-largea}, with the Pad\'e approximant for $a_0(M)$ analogous to eq.~\eqref{eq:a0Pade} but for $\alpha=0$, is  shown as a dotted curve.
}
\label{fig:bindingenergyLO}
\end{figure}

If $\gamma_0 > 0$, each of the transition amplitudes ${\cal A}_{ij}(E)$ given by the matrix in eq.~\eqref{eq:AmatrixLOL0} has a pole at a real energy $E$ below the neutral-wino-pair threshold. The resonance is associated with a wino-pair bound state that we denote by $(ww)$. The energy of the bound state can be expressed as $E = -\gamma^2/M$, where the binding momentum $\gamma$ is a positive solution to the equation
\begin{equation}
0=  \gamma - \gamma_0   + t_\phi^2 \, \big[\sqrt{\Delta^2+\gamma^2} - \Delta \big].
\label{eq:gammaLO-eq}
\end{equation}
We have used eq.~\eqref{eq:gammau-a0LO} to eliminate $\gamma_u$ in favor of the inverse scattering length $\gamma_0$. This equation can be transformed into a quadratic equation for $\gamma$ with two roots. The correct root is the one that approaches\ 0 as $\gamma_0$ decreases to $0^+$:
\begin{equation}
\gamma = \frac{t_\phi^2\, \Delta + \gamma_0 - 
 t_\phi^2\sqrt{\Delta^2 + 2 t_\phi^2\, \Delta \gamma_0 + \gamma_0^2}}{1 - t_\phi^4}.
\label{eq:gammaLO}
\end{equation}
This expression is a smooth function of $t_\phi$ even at $t_\phi=1$. In figure~\ref{fig:bindingenergyLO}, the prediction for the binding energy in ZREFT at LO is compared to the result from NREFT. The error in ZREFT at LO is less than 5\% for $M-M_* <0.15$~TeV. The universal approximation in eq.~\eqref{eq:Eww-largea} with the Pad\'e approximant for $a_0(M)$ analogous to eq.~\eqref{eq:a0Pade} but for $\alpha=0$ provides a better qualitative fit out to larger values of $M-M_*$.

The amplitudes for the transitions from a pair of winos to the wino-pair bound state $(ww)$  can be deduced from the transition amplitudes $\mathcal{A}_{00}(E)$ and $\mathcal{A}_{11}(E)$ in eq.~\eqref{eq:AmatrixLOL0}. Both of these amplitudes have a pole in the energy at  $E = - \gamma^2/M$, where $\gamma$ satisfies eq.~\eqref{eq:gammaLO-eq}. The residues of the poles are $-\mathcal{Z}_0$ and $-\mathcal{Z}_1$, respectively, where
\begin{subequations}
\label{eq:BoundState}
\begin{eqnarray}
\mathcal{Z}_0 &=&
\frac{16 \pi \gamma\,\sqrt{\Delta^2+\gamma^2}}
       {M^2 \big[\sqrt{\Delta^2+\gamma^2}+ t_\phi^2\,  \gamma \big]}, \qquad
\label{eq:Z0}
\\
\mathcal{Z}_1 &=& \frac{t_\phi^2}{2}\, \mathcal{Z}_0.
\label{eq:Z1}
\end{eqnarray}
\label{eq:Z01}%
\end{subequations}
The amplitudes for the transition from a neutral-wino pair to $(ww)$ and from a charged-wino pair to $(ww)$ are  $- i \mathcal{Z}_0^{1/2}$ and $- i \mathcal{Z}_1^{1/2}$, respectively. They are represented diagrammatically by the blobs in figure~\ref{fig:BoundState}. As $M$ approaches $M_*$, the residue factors scale as $\mathcal{Z}_i  \sim \gamma/M^2$.

The bound state $(ww)$ is a superposition of a neutral-wino  pair $w^0 w^0$ and  a charged-wino pair $w^+ w^-$. The residues $\mathcal{Z}_0$ and $\mathcal{Z}_1$ in eqs.~\eqref{eq:Z01} are proportional to the probabilities for the $w^0 w^0$ and $w^+ w^-$ components of the bound state, respectively. The probabilities for $w^0 w^0$ and  $w^+ w^-$ are  $\cos^2\phi$ and $\sin^2\phi$, respectively. The extra factor of $1/2$ in $\mathcal{Z}_1/\mathcal{Z}_0$ is because the neutral wino constituents are identical fermions. Given the numerical value $\tan\phi=0.832$ from the LO fit, 
the ratio is $\mathcal{Z}_1/\mathcal{Z}_0 = 0.346$.

\subsection{Scale-invariant limit}
\label{sec:splittingLO}

As discussed in section~\ref{sec:noCoulomb}, there is a critical point in the parameter space for NREFT without electromagnetism at which wino interactions have the remarkable property of scale invariance in the low-energy limit.  At the scale-invariant point, the wino mass splitting $\delta$ is 0 and the neutral-wino scattering length $a_0 = 1/\gamma_0$ is infinite. The T-matrix has the form in eq.~\eqref{eq:Tfp3}, which depends on the mixing angle $\phi$ and is proportional to $E^{-1/2}$.  The power-law dependence on $E$ reflects the scale invariance of the interactions. The scale invariance allows the system to be described by a renormalization group fixed point.

At a generic point in the parameter space for NREFT with $\alpha = 0$, the real part of the inverse T-matrix has a range expansion in even-integer powers of the relative momentum $p = \sqrt{ME}$, as in eq.~\eqref{eq:T00NRinv}. The coefficients in the expansion have dimensions. Dimensionless coefficients can be defined by multiplying the coefficients by appropriate powers of $\Delta = \sqrt{2 M \delta}$. At the  scale-invariant point, all the coefficients in the range expansion must be either zero or $\infty$. The results in section~\ref{sec:noCoulomb} reveal that  the dimensionless coefficients have well-behaved limits as you approach the scale-invariant point.

The scale-invariant point can be reached by first tuning the mass $M$ to the unitarity point $M_*(\delta)$, and then taking the limit $\delta \to 0$. At the unitarity point, the inverse scattering length vanishes: $\gamma_0 = 0$. Each of the higher coefficients in the range expansion in eq.~\eqref{eq:T00NRinv} can be expressed as a negative-integer power of $\Delta_* = \sqrt{2 M_*(\delta)\,  \delta}$ multiplied by a dimensionless coefficient. In the scale-invariant limit $\delta \to 0$, the dimensionless coefficients are pure numbers. They can be determined by solving the Schr\"odinger equation for NREFT. The dimensionless coefficients for the effective range $r_0$ and the shape parameter $s_0$ are given in eqs.~\eqref{eq:r0dr0s0-delta}.  They have the same numerical values $-1.552$. The predictions for the dimensionless coefficients for  $r_0$ and $s_0$ in ZREFT can be obtained from eqs.~\eqref{eq:r0s0LO}. They have the same values $-\tan^2\phi$.  This explains the equality of the two dimensionless coefficients in eqs.~\eqref{eq:r0dr0s0-delta}. By matching the dimensionless coefficients, we find that the mixing angle at the scale-invariant point is given by $\tan \phi = 1.246$.  This is about 1.5 times larger than the value 0.832 obtained in section~\ref{sec:MatchingLO} by matching $r_0$ at unitarity for $\delta = 170$~MeV.  The difference can be attributed to relatively weak dependence of the mixing angle $\phi$ on $M$ and $\delta$.

Complex scattering lengths $a_{i\to j}$ are defined at the charged-wino-pair threshold in eq.~\eqref{eq:aijdef}. At the unitarity mass for the mass splitting $\delta$, they can be expressed as $1/\Delta_* = [2 M_*(\delta)\,  \delta]^{-1/2}$ multiplied by a dimensionless coefficient. In the scale-invariant limit $\delta \to 0$, the dimensionless coefficients are pure numbers. The results for these complex numbers in NREFT are given in eqs.~\eqref{eq:Reaij-delta}. The predictions for the dimensionless coefficients in ZREFT can be obtained from eqs.~\eqref{eq:aijLO} by setting $\gamma_0 = 0$. The dimensionless coefficients for $a_{0\to 0}$, $a_{0\to 1}$, and $a_{1\to 1}$ are predicted to be $1/(t_\phi+i)$, $t_\phi/(t_\phi+i)$, and $t_\phi^2/(t_\phi+i)$, where $t_\phi = \tan \phi$. The results from NREFT in eqs.~\eqref{eq:Reaij-delta} are consistent with these predictions with the same value $\tan \phi = 1.246$ determined from the dimensionless coefficient for $r_0$. This explains the simple relations between the real and imaginary parts of $a_{0\to 0}$, $a_{0\to 1}$, and $a_{1\to 1}$ that were noted after eqs.~\eqref{eq:Reaij-delta}. 

\section{ZREFT at NLO}
\label{sec:ZREFTNLO}

In this section, we consider ZREFT without electromagnetism  at NLO. We determine the three adjustable parameters by matching to NREFT with $\alpha = 0$. We show that ZREFT at NLO gives systematic improvements of the predictions for wino cross sections in the threshold region.

The interaction parameters of ZREFT at NLO are $\phi$, $\gamma_u$, $r_u$, and $a_v$. The inverse of the matrix of transition amplitudes at NLO is obtained by setting the terms that are not shown explicitly in eq.~\eqref{eq:AinverseZREFT} to 0 and also setting $r_m=0$:
\begin{eqnarray}
\label{eq:AinverseNNLO}
\bm{\mathcal{A}}(E)^{-1} = \frac{1}{8\pi} \bm{M}^{1/2} 
\bigg[ \big(- \gamma_u + \tfrac12 r_u p^2 \big) \bm{\mathcal{P}}_u(\phi)
 + \big(-1/a_v  \big) \bm{\mathcal{P}}_v(\phi)  
+ \bm{\kappa}(E) \bigg]
\bm{M}^{1/2},
\end{eqnarray}
where $p^2 = ME$ and $\bm{\kappa}(E)$ is the diagonal $2\times 2$ matrix in eq.~\eqref{eq:kappamatrix}. The matrix $\bm{\mathcal{A}}(E)$ of transition amplitudes is obtained by inverting the matrix in eq.~\eqref{eq:AinverseNNLO}. The resulting expression is complicated, but its expansion to NLO in $Q/\Lambda$ is relatively simple:
\begin{eqnarray}
\label{eq:AmatrixNLO}
\bm{\mathcal{A}}(E) &=&
8\pi \bm{M}^{-1/2}\Bigg[\left(\frac{1}{L_u(E)} - r_u  \frac{p^2}{2L_u(E)^2}  \right) \bm{\mathcal{P}}_u(\phi) 
\nonumber\\
&&\hspace{3cm}
- a_v \, \bm{V}(\phi,E)\,\bm{V}(\phi,E)^T  + \ldots \Bigg] \bm{M}^{-1/2},
\end{eqnarray}
where $L_u(E)$ is defined in eq.~\eqref{eq:Ku} and $\bm{V}(\phi,E)$ is the 2-component column vector 
\begin{equation}
\bm{V}(\phi,E)=\frac{1}{L_u(E)}
\binom{-\sin \phi\,  [-\gamma_u +\kappa_1(E)]}{~~\cos \phi\,  [-\gamma_u +\kappa_0(E)]}.
\end{equation}
In the high-energy limit, this column vector approaches the unit vector $\bm{v}(\phi)$ defined in eq.~\eqref{eq:u,v-def}. Thus in the high-energy limit, the $\bm{V}\bm{V}^T$ term in eq.~\eqref{eq:AmatrixNLO} describes S-wave scattering in the channel $\bm{\mathcal{P}}_v$ with a small scattering length $a_v$. In the coefficient  of  $\bm{\mathcal{P}}_u$ in eq.~\eqref{eq:AmatrixNLO}, the NLO term can be absorbed into the LO term by replacing $L_u(E)$ by $L_u(E) + \tfrac12 r_u  p^2$.

\subsection{Wino-wino scattering}
\label{sec:CrossSectionNLO}

The T-matrix element ${\cal T}_{00}(E)$ at NLO is obtained by inverting the matrix in eq.~\eqref{eq:AinverseNNLO} and evaluating it at a positive energy $E$. The neutral-wino scattering length $a_0$ is defined by ${\cal T}_{00}(E)$ at the threshold $E=0$ as in eq.~\eqref{eq:T00-a0}. The inverse scattering length $\gamma_0 \equiv 1/a_0$ is
\begin{equation}
\gamma_0  =
\frac{(1 + t_\phi^2)\, \gamma_u - t_\phi^2\,  \Delta - a_v\Delta \gamma_u }
{1 + a_v \big[t_\phi^2 \,  \gamma_u - (1 + t_\phi^2) \, \Delta\big] },
\label{eq:a0-gammauNLO}
\end{equation}
where $t_\phi = \tan \phi$. This equation can be inverted to obtain  $\gamma_u$ as a function of $\gamma_0$:
\begin{equation}
\gamma_u = \frac{t_\phi^2\,  \Delta + \gamma_0 - (1 +  t_\phi^2) a_v \Delta \gamma_0}
{1 +  t_\phi^2 - a_v [\Delta +  t_\phi^2 \gamma_0]}.
\label{eq:gammau-a0NLO}
\end{equation}
Large cancellations associated with the small momentum scale $\gamma_0$ can be avoided by eliminating $\gamma_u$  in favor of $\gamma_0$. The inverse scattering length  $\gamma_0= 1/a_0$ can be accurately approximated by a Pad\'e approximant obtained from  eq.~\eqref{eq:a0Pade} by replacing the parameters by those for $\alpha = 0$, which are given in the paragraph after eq.~\eqref{eq:a0Pade}.

For center-of-mass energy in the range $0 \leq E < 2\delta$ below the charged-wino-pair threshold, only the neutral-wino-pair channel is open. The T-matrix element at  energy $E=p^2/M$ is given by
\begin{eqnarray}
\frac{8 \pi}{M\, {\cal T}_{00}(E)} 
= \frac{A_0 + A_1\,  p^2 + A_2\,  (\kappa_1 - \Delta) + A_3\,  p^2 (\kappa_1 - \Delta)}{B_0 + B_1 \, p^2+ B_2\,  (\kappa_1 - \Delta)} 
-i\, p,
\label{eq:T00MLONP}
\end{eqnarray}
where $\kappa_1(E)$ is given in eq.~\eqref{eq:kappa1}. For $0<E<2\delta$, $\kappa_1$ is real so the imaginary part is simply $-p$. The coefficients in the numerator of the first term on the right side of eq.~\eqref{eq:T00MLONP} are
\begin{subequations}
\begin{eqnarray}
A_0 &=& -2(1+t_\phi^2) \big[1-a_v \Delta \big]^2 \gamma_0,
\\
A_1 &=& 
\big[ (1+t_\phi^2) - a_v\Delta \big] \big[ (1+t_\phi^2) - a_v(\Delta +t_\phi^2\gamma_0) \big] r_u ,
\\
A_2 &=&
2(1+t_\phi^2)\big[t_\phi^2 + (1-t_\phi^2 - a_v\Delta)a_v\gamma_0 \big] ,
\\
A_3 &=& -\big[1+t_\phi^2-a_v(\Delta+t_\phi^2\gamma_0) \big] a_v r_u.
\end{eqnarray}
\end{subequations}
The coefficients in the denominator of the first term on the right side of eq.~\eqref{eq:T00MLONP} are
\begin{subequations}
\begin{eqnarray}
B_0 &=& 2 (1+t_\phi^2) \big[1-a_v \Delta \big]^2,
\\
B_1 &=& 
-t_\phi^2 \big[ (1+t_\phi^2)- a_v(\Delta+t_\phi^2\gamma_0) \big] a_v r_u,	
\\
B_2 &=& -2(1+t_\phi^2) \big[ (1+t_\phi^2)- a_v(\Delta+t_\phi^2\gamma_0) \big] a_v.	
\end{eqnarray}
\end{subequations}

For $E < 2 \delta$, the real part of $1/{\cal T}_{00}(E)$ in eq.~\eqref{eq:T00MLONP} can be expanded in even-integer powers of the relative momentum $p=\sqrt{M E}$:
\begin{eqnarray}
\frac{8\pi}{M}\text{Re}\frac{1}{{\cal T}_{00}(E)} =
-\gamma_0  + \tfrac12 r_0 \, p^2+ \tfrac18 s_0 \, p^4  + {\cal O}(p^6).
\label{eq:ReT00NLOinv}
\end{eqnarray}
The leading term $-\gamma_0$ vanishes at the unitarity mass $M_*(\delta)$. The coefficients of $p^2$ and $p^4$ are  the {\it effective range} $r_0$ and the  {\it shape parameter} $s_0$. The effective range at NLO is 
\begin{eqnarray}
r_0(M,\delta) =
\frac{- t_\phi^2 (1+t_\phi^2)(1 - a_v \gamma_0)^2 
+\big[1+ t_\phi^2 - a_v( \Delta+t_\phi^2 \gamma_0) \big]^2 r_u \Delta}
{(1+t_\phi^2)  (1 - a_v \Delta)^2}  \Delta^{-1}.
\label{eq:r0NLO}
\end{eqnarray}
The right side depends on $M$ and $\delta$ through $\Delta = \sqrt{2 M \delta}$ and through $\gamma_0(M,\delta)$.

For center-of-mass energy in the range $ E > 2\delta$ above the charged-wino-pair threshold, the $w^0\, w^0$ and $w^+ w^-$ channels are both open. The T-matrix elements ${\cal T}_{ij}(E)$ at NLO are obtained by inverting the matrix in eq.~\eqref{eq:AinverseNNLO}. The scattering lengths $a_{i \to j}(M,\delta)$ that determine the cross sections at the charged-wino-pair threshold $E = 2 \delta$ can be obtained from the T-matrix elements ${\cal T}_{ij}(E)$ using the definitions in eqs.~\eqref{eq:T-aijdef}. The expressions for the inverse scattering lengths $1/a_{i \to j}$ are simpler than those for the scattering lengths $a_{i \to j}$. For example, the predictions at NLO for the real and imaginary parts of $1/a_{0 \to 0}$ are
\begin{subequations}
\label{eq:a00NLO}
\begin{eqnarray}
\mathrm{Re}\frac{1}{a_{0 \to 0}(M_*)} &&
\nonumber\\
&&\hspace{-3cm}=
\frac{(1+t_\phi^2) \Big[(t_\phi^2  \Delta +\gamma_0)  - (1+t_\phi^2)a_v\Delta\gamma_0
- \frac12 \big[1+t_\phi^2-a_v(\Delta+t_\phi^2 \gamma_0) \big]r_u\Delta^2\Big] }
{(1+t_\phi^2) \big[1 -  (1- t_\phi^2  +t_\phi^2 a_v \gamma_0) a_v\Delta\big]
- \frac12  t_\phi^2 \big[1+t_\phi^2-a_v(\Delta+t_\phi^2 \gamma_0) \big]a_v r_u \Delta^2} , ~~~
\label{eq:Rea00NLO}
\\
\mathrm{Im}\frac{1}{a_{0 \to 0}(M_*)}& =& \Delta.
\label{eq:Ima00NLO}
\end{eqnarray}
\end{subequations}
The simple form of the imaginary part is required by unitarity. The NLO expressions for $a_{0 \to 1}$ and $a_{1 \to 1}$ are  more complicated.

The analytic expressions for the NLO transition amplitude $\bm{{\cal A}}(E)$ obtained by inverting $\bm{{\cal A}}(E)^{-1}$ in eq.~\eqref{eq:AinverseNNLO} and then eliminating $\gamma_u$ in favor of $\gamma_0$ using eq.~\eqref{eq:gammau-a0NLO} are rather complicated. The matrix $\bm{M}^{1/2} \bm{{\cal A}}(E) \bm{M}^{1/2}$ is order $1/Q$, and its relative error is order $Q^2/\Lambda^2$. The expression for  $\bm{{\cal A}}(E)$ can be simplified without any parametric increase in the error  by expanding to first order in $Q/\Lambda$ and truncating the expansion. We have found that the truncated expansion has surprisingly large  numerical errors. For $\delta = 170$~MeV, the errors are so large that the NLO approximation is actually worse than the LO  approximation.

We proceed to describe possible truncated expansions of observables obtained from the NLO transition amplitude $\bm{{\cal A}}(E)$. We use the effective range $r_0$ at unitarity to illustrate each of the possibilities. We label the result  for an observable that is obtained exactly from the NLO transition amplitude without any expansion by ``NP'' (for nonperturbative). The NP approximation to the NLO effective range at unitarity is obtained by setting $\gamma_0 = 0$ in eq.~\eqref{eq:r0NLO}:
\begin{equation}
\text{NP:} \quad
r_0(M_*) =
\frac{- t_\phi^2 (1+t_\phi^2)
+ (1+ t_\phi^2 - a_v \Delta_* )^2 r_u \Delta_*}
{(1+t_\phi^2)  (1 - a_v \Delta_*)^2}  \Delta_*^{-1}.
\label{eq:r0NLONP}
\end{equation}
For brevity, we have suppressed the dependence of both sides on $\delta$. On the left side, the single argument $(M_*)$ should be interpreted as the pair of arguments $(M_*(\delta),\delta)$. On the right side, we have suppressed the argument $\delta$ of $\Delta_*(\delta) = \sqrt{2 M_*(\delta)\, \delta}$. We label the result for an observable that is obtained by truncating its expansion in powers of $Q/\Lambda$ after the $n^\text{th}$-order term by ``P$n$'' (for $n^\text{th}$-order perturbative). The P1 approximation to the NLO effective range at unitarity is obtained by expanding the expression in eq.~\eqref{eq:r0NLONP} to $1^\text{st}$ order in $a_v$ and $r_u$:
\begin{equation}
\text{P1:} \quad
r_0(M_*) =
\Big( - t_\phi^2  + \big[- 2 t_\phi^2 \, a_v   + (1+t_\phi^2) \, r_u  \big]\Delta_*
 \Big)  \Delta_*^{-1}.
\label{eq:r0NLOP1}
\end{equation}
The P2 approximation to $r_0(M_*)$ is obtained by expanding the expression in eq.~\eqref{eq:r0NLONP} to $2^\text{nd}$ order in $a_v$ and $r_u$:
\begin{equation}
\text{P2:} \quad
r_0(M_*) =
\Big( - t_\phi^2  + \big[- 2 t_\phi^2 \, a_v   + (1+t_\phi^2) \, r_u  \big]\Delta_*
+ \big[ -  3 t_\phi^2 \, a_v^2 + 2 t_\phi^2 \, a_v r_u \big]\Delta_*^2    \Big)  \Delta_*^{-1}.
\label{eq:r0NLOP2}
\end{equation}
The P3 approximation to $r_0(M_*)$ is obtained by expanding the expression in eq.~\eqref{eq:r0NLONP} to $3^\text{rd}$ order in $a_v$ and $r_u$. The sequence defined by the P$n$ approximations to an observable are partial sums for a power series in $a_v$ and $r_u$. In the case of $r_0(M_*)$, the radius of convergence of the power series is determined by the factor $(1 - a_v \Delta_*)^{-2}$ in eq.~\eqref{eq:r0NLONP}, which has a double pole in the variable $\Delta_*$. The power series converges to the NP approximation to $r_0(M_*)$ in eq.~\eqref{eq:r0NLONP} as $n \to \infty$ if  $|a_v \Delta_*| < 1$, and it diverges if $|a_v \Delta_*| > 1$.

\subsection{Matching with NREFT}
\label{sec:MatchingNREFT}

The scattering parameters of ZREFT at NLO are $\phi$, $\gamma_0$, $a_v$, and $r_u$. The four scattering parameters can be determined by matching T-matrix elements in ZREFT with low-energy T-matrix elements in NREFT. We will carry out the matching without the Coulomb potential in both ZREFT and NREFT.

If ZREFT was used as a phenomenological description of a real physical system with a known particle mass $M$ and a  known  mass splitting  $\delta$, the quantities used for matching would have to be defined by the dependence of the T-matrix elements on the energy  $E$. In this case, it would be better to use matching quantities with lower energy. In our case, NREFT provides a microscopic description for the system in which the T-matrix elements can be calculated as functions of $M$ and $\delta$ as well as $E$. Thus the quantities used for matching can be defined by the dependence of T-matrix elements on $M$, $\delta$, and $E$. It is better to use matching quantities closer to the renormalization-group fixed point, which is equal scattering thresholds ($\delta = 0$) and with the mass tuned to unitarity: $M = M_*(\delta = 0)$. We choose to carry out the matching with $\delta = 170$~MeV and with the mass tuned to unitarity: $M = M_*(\delta)$.

The lowest possible energy for matching T-matrix elements is the neutral-wino-pair threshold $E=0$. The coefficients in the low-energy expansion of  the neutral-wino elastic scattering amplitude ${\cal T}_{00}(E)$ in powers of $p = \sqrt{ME}$ are possible matching quantities that are defined essentially at $E=0$. The low-energy expansion of $1/{\cal T}_{00}(E)$ is simpler than that of ${\cal T}_{00}(E)$, because its imaginary part is determined by unitarity and its real part has a low-energy expansion in integer powers of $p^2 = ME$. A particularly convenient  set of matching quantities at $E=0$ are the coefficients in the low-energy expansion of the real part of $1/{\cal T}_{00}(E)$, which is given in eq.~\eqref{eq:ReT00NLOinv}. Four possible matching quantities are the inverse scattering length $\gamma_0$, the effective range $r_0$, its derivative $dr_0/d\gamma_0$ with respect to  the inverse scattering length, and the shape parameter $s_0$. In NREFT at $\delta = 170$~MeV and at unitarity, the inverse scattering length is $\gamma_0 = 0$ and the other three matching quantities are given in eqs.~\eqref{eq:r0,dr0,s0*}.

\begin{table}[t]
\begin{center}
\begin{tabular}{|l|ccc|}
\hline
             & ~~$\tan\phi$~~ & ~~~$a_v m_W$~~~ & ~~~$r_u m_W$~~~  \\
\hline
\hline
LO      & ~0.832~  &         0        &         0         \\
\hline
NLO(P1)   &    1.401    &     0.325     &    1.531    \\
NLO(P2)    &    1.132    &     0.521     &    1.162      \\
NLO(P3)   &    1.114    &     0.533     &    1.125      \\
NLO(P4)   &    1.109    &     0.535     &    1.116      \\
NLO(NP)~   &    1.108    &     0.536     &    1.114    \\
\hline
\end{tabular}
\end{center}
\caption{
Parameters of ZREFT for  $\alpha = 0$ and $\delta = 170$~MeV at unitarity. The parameters $a_v$ and $r_u$, which have dimensions of length, are made dimensionless by multiplying by $m_W= 80.4$~GeV. The LO parameter is determined by matching the effective range $r_0$ with results from NREFT. The NLO parameters are determined by also matching  $dr_0/d\gamma_0$ and $s_0$.
}
\label{tab:Parameters}
\end{table}

We choose to determine the parameters of ZREFT at NLO using results from NREFT at $\delta= 170$~MeV and at unitarity, where $\gamma_0 = 0$. We need to match three additional quantities to determine the other three parameters $\phi$, $a_v$, and $r_u$. We choose the three matching quantities to be the effective range $r_0$, its derivative with respect to $\gamma_0$, and the shape parameter $s_0$. The three matching quantities in NREFT are given in eqs.~\eqref{eq:r0,dr0,s0*}. We obtain the parameters labeled NLO(NP) in table~\ref{tab:Parameters} by matching the nonperturbative NLO expressions for $r_0$, $dr_0/\gamma_0$, and $s_0$ to the results from NREFT. For the perturbative truncation P$n$, we obtain the parameters labeled NLO(P$n$) by matching the $n^\text{th}$ order truncated expansions  of the NLO expressions for $r_0$, $dr_0/\gamma_0$, and $s_0$ to the results from NREFT. The NLO parameters for the first four perturbative truncations are given in table~\ref{tab:Parameters}.

\begin{table}[t]
\begin{center}
\begin{tabular}{|l|ccc|c|}
\hline
& $a_{0 \to 0}$  & $a_{0 \to 1}$  & $a_{1 \to 1}$    & ~$\Delta a$~  \\
\hline
\hline
LO            &  ~$0.468-0.676\,i$~ & ~$0.390-0.562\,i$~ & ~$0.324-0.468\,i$~ & ~0.661~  \\
\hline
NLO(P1)  &  $0.540 - 0.393\,i$  &   $0.616-0.479\,i$   &   $1.041-0.571\,i$   & ~0.335~ \\
NLO(P2)  &  $0.492 - 0.612\,i$  &   $0.464-0.569\,i$   &   $0.857-0.536\,i$   &  0.142  \\
NLO(P3)  &  $0.486 - 0.612\,i$  &   $0.454-0.569\,i$   &   $0.843-0.529\,i$  &   0.150  \\
NLO(P4)  &  $0.487 - 0.614\,i$  &   $0.451-0.569\,i$   &   $0.840-0.526\,i$   &  0.152  \\
NLO(NP)~  &  $0.487 - 0.614\,i$  &   $0.450-0.568\,i$   &   $0.839-0.526\,i$  & ~0.153~ \\
\hline  
NREFT   &  $0.483 - 0.629\,i$  &   $0.424 -0.553\,i$  &   $0.982-0.486\,i$  & 0  \\
\hline
\end{tabular}
\end{center}
\caption{Predictions of ZREFT for complex scattering lengths at the charged-wino-pair threshold. The predictions for $\alpha = 0$ and $\delta = 170$~MeV at unitarity are made using the parameters in table~\ref{tab:Parameters}. The scattering lengths are made dimensionless by multiplying them by $\Delta_*= 31.3$~GeV. The row labeled LO gives the predictions at leading order. The rows labeled NLO give the predictions at next-to-leading order using successive perturbative truncations (P1, P2, P3, P4) and using nonperturbative NLO results (NP). The last row labeled NREFT gives the actual scattering lengths calculated using NREFT. The last column labeled $\Delta a$ gives the square root of the sum of the squares of the errors.
}
\label{tab:Predictions}
\end{table}

Having determined the NLO parameters, we can predict the complex scattering lengths $a_{i \to j}$ that determine the cross sections at the charged-wino-pair threshold. We obtain the prediction  for $a_{i \to j}$ labeled NLO(NP)  in table~\ref{tab:Predictions} by inserting the NLO(NP) parameters in table~\ref{tab:Parameters} into the nonpertubative NLO expression for $a_{i \to j}$. For the perturbative truncation (P$n$), we obtain the prediction for $a_{i \to j}$ labeled NLO(P$n$) by inserting the NLO(P$n$)  parameters in table~\ref{tab:Parameters} into the $n^\text{th}$ order truncated expansion  of the NLO expression for $a_{i \to j}$. The predictions are shown in table~\ref{tab:Predictions} for the first four perturbative truncations (P1, P2, P3, P4). The LO predictions and the correct results calculated in NREFT are also shown in table~\ref{tab:Predictions}. Recall that at LO, the relative errors in the predictions for all the real and imaginary parts of $a_{i \to j}$ were at most 8\%, with the exception of the  Re$(a_{1 \to 1})$, which had a large relative error of 67\%. The NLO(NP) predictions  are consistent with expectations for a systematically improvable approximation method. The  67\% error in the real part of $a_{1 \to 1}$ is reduced to 15\% at the expense of an increase in the error in its imaginary part  from 4\%  at LO to 8\%. The errors in the other four predictions are all  less than 6\%. In  contrast, the  NLO(P1)  predictions do not exhibit the expected improvements. The  67\% error in Re$(a_{1 \to 1})$ is reduced to 6\%, but the errors in the other five predictions are larger than at LO. The error in Im$(a_{0 \to 0})$ is increased from 7\% at LO to 38\%. The error in Re$(a_{0 \to 1})$ is increased from 8\% at LO to 45\%. Table~\ref{tab:Predictions} shows that as the order $n$ of the perturbative truncation is increased, the NLO(P$n$) predictions for $a_{i \to j}$ converge quickly to the NLO(NP) predictions.

\begin{figure}[t]
\centering
\includegraphics[width=0.8\linewidth]{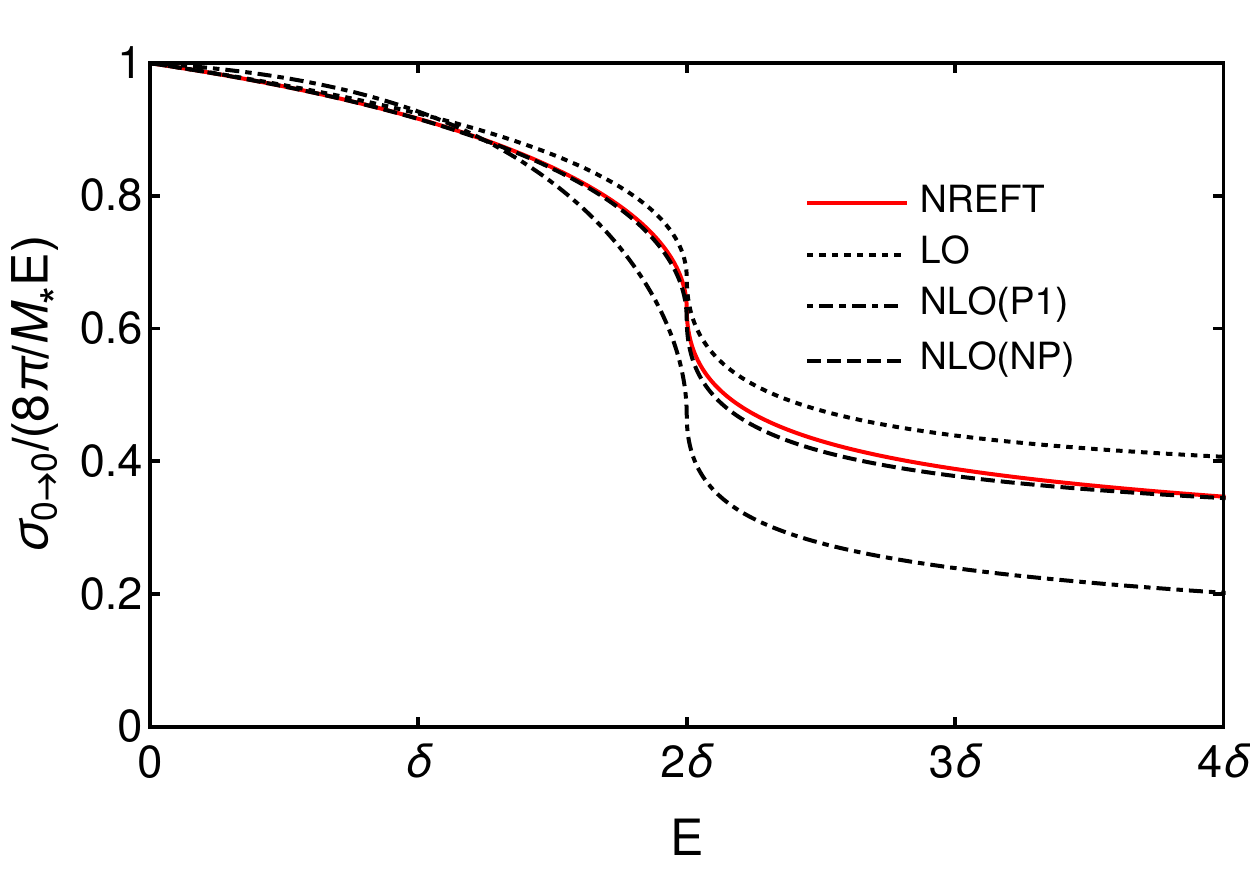}
\caption{The neutral-wino elastic cross section $\sigma_{0 \to 0}(E)$ divided by the S-wave unitarity bound as a function of the energy $E$. The cross section at the unitarity mass $M_* = 2.88$~TeV for $\delta = 170$~MeV and $\alpha=0$ is shown for NREFT (solid curve) and for ZREFT at LO (dotted curve), NLO(P1) (dot-dashed curve), and NLO(NP) (dashed curve).
}
\label{fig:sigma00vsE-NLO}
\end{figure}

Given the ZREFT scattering parameters at NLO in table~\ref{tab:Parameters}, we can compare the energy dependence of the cross-sections predicted by ZREFT with the actual results calculated in NREFT. For the NLO prediction of ZREFT of the T-matrix element $ {\cal T}_{00}(E)$, we insert the NLO parameters into the expression in eq.~\eqref{eq:T00MLONP}, which is nonperturbative in $a_v$ and $r_u$. In figure~\ref{fig:sigma00vsE-NLO}, the NLO(NP) and NLO(P1) predictions for the neutral-wino elastic cross section are compared to the exact result from NREFT and to the LO prediction. The cross sections are divided by the S-wave unitarity bound in order to facilitate comparisons of the predictions in the low-energy limit. As $E$ increases from 0, the NLO(NP) prediction tracks the  NREFT result more accurately than the LO prediction. The NLO(P1) prediction has the correct limit at $E=0$ but the wrong slope. Above the charged-wino-pair threshold, the error in the NLO(NP) prediction is much smaller than the error in the LO prediction, while the error  in the NLO(P1) prediction is much larger. We conclude that the NLO(NP) prediction is much more accurate than the LO prediction throughout the threshold region, and that the NLO(P1) approximation gives rather poor predictions.

\begin{figure}[t]
\centering
\includegraphics[width=0.48\linewidth]{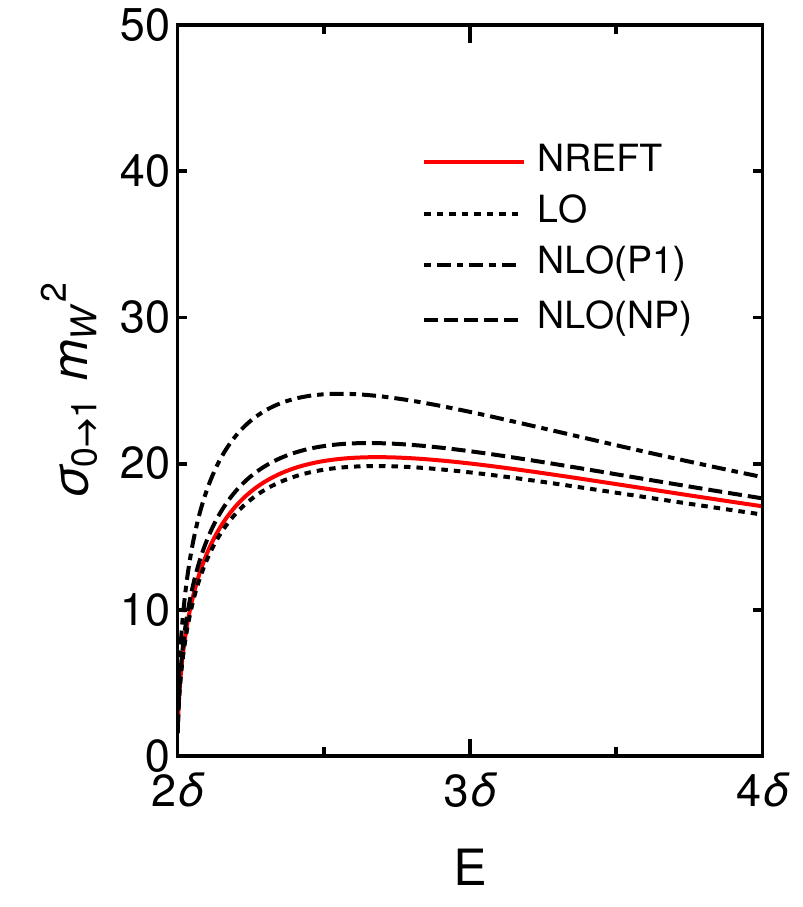}
~
\includegraphics[width=0.48\linewidth]{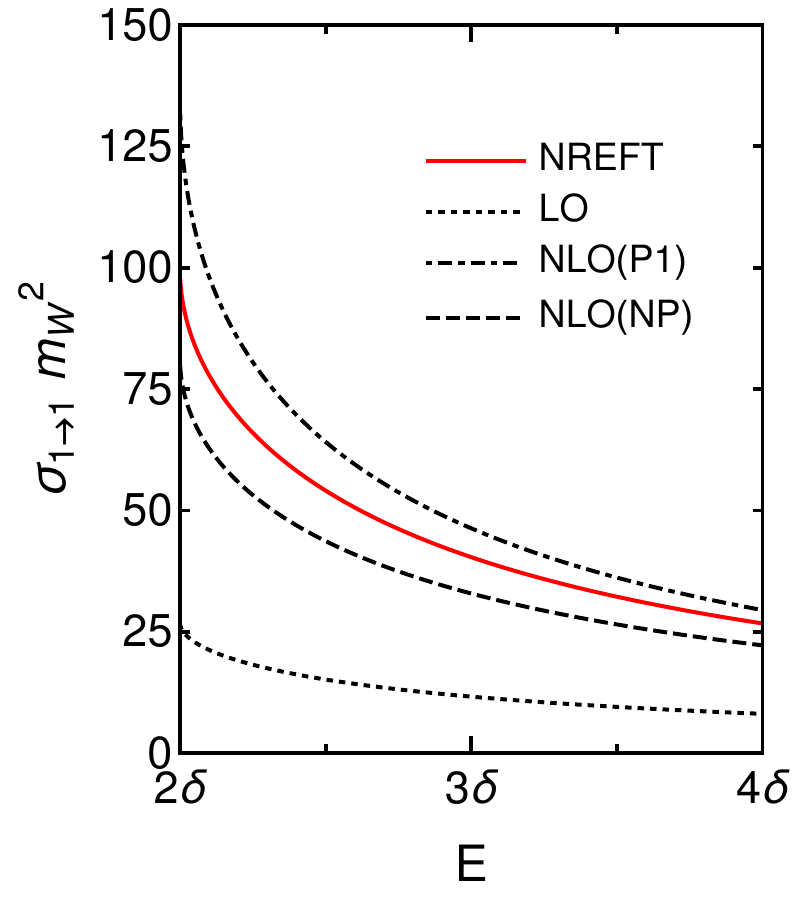}
\caption{
The neutral-to-charged transition cross section $\sigma_{0 \to 1}$  (left panel) and the charged-wino elastic cross section $\sigma_{1 \to 1}$  (right panel) as functions of the energy $E$. The cross sections at 
the unitarity mass $M_* = 2.88$~TeV for $\delta = 170$~MeV and $\alpha=0$ are shown for NREFT (solid curve) and for ZREFT at LO (dotted curve), NLO(P1) (dot-dashed curve), and NLO(NP) (dashed curve).
 }
\label{fig:sigma01,11vsE-NLO}
\end{figure}

In figure~\ref{fig:sigma01,11vsE-NLO}, the  NLO(NP) and NLO(P1) predictions for the neutral-to-charged transition cross section $\sigma_{0 \to 1}$ and the charged-wino elastic cross section $\sigma_{1 \to 1}$ are compared to the exact results from NREFT and to the LO predictions. For $\sigma_{0 \to 1}$, the error in the NLO(NP) prediction is  a little larger than the error in the LO prediction, while  the error in the NLO(P1) prediction is much larger. The relatively small error in the  LO prediction for $|\mathcal{T}_{01}|^2$ is the result of a compensation between significantly larger errors in the real and imaginary parts of $\mathcal{T}_{01}$. The errors in the NLO(NP) predictions for Re($\mathcal{T}_{01}$) and Im($\mathcal{T}_{01}$) are significantly smaller than the errors in the LO predictions. For $\sigma_{1 \to 1}$, the errors in the NLO(NP) prediction and the NLO(P1) prediction are comparable, while the error in the LO prediction is much larger. The relatively small error in the  NLO(P1) prediction for $|\mathcal{T}_{11}|^2$ is the result of a compensation between significantly larger errors in the real and imaginary parts of $\mathcal{T}_{11}$. The errors in the NLO(NP) predictions for Re($\mathcal{T}_{11}$) and Im($\mathcal{T}_{11}$) are significantly smaller than the errors in the NLO(P1) predictions. We conclude that the NLO(NP) predictions are much more accurate than the LO predictions throughout the threshold region, and they are also significantly more accurate than the NLO(P1) predictions.

\subsection{Wino-pair bound state}
\label{sec:BindingNLO}

For neutral-wino mass $M$ above the unitarity mass $M_*(\delta)$, there is a wino-pair bound state  $(ww)$. Each entry of the matrix $\bm{{\cal A}}(E)$ of transition amplitudes has a pole at a real energy $E= - \gamma^2/M$ below the neutral-wino-pair threshold. The determinant of the inverse matrix $\bm{{\cal A}}(E)^{-1}$ must therefore vanish at $E= - \gamma^2/M$. The matrix $\bm{{\cal A}}(E)^{-1}$ at  NLO is given in eq.~\eqref{eq:AinverseNNLO}. The  equation for the binding momentum $\gamma$ at NLO is
\begin{eqnarray}
0 &=& 2(1+t_\phi^2)(1 - a_v  \Delta)(1 - a_v \kappa_1) (\gamma - \gamma_0)   
+ 2 t_\phi^2 (1+t_\phi^2)   (1 - a_v  \gamma_0)(1 - a_v \gamma) (\kappa_1 - \Delta)
\nonumber\\
&&-  \big[ 1 + t_\phi^2 -  a_v (\Delta+ t_\phi^2 \gamma_0) \big] 
\big[ 1 + t_\phi^2 -  a_v (\kappa_1+ t_\phi^2 \gamma) \big] r_u \gamma^2 ,
\label{eq:gammaNLO-eq}
\end{eqnarray}
where $\kappa_1 = \sqrt{\Delta^2+\gamma^2}$ and $\Delta = \sqrt{2 M \delta}$. The binding momentum $\gamma(M,\delta)$  is the positive solution to this equation that approaches 0 as $\gamma_0$ approaches $0^+$. Equation~\eqref{eq:gammaNLO-eq} is linear in $\kappa_1$ and cubic in $\gamma$. By solving for $\kappa_1$ and squaring, it can be transformed into a sixth order polynomial equation for $\gamma$.

\begin{figure}[t]
\centering
\includegraphics[width=0.8\linewidth]{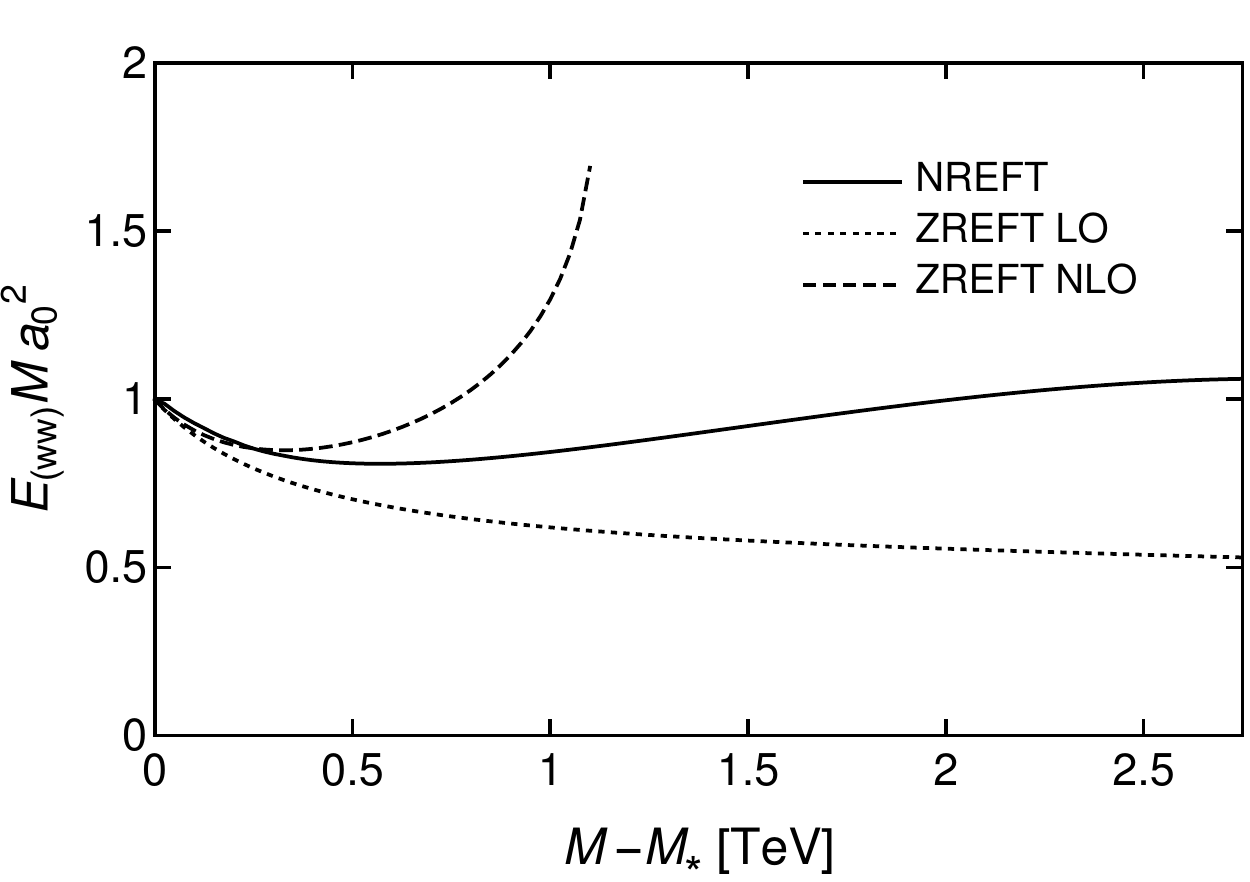}
\caption{The binding energy $E_{(ww)}$ of the wino-pair bound state divided by the universal approximation in eq.~\eqref{eq:Eww-largea} as a function of the wino mass $M$. The binding energy for $\alpha = 0$ is shown for NREFT  (solid line) and  for ZREFT at LO (dotted line) and at NLO (dashed line). In the universal approximation and in ZREFT, $a_0 = 1/\gamma_0$ is given by the Pad\'e approximant analogous to eq.~\eqref{eq:a0Pade} but for $\alpha=0$.
}
\label{fig:bindingenergyNLO}
\end{figure}
 
The binding energy $E_{(ww)} = \gamma^2/M$ of the wino-pair bound state in ZREFT at NLO can be obtained by solving equation~\eqref{eq:gammaNLO-eq} for $\gamma$. In figure~\ref{fig:bindingenergyNLO}, the prediction for  $E_{(ww)}$ for $\delta = 170$~MeV at NLO is compared as a function of $M-M_*$ to the prediction at LO and the result from NREFT. The binding energies are divided by the universal approximation $\gamma_0^2/M$ in eq.~\eqref{eq:Eww-largea} in order to make the differences as $M \to M_*$ more visible. The universal approximation and the prediction of ZREFT at NLO depend on the scattering length $a_0 = 1/\gamma_0$, which is given by the Pad\'e approximant analogous to eq.~\eqref{eq:a0Pade} but with the parameters for $\alpha=0$, which are given in the paragraph after eq.~\eqref{eq:a0Pade}. The binding energies at LO and NLO agree with NREFT very close to the resonance, since they all have the same intercept at $M=M_*$. For small $M-M_*$. the percentage error is significantly smaller at NLO than at LO. At NLO, the error in $E_{(ww)}/(\gamma_0^2/M)$ is less than 5\% for $M-M_* < 0.5$~TeV. At LO, the error is less than 5\% for $M-M_* < 0.15$~TeV. The error at NLO becomes larger than at LO when $M-M_*$ exceeds 1.15~TeV. At NLO, the binding energy $E_{(ww)}$ is only defined for $M-M_* < 1.18$~TeV, because the solution to equation~\eqref{eq:gammaNLO-eq} becomes complex and therefore unphysical for larger $M$. Parametric improvement in the dependence of the binding energy on $M-M_*$ could be obtained by choosing the scattering parameters of ZREFT to depend on $M$.

\section{Double radiative formation of bound state}
\label{sec:DoubleRadiative}

\begin{figure}[t]
\centering
\includegraphics[width=0.4\linewidth]{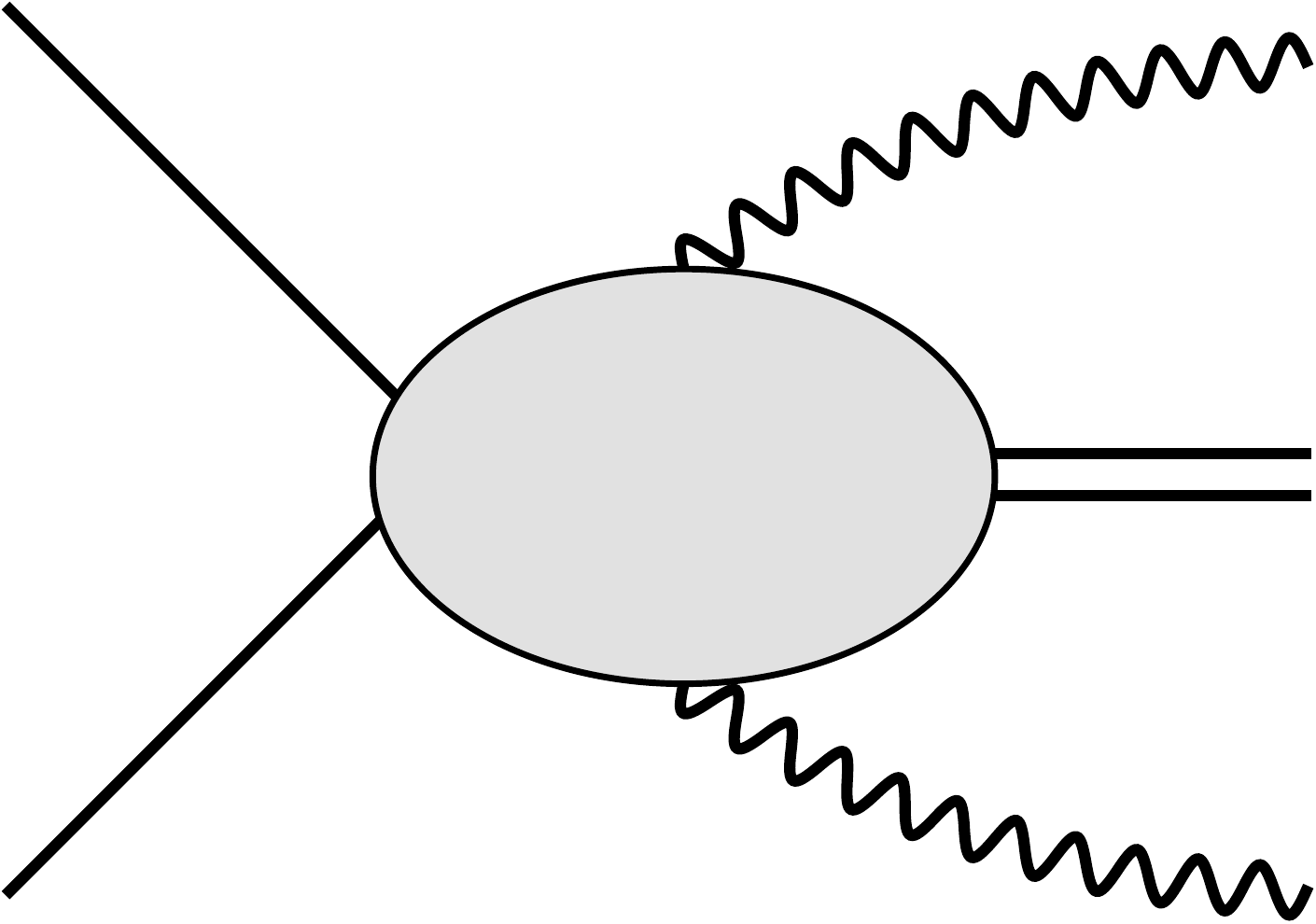}
\caption{Schematic diagram for the double radiative transition $w^0 w^0 \rightarrow (ww) + \gamma\gamma$. The double solid line  represents the wino-pair bound state $(ww)$.}
\label{fig:generalprocess} 
\end{figure}

In order to form the wino-pair bound state in the scattering of two neutral winos, it is necessary to radiate photons in order to conserve energy and momentum. The radiation of a single photon is not allowed by the quantum numbers. The contribution to the formation amplitude that is leading order in the electromagnetic interaction involves the radiation of two photons. Although the electromagnetic interactions can be treated perturbatively, the zero-range interactions of the winos must be treated nonperturbatively. The process $w^0 w^0 \rightarrow (ww) + \gamma\gamma$ is illustrated diagramatically in figure~\ref{fig:generalprocess}. The blob represents the sum of arbitrarily many one-loop diagrams with either a neutral-wino pair or a pair of charged-wino lines. The photons must attach to charged-wino lines.

The matrix element for the two-photon radiative transition to the wino-pair bound state is calculated in Appendix~\ref{app:TwoPhoton}. There are eight diagrams in which the two virtual photons are attached to charged-wino lines in the same loop, which are shown in figure~\ref{fig:OneLoop}. There are eight diagrams in which the two virtual photons are attached to charged-wino lines in separate loops, which are shown in figure~\ref{fig:TwoLoop}. Their sum vanishes due to cancellations among pairs of diagrams.

If the total energy $E$  of the colliding winos in the center-of-momentum frame is small compared to $2\delta$, the matrix element $\cal{M}$ can be  expanded in powers of $E$, $q_1$, and $q_2$, where $q_1$ and $q_2$ are the energies of the two photons, and in powers of $\delta/M$. The leading terms in the expansions in $E/\delta$ and  in $\delta/M$ are given in eq.~\eqref{eq:Mexpand}. The leading term of $\cal{M}$ is
\begin{equation}
\label{eq:Mleading}
\mathcal{M} \approx
\frac{\alpha}{16M^2} \left( \frac{M}{2 \delta} \right)^{5/2} \mathcal{Z}_1^{1/2}\mathcal{A}_{01}(E) \,
q_1 q_2 \, \bm{\varepsilon}_1^* \!\cdot\! \bm{\varepsilon}_2^*.
\end{equation}
In ZREFT at LO, the residue factor $\mathcal{Z}_1$ is given in eq.~\eqref{eq:Z1}. If the binding energy $\gamma^2/M$ is small compared to $\delta$, the residue factor reduces to
\begin{equation}
\label{eq:Z1approx}
Z_1 \approx
\frac{8 \pi \tan^2\phi\, \gamma}{M^2}.
\end{equation}
In ZREFT at LO, the amplitude $\mathcal{A}_{01}(E)$ is equal to $\mathcal{T}_{00}(E)\, \tan \phi/\sqrt{2}$. For energy in the range $0<E<2\delta$, the T-matrix element is given in eq.~\eqref{eq:T00LOloE}. For very low energy satisfying $E\ll 2\delta$, the amplitude is
\begin{equation}
\mathcal{A}_{01}(E) \approx \frac{4 \sqrt2\,\pi \tan\phi/M}{-\gamma_0    - i \sqrt{ME}}.
\label{eq:A0loE}
\end{equation}

To obtain the cross section for the formation of the wino-pair bound state, the matrix element $\mathcal{M}$ in eq.~\eqref{eq:Mtotal} must be squared, summed over the photon polarizations, and then integrated over the 3-body phase space:
\begin{equation}
v_\mathrm{rel} \, \sigma[w^0 w^0 \to (ww) + \gamma \gamma] = 
\sum_{\rm spins} \int |\mathcal{M}|^2\, d\Pi_3,
\label{eq:sig-(ww)}
\end{equation}
where $v_\mathrm{rel} = 2 v_0(E)$ is the relative velocity of the two colliding winos. The sum over spins gives an angular factor:
\begin{equation}
\sum_{\rm spins} (\bm{\varepsilon}_1^* \cdot \bm{\varepsilon}_2^*) (\bm{\varepsilon}_1 \cdot \bm{\varepsilon}_2) = 1 + \cos^2 \theta_{12},
\label{sum-spins}
\end{equation}
where $\theta_{12}$ is the angle between the two photons. In the 3-body phase space integral, we can use the momentum delta function to integrate over the momentum of the bound state. In the energy delta function, we can neglect the recoil energy $(\bm{q}_1 + \bm{q}_2)^2/4M$ of the bound state. The 3-body differential phase space then reduces to
\begin{equation}
d\Pi_3 = 
\frac{1}{4(2\pi)^3} q_1 dq_1\,  q_2 dq_2\,  d\cos\theta_{12} \,  \delta(q_1 + q_2 - \gamma^2/M - E).
\label{eq:dPi3}
\end{equation}
We have included a factor of 1/2 to compensate for overcounting by integrating over all momenta of the two identical photons. The angular integral is trivial. The energy delta function can be used to integrate over $q_2$. The integral over $q_1$ can be evaluated analytically.

At energies satisfying $E \ll 2 \delta$, we obtain a relatively simple result for the reaction rate:
\begin{equation}
v_\mathrm{rel}\sigma[w^0 w^0 \to (ww) + \gamma \gamma] = 
\frac{\alpha^2 \tan^4\phi\,\delta\, \gamma}{840\, M^4}
\left( \frac{2 \delta}{E + \gamma_0^2/M} \right) \left( \frac{E + \gamma^2/M}{2 \delta} \right)^7.
\label{eq:sig-(ww)loE}
\end{equation}
In the scaling region $\gamma_0^2/M \ll E \ll 2 \delta$, it reduces to
\begin{equation}
v_\mathrm{rel}\sigma[w^0 w^0 \to (ww) + \gamma \gamma] \approx
\frac{\alpha^2 \tan^4\phi\, \delta \, \gamma}{840\,  M^4}
\left( \frac{E}{2 \delta} \right)^6.
\label{eq:sig-(ww)scale}
\end{equation}
As $E$ decreases, the cross section decreases rapidly as $E^6$. When $E$ decreases to below $\gamma_0^2/M$, the cross section saturates at a value obtained by replacing $E$ in eq.~\eqref{eq:sig-(ww)scale} by $\gamma_0^2/M$. For collision energy $E=\tfrac14 M v_\mathrm{rel}^2$, the formation rate in eq.~\eqref{eq:sig-(ww)scale} scales as $\alpha^2(\gamma/\sqrt{M\delta})(M/\delta)^{9/2}M^{-2}\, v_\mathrm{rel}^{12}$. In comparison, the annihilation rate into electroweak gauge bosons scales as $\alpha_2^2M^{-2}\,v_\mathrm{rel}^{-2}$ \cite{Hisano:2004ds}. For $v_\mathrm{rel}=10^{-3}$, the relative suppression factor of $v_\mathrm{rel}^{14}$ in the formation rate overwhelms the factor of $(M/\delta)^{9/2}$. The formation rate for the bound state is many orders of magnitude smaller than the annihilation rate, and is therefore not phenomenologically relevant. However, this example calculation illustrates the usefulness of the ZREFT formalism to obtain analytic results for reaction rates which are difficult to calculate in the NREFT framework.

\section{Summary}
\label{sec:Conclusion}

We have developed a zero-range effective field theory (ZREFT) to describe winos whose mass is near a critical value for an S-wave resonance at the neutral-wino-pair threshold. The effects of the exchange of weak gauge bosons between winos is reproduced by contact interactions between the winos that must be  treated nonperturbatively. The electromagnetic interactions of the  charged winos are taken into account through local couplings  to the electromagnetic field. ZREFT is applicable to winos with momenta smaller than $m_W=80.4$~GeV.

An alternative  nonrelativistic effective field theory for winos that we call NREFT was first introduced by Hisano et al.\ to calculate the ``Sommerfeld enhancement'' of the wino-pair-annihilation rate \cite{Hisano:2002fk}. In NREFT, low-energy winos interact instantaneously at a distance through a potential generated by the exchange of  weak gauge bosons, and charged winos also have local couplings  to the electromagnetic field. Calculations in NREFT require the numerical solution of a coupled-channel Schr\"odinger equation. The power of NREFT has recently been demonstrated by a calculation of the capture rates of two neutral winos into wino-pair bound states through the radiation of a photon \cite{Asadi:2016ybp}.

NREFT is more broadly applicable than ZREFT. NREFT can describe winos with any mass $M$, while ZREFT is only applicable if the wino mass is in a window around a critical mass for an S-wave resonance at the neutral-wino-pair threshold. If the wino mass splitting is $\delta = 170$~MeV, the first such unitarity mass is $M_* = 2.39$~TeV, and the window for the applicability of ZREFT is $M$ from about 1.8~TeV to about 4.6~TeV. NREFT describes nonrelativistic winos, while ZREFT can only describe winos with relative momentum less than $m_W$. This limitation of ZREFT may not be significant for most applications to wino dark matter. NREFT can describe the interactions of a pair of winos in any angular-momentum channel, while ZREFT can only describe S-wave interactions. A resonance in the S-wave channel can have larger amplitude and larger width than a resonance in a channel with angular-momentum suppression factors, so it can have a particularly large impact on dark matter. Despite its more limited range of applicability, ZREFT has distinct advantages over NREFT. In particular, two-body observables can be calculated analytically in  ZREFT. This makes it easier to explore the impact of an S-wave near-threshold resonance on dark matter.

In the absence of electromagnetism, ZREFT is a systematically improvable effective field theory. The improvability is guaranteed by identifying a point in the parameter space in which the S-wave interactions of winos are scale invariant in the low-energy limit, and can therefore be described by an effective field theory that is a renormalization-group fixed point. This limit can be obtained by first tuning the wino mass $M$ with fixed wino-mass  splitting $\delta$ to a critical point $M_*(\delta)$ where the neutral-wino scattering length diverges, and then taking the limit $\delta \to 0$.  The RG fixed point describes neutral and charged winos with equal masses, with scattering that saturates the S-wave unitarity bound in a channel that is a linear combination of $w^0 w^0$ and $w^+ w^-$, and with no scattering in the orthogonal channel. The channel in which scattering saturates the unitarity bound is specified by a mixing angle $\phi$. The systematic improvement of ZREFT with $\alpha=0$ is obtained by including deformations of the  RG fixed point with increasingly higher scaling dimensions.

The parameters of ZREFT are the wino mass $M$, the wino mass spitting $\delta$, and scattering parameters. The scattering parameters can be determined by matching low-energy scattering amplitudes in NREFT calculated by solving the Schr\"odinger equation numerically with the scattering amplitudes in ZREFT obtained by solving the Lippmann-Schwinger equations analytically. In ZREFT at LO, the scattering parameters are the mixing angle $\phi$, the neutral-wino scattering length $a_0$, and the electromagnetic coupling constant $\alpha = 1/137$. For NREFT with $\delta = 170$~MeV, $\alpha = 1/137$, and $M$ near the first unitarity mass $M_*=2.39$~TeV, the neutral-wino scattering length $a_0$ is accurately approximated by the Pad\'e approximant in eq.~\eqref{eq:a0Pade}, as illustrated in figure~\ref{fig:Real_a_vsM}. The mixing angle $\phi$ can be determined by matching the effective range $r_0$ for neutral winos at the unitarity mass as in eq.~\eqref{eq:matchr0}. 
For NREFT with $\delta = 170$~MeV, $\alpha = 0$, and the unitarity mass $M_*=2.88$~TeV, the matching of $r_0$ gives $\tan \phi = 0.832$. More accurate matching requires the calculation of $r_0$ in ZREFT at LO including Coulomb resummation, which is presented in a companion paper \cite{BJZ-Coulomb}.
The accuracy of ZREFT at LO with $\alpha = 0$ is studied in section~\ref{sec:ZREFTLO}. The T-matrix elements for wino-wino scattering are given analytically in eqs.~\eqref{eq:T00LOloE} and \eqref{eq:T01,11LO}. The scattering parameter $\phi$ is determined by matching $r_0$ from NREFT for $\delta = 170$~MeV and the unitarity mass $M_*=2.88$~TeV. As illustrated in figures~\ref{fig:sigma00vsE-LO} and \ref{fig:sigma01,11vsE-LO}, ZREFT at LO gives a good approximation to the neutral-wino elastic cross section $\sigma_{0\to0}$ and the neutral-to-charged transition cross section $\sigma_{0\to1}$  for energy $E$ in the  wino-pair threshold region, but it underpredicts the charged-wino elastic  cross section $\sigma_{1\to1}$  by about a factor of 3. This large discrepancy motivated a study of ZREFT at NLO.

The accuracy of ZREFT at NLO with $\alpha = 0$ is studied in section~\ref{sec:ZREFTNLO}. The scattering parameters $\phi$, $a_v$, and $r_u$ are determined by matching $r_0$, $dr_0/d\gamma_0$, and $s_0$ from NREFT for $\delta = 170$~MeV and the unitarity mass $M_*=2.88$~TeV. As illustrated in table~\ref{tab:Predictions}, ZREFT at NLO gives significant improvements over LO in the predictions for the complex scattering lengths $a_{0\to0}$, $a_{0\to1}$, $a_{1\to1}$. As illustrated in figures~\ref{fig:sigma00vsE-NLO} and \ref{fig:sigma01,11vsE-NLO}, ZREFT at NLO gives  significant improvements over LO in the cross sections $\sigma_{0\to0}$, $\sigma_{0\to1}$, and $\sigma_{1\to1}$. In particular, the error in  $\sigma_{1\to1}$ is reduced dramatically. To obtain the improvement, it was essential to avoid simplifying the predictions of ZREFT at NLO by expanding them in powers of $a_v$ and $r_u$. 

We illustrated the power of ZREFT by calculating the rate for the formation of the wino-pair bound state in the collision of two neutral winos through a double radiative transition in which two soft photons are emitted. For collision energy $E \ll 2 \delta$, the cross section is given in eq.~\eqref{eq:sig-(ww)loE}. The reaction rate is many orders of magnitude too small to be of any phenomenological relevance.

Neutral winos with energy much less than $2\delta$ can be described by a simpler ZREFT with only neutral wino fields. The ZREFT for neutral winos only is applicable if the neutral-wino scattering length is larger than $(2 M \delta)^{-1/2}$ as well as $1/m_W$. If the wino mass splitting is $\delta = 170$~MeV, the range around the first unitarity mass $M_* = 2.39$~TeV is $M$ from about 2.1~TeV to about 2.9~TeV. For the ZREFT for neutral winos only, the only interaction parameter at LO is $a_0$. We referred to the LO predictions as the {\it universal approximation}. The universal approximation for the neutral-wino elastic cross section is given in eq.~\eqref{eq:sigma00-largea}. If $M> M_*$ so that there is a bound state, the universal approximation to its binding energy is given in eq.~\eqref{eq:Eww-largea}. We can get accurate quantitative approximations to the cross section and the binding energy in the universal region of $M$ near $M_* = 2.39$~TeV by inserting the Pad\'e approximant for the scattering length in eq.~\eqref{eq:a0Pade}.

The Coulomb interaction between the $w^+$ and $w^-$ is very important near the charged-wino-pair threshold at $E=2 \delta$. For the neutral-wino elastic cross section $\sigma_{0\to0}$, the Coulomb interaction produces resonances below the charged-wino-pair threshold that can be seen in figure~\ref{fig:sigma00-NREFT}. For the neutral-to-charged transition cross section $\sigma_{0\to1}$ and the charged-wino elastic cross section $\sigma_{1\to1}$, the Coulomb interaction produces the Sommerfeld factors that dramatically increase the  cross sections near the charged-wino-pair threshold, as can be seen in figure~\ref{fig:sigma01,11-NREFT}. For $w^+$ and $w^-$ with relative velocity of order $\alpha$ or smaller, the Coulomb exchange diagrams must be summed all orders in $\alpha$. This Coulomb resummation can be carried out analytically in ZREFT. It was first carried out for proton-proton scattering in pionless EFT by Kong and Ravndal \cite{Kong:1999sf}. It was carried out in the ZREFT for the two-channel nuclear physics problem of \textit{p}\,$^7$Li and \textit{n}\,$^7$Be by Lensky and Birse \cite{Lensky:2011he}. Coulomb resummation in the ZREFT for winos is carried out in a companion paper \cite{BJZ-Coulomb}. It provides accurate analytic approximations for the two-wino sector at all relative momenta smaller than $m_W$.

One of the primary motivations for the development of ZREFT for winos was the ``Sommerfeld enhancement'' of the annihilation of a pair of neutral winos into electroweak gauge bosons. 
Wino-pair annihilation not only provides additional wino-wino scattering channels, but it also affects other aspects of the few-body physics for low-energy winos. The effects of wino-pair annihilation are usually suppressed by $\alpha_2^2 m_W^2/M^2$, which is roughly $10^{-6}$ for $M$ in the TeV region, but they can be dramatic near  a unitarity mass. For example, the neutral-wino elastic cross section does not actually diverge at a unitarity mass, but instead has a very narrow peak as a function of $M$  \cite{Blum:2016nrz}. The finite maximum cross section comes from unitarization of the wino-pair annihilation, which has not been taken into account in most previous  calculations of the Sommerfeld enhancement factor. A naive estimate of the maximum cross section is 6 orders of magnitude higher than the cross section above the charged-wino pair threshold. The effects of wino-pair annihilation on low-energy winos can be taken into account in ZREFT by analytically continuing real interaction parameters to complex values. ZREFT can be used to provide analytic results for low-energy wino-wino cross sections and for inclusive wino-pair annihilation rates, including the effects of the unitarization of wino-pair annihilation. The results will be presented in a companion paper \cite{BJZ-Annihilation}.

The ZREFT for  winos with an S-wave resonance near the neutral-wino pair threshold can be generalized to other wimp models that include a dark-matter candidate and have small mass splittings. One important example is higgsino dark matter. The NREFT for higgsinos was used by Hisano et al.\ to calculate the Sommerfeld enhancement of the pair annihilation of the dark-matter particle \cite{Hisano:2002fk,Hisano:2003ec,Hisano:2004ds}. In the spin-singlet S-wave channel, there are three coupled channels. Two angles are therefore required to identify the resonant channel. In the ZREFT at LO, there is one additional interaction parameter, which can be eliminated in favor of the scattering length for the dark-matter particle. The angles and the scattering length can be calculated using NREFT. The development of the ZREFT for  resonant Higgsinos should greatly facilitate the exploration of the effects of S-wave resonances on Higgsino  dark matter.

Another application of  the ZREFT for winos is to models of strongly interacting dark matter in which the dark-matter particle is a member of an $SU(2)$ triplet in a dark sector. Dramatic velocity-dependence of the low-energy cross section can be produced by an S-wave resonance near the pair threshold. ZREFT provides a fairly constrained framework for determining the effects of the dark-matter self-interactions on the small-scale structure of the universe.

\begin{acknowledgments}
This research project was stimulated by a discussion with M.~Baumgart. We thank E.~Tiesinga for providing us with a Fortran code for solving the coupled-channel Schr\"odinger equation for scattering. We thank M.~Baumgart, J.~Beacom, M.~Beneke, R.~Laha, and A.~Peter for useful comments on dark matter. We thank S.~K\"onig for useful discussions of Coulomb effects in zero-range effective field theory. We thank T.~Slatyer for helpful discussions on the partial wave expansion for the Coulomb interaction. This work was supported in part by the Department of Energy under grant DE-SC0011726.
\end{acknowledgments}

\appendix
\section{Lippmann-Schwinger equation}
\label{app:LSeq}

\begin{figure}[t]
\centering
\includegraphics[width=0.9\linewidth]{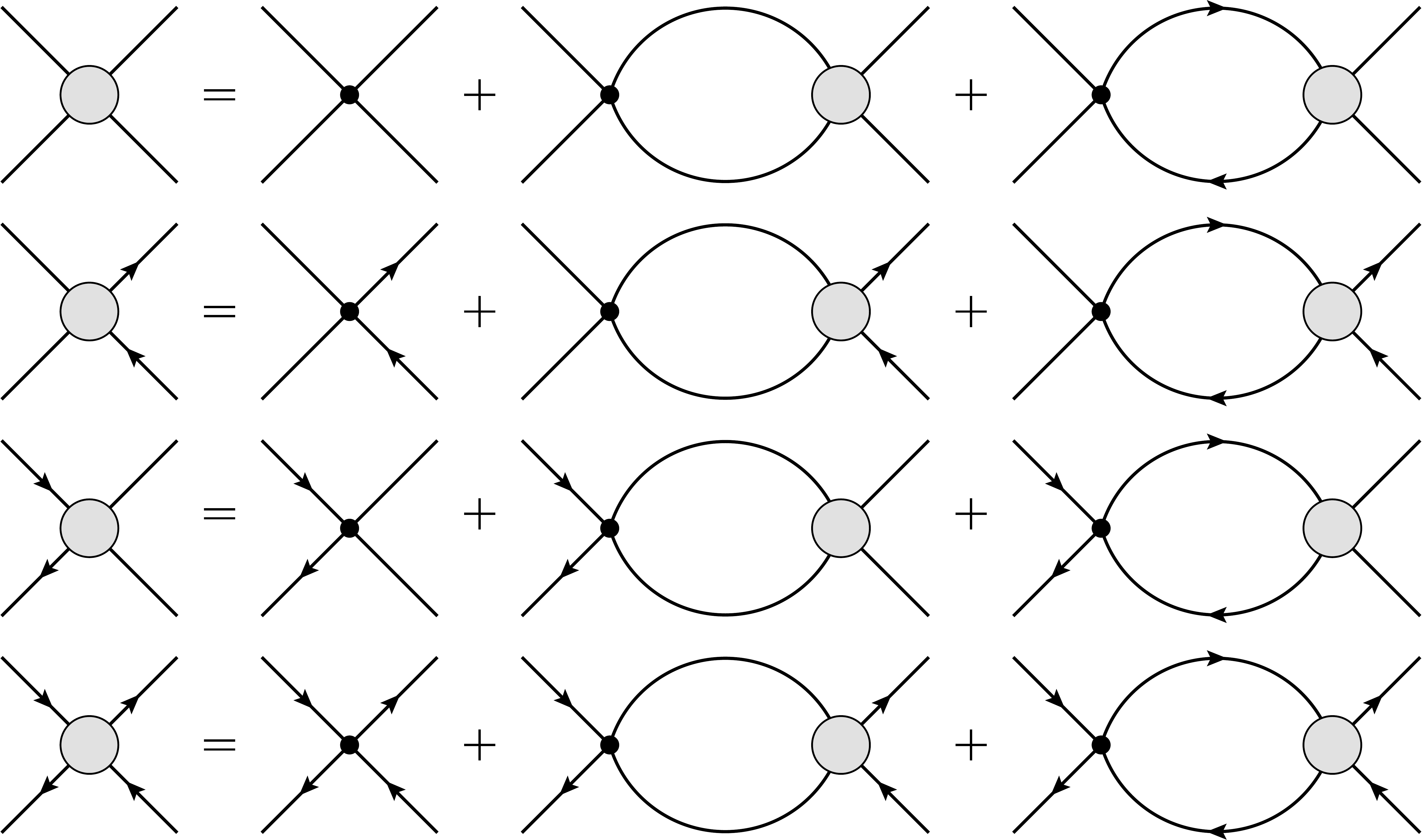}
\caption{Diagrammatic representation of the coupled-channel Lippmann-Schwinger integral equations for the transition amplitudes ${\cal A}_{00}(E)$, ${\cal A}_{01}(E)$, ${\cal A}_{10}(E)$, and ${\cal A}_{11}(E)$.
}
\label{fig:LSeq}
\end{figure}

In the zero-range effective field theory, there are two wino-wino scattering channels: a pair of neutral winos in the spin-singlet channel, which we label by 0, and a pair of charged winos in the spin-singlet channel, which we label by 1. The scattering thresholds are $E=0$ for the neutral channel and $E=2\delta$ for the charged channel. The transition amplitudes have the same Pauli spinor structure as the zero-range interaction vertices. They can be expressed as $\mathcal{A}_{ij}(E)$ multiplied by the spin-singlet projector $\tfrac12(\delta^{ac}\delta^{bd} - \delta^{ad} \delta^{bc})$, where $i$ and $j$ are the incoming and outgoing channels, $a$ and $b$ are Pauli spinor indices for the incoming lines, and $c$ and $d$ are Pauli spinor indices for the outgoing lines. The transition amplitudes $\mathcal{A}_{ij}(E)$ are functions of the total energy $E$ in the center-of-mass frame;  they do not depend separately on the energies and momenta of the incoming and outgoing lines. The transition amplitudes for the two coupled channels can be organized into a $2 \times 2$ matrix:
\begin{equation}
\bm{\mathcal{A}}(E) = 
\begin{pmatrix} 
\mathcal{A}_{00}(E) &\mathcal{A}_{01}(E)\\ \mathcal{A}_{10}(E) & \mathcal{A}_{11}(E) 
\end{pmatrix} .
\label{eq:amplitudematrix}
\end{equation}
The matrix is symmetric: $\mathcal{A}_{10}(E) = \mathcal{A}_{01}(E)$.

The transition amplitudes can be calculated by solving the Lippmann-Schwinger integral equations, which can be expressed as the diagrammatic equations in figure~\eqref{fig:LSeq}. In the momentum representation, the  Lippmann-Schwinger equations can be expressed as a matrix equation:
\begin{equation}
\label{eq:LSequation}
\bm{\mathcal{A}}(E) = -\bm{\lambda}
+\bm{\lambda}\;\bm{I}(E)\;\bm{\mathcal{A}}(E),
\end{equation}
where $\bm{\lambda}$ is a symmetric matrix of bare coupling constants,
\begin{equation}
\bm{\lambda} = \begin{pmatrix}
\lambda_{00} & \lambda_{01} \\ \lambda_{01} & \lambda_{11}
\end{pmatrix} ,
\label{eq:LOcontactmatrix}
\end{equation}
and $\bm{I}(E)$ is a diagonal matrix of loop integrals:
\begin{equation}
\bm{I}(E) = \begin{pmatrix}
\tfrac12 I_0(E) & 0 \\ 0 & I_1(E)
\end{pmatrix} .
\label{eq:looptmatrix}
\end{equation}
The loop integral $I_{1}(E)$ in the center-of-mass frame is
\begin{equation}
I_{1}(E) = i\int \! \frac{dk_0}{2\pi} \int \!\! \frac{d^3k}{(2\pi)^3}
\frac{1}{k_0- k^2/(2M)-\delta+i\epsilon}\frac{1}{E-k_0-k^2/(2M)-\delta+i\epsilon}.
\label{eq:I1intk0}
\end{equation}
The loop integral $I_0(E)$ is obtained by setting $\delta = 0$. The $k_0$ integral in eq.~\eqref{eq:I1intk0} can be evaluated by contours. The resulting integral over $\bm{k}$ is linearly ultraviolet divergent. It can be regularized with dimensional regularization in $d$ spatial dimensions:
\begin{equation}
I_{1}(E) =-M \left( \frac{\Lambda}{2} \right)^{3-d} \int \!\! \frac{d^dk}{(2\pi)^d}
\frac{1}{k^2 - M(E - 2\delta)-i\epsilon},
\label{eq:I1intk}
\end{equation}
where $\Lambda$ is an arbitrary renormalization scale. The integral can be evaluated analytically:
\begin{equation}
I_{1}(E) =-M  \left( \frac{\Lambda}{2} \right)^{3-d} \frac{\Gamma\big((2-d)/2\big)}{(4 \pi)^{d/2}}
\big[ -M(E-2\delta)- i \epsilon) \big]^{(d-2)/2}.
\label{eq:I1int}
\end{equation}
The linear ultraviolet divergence in $d=3$ spacial dimensions appears as a pole in $d-2$ with residue $M \Lambda/4\pi$. The integral can be renormalized by  {\it power divergence subtraction} \cite{Kaplan:1998tg}, in which the limit $d \to 3$ is taken after subtracting the pole in $d-2$. The resulting loop integrals are
\begin{subequations}
\begin{eqnarray}
I_0(E) &=&
 - \frac{M}{4\pi}\big[ \Lambda- \kappa_0(E)\big] ,
\\
I_1(E) &=&
 - \frac{M}{4\pi}\big[ \Lambda- \kappa_1(E) \big] ,
\end{eqnarray}
\end{subequations}
where $\kappa_0(E)$ and $\kappa_1(E)$ are the  functions of the complex energy $E$ defined in eqs.~\eqref{eq:kappa01}. The same results for the loop integrals can be obtained by imposing a sharp momentum cutoff $|\bm{k}| < \frac{\pi}{2}\Lambda$.

To solve the integral equation in eq.~\eqref{eq:LSequation}, we multiply by $\bm{\mathcal{A}}^{-1}$ on the right and $\bm{\lambda}^{-1}$ on the left and then rearrange:
\begin{equation}
\bm{\mathcal{A}}^{-1}(E) = -\bm{\lambda}^{-1} +\bm{I}(E) .
\label{eq:LSsolution}
\end{equation}
The dependence of the amplitudes $\mathcal{A}_{ij}(E)$ on the renormalization scale can be eliminated by choosing the bare parameters $\lambda_{ij}$ to depend on $\Lambda$ in such a way that
\begin{equation}
\bm{\lambda}^{-1} = 
\frac{M}{8 \pi} 
\begin{pmatrix} \gamma_{00}  &\sqrt{2}\, \gamma_{01} \\ \sqrt{2}\,\gamma_{01} & 2\gamma_{11}  \end{pmatrix} 
- \frac{M\Lambda}{8\pi} \begin{pmatrix} ~1~  & ~0~ \\ 0 & 2  \end{pmatrix} .
\label{eq:lambda-gamma}
\end{equation}
This defines physical scattering parameters $\gamma_{00}$, $\gamma_{01}$, and  $\gamma_{11}$ with dimensions of momentum. Substituting these relations into eq.~\eqref{eq:LSsolution}, we have
\begin{equation}
\label{eq:Ainverse2}
\bm{\mathcal{A}}^{-1}(E) = \frac{M}{8\pi}
\begin{pmatrix} -\gamma_{00} + \kappa_0(E)  & -\sqrt{2}\,\gamma_{01} \\ 
 -\sqrt{2}\,\gamma_{01} & 2\big[- \gamma_{11} +\kappa_1(E) \big] 
\end{pmatrix} .
\end{equation}
The inverse of this matrix gives the transition amplitudes for the two coupled channels:
\begin{equation}
\label{eq:Amatrix}
\bm{\mathcal{A}}(E) = \frac{4\pi}{M D(E)}
\begin{pmatrix} 2\big[-\gamma_{11} +\kappa_1(E)\big]  & \sqrt{2}\,\gamma_{01} \\ 
 \sqrt{2}\,\gamma_{01} &   - \gamma_{00} + \kappa_0(E)
\end{pmatrix} ,
\end{equation}
where the function of $E$ in the denominator is
\begin{equation}
\label{eq:Aden}
D(E) =
 \left[ \gamma_{00} - \kappa_0(E) \right] \left[ \gamma_{11} - \kappa_1(E) \right]
- \gamma_{01}^2.
\end{equation}

The T-matrix elements obtained by evaluating these transition amplitudes on the appropriate energy shells are exactly unitary if the scattering parameters $\gamma_{00}$, $\gamma_{01}$, and $\gamma_{11}$ are real.  The unitarity equations can be obtained by expressing the imaginary part of the matrix $\bm{\mathcal{A}}(E+ i \epsilon)$ in a form consistent with the unitarity cutting rules:
\begin{equation}
{\rm Im}\bm{\mathcal{A}}(E+ i \epsilon) = 
- \bm{\mathcal{A}}(E) \left( {\rm Im}\bm{\mathcal{A}}^{-1}(E) \right) \bm{\mathcal{A}}^*(E).
\end{equation}
Using the expression for $\bm{\mathcal{A}}^{-1}$ in eq.~\eqref{eq:Ainverse2}, this can be expressed as the sum of two terms:
\begin{equation}
{\rm Im}\bm{\mathcal{A}}(E+ i \epsilon) = 
\frac{M}{8 \pi} \bm{\mathcal{A}}(E) \left[ 
{\rm Im} \begin{pmatrix} \gamma_{00}  & \sqrt{2}\,\gamma_{01} \\ 
                          \sqrt{2}\,\gamma_{01} & 2\gamma_{11}  \end{pmatrix} 
- {\rm Im}\begin{pmatrix} \kappa_0(E)  & 0 \\ 0 & 2\kappa_1(E)  \end{pmatrix} 
 \right] \bm{\mathcal{A}}^*(E).
\label{A-unitarity}
\end{equation}
If the scattering parameters $\gamma_{00}$, $\gamma_{01}$, and $\gamma_{11}$ are complex, the first term in the square brackets corresponds to deeply inelastic scattering channels, such as annihilation into pairs of electroweak gauge bosons.

\section{Matrix element for transition to bound state}
\label{app:TwoPhoton}

In this appendix, we use ZREFT to calculate the matrix element for the formation of the wino-pair bound state in the collision of two neutral winos by the radiation of two soft photons. The reaction is $w^0 w^0 \to (ww) + \gamma \gamma$, where $(ww)$ is the wino-pair bound state. We work in the center-of-momentum frame of the incoming neutral winos, and take their total energy to be $E$. The outgoing photons have momenta $\bm{q}_1$ and $\bm{q}_2$ and polarization vectors $\bm{\varepsilon}_1$ and $\bm{\varepsilon}_2$. By energy-momentum conservation, the bound state has momentum $-(\bm{q}_1 + \bm{q}_2)$ and energy $E - q_1- q_2 - \gamma^2/M$, where $\gamma$ is the binding momentum that satisfies eq.~\eqref{eq:gammaLO-eq}.

The two soft photons are emitted from charged-wino lines in loop diagrams. The zero-range interactions of the winos must be summed to all orders before and after each photon interaction. Thus each term in the matrix element has a factor of $i\mathcal{A}_{01}(E)$ from the transition of the incoming $w^0 w^0$ to an intermediate $w^+ w^-$ pair and a factor of $-i \mathcal{Z}_1^{1/2}$ from the transition of an intermediate $w^+ w^-$ pair to the bound state $(ww)$.

\begin{figure}[t]
\centering
\includegraphics[width=0.8\linewidth]{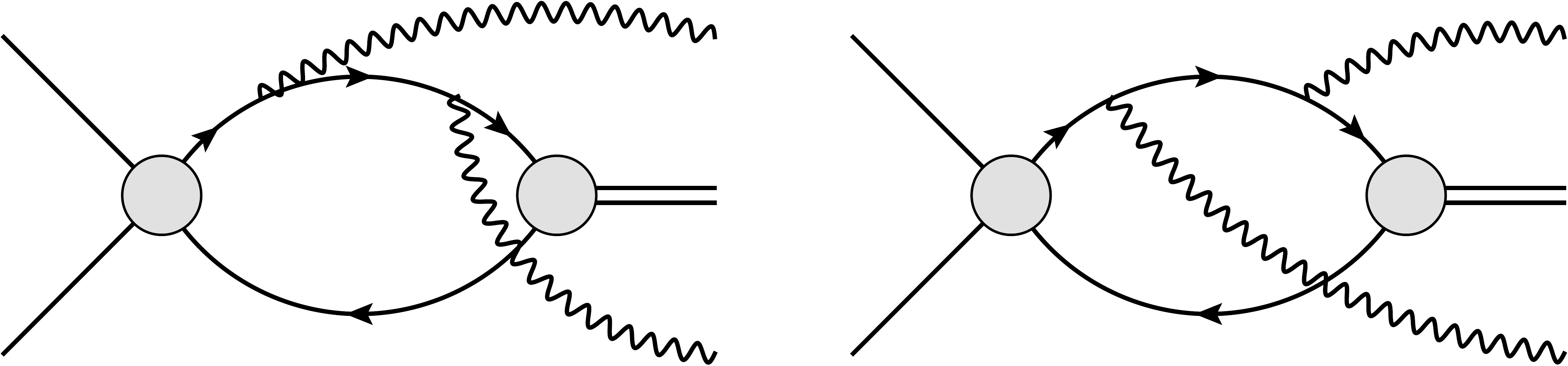}
\\[0.25cm]
\includegraphics[width=0.8\linewidth]{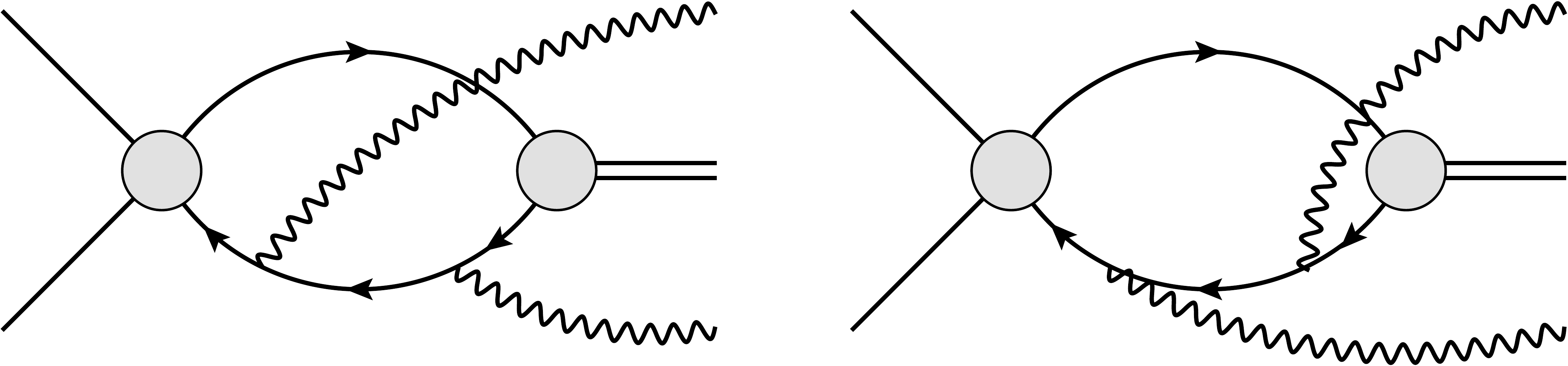}
\\[0.25cm]
\includegraphics[width=0.8\linewidth]{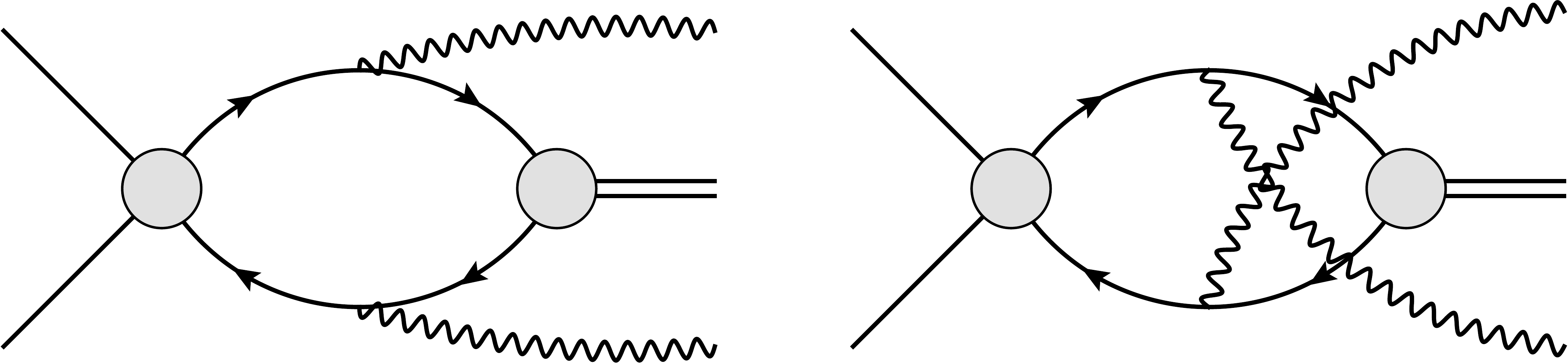}
\\[0.25cm]
\includegraphics[width=0.8\linewidth]{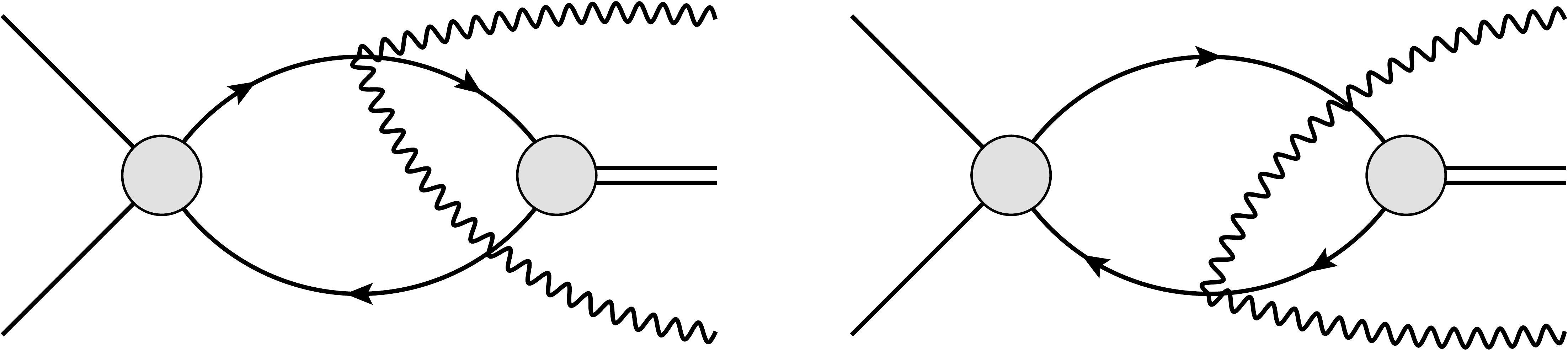}
\caption{Diagrams for $w^0 w^0 \to (ww) + \gamma\gamma$
in which the two photons are attached to charged-wino lines in the same loop.
}
\label{fig:OneLoop}
\end{figure}

There are eight diagrams in which the two virtual photons are attached to charged-wino lines in the same loop. These diagrams are shown in figure~\ref{fig:OneLoop}. After integrating over the loop energy, the diagrams can be expressed as momentum integrals with denominators such as
\begin{subequations}
\begin{eqnarray}
\label{eq:denom}
D_0 &=& 2 \delta - E + k^2/M - i \epsilon,
\\
D_1 &=& 2 \delta - E + q_1+ k^2/(2M) + (\bm{k} + \bm{q}_1)^2/(2M)- i \epsilon,
\\
D_{12} &=&2 \delta - E + q_1+q_2 + k^2/(2M) + (\bm{k} + \bm{q}_1 + \bm{q}_2)^2/(2M) - i \epsilon,
\\
D_{12}' &=&2 \delta - E + q_1+q_2 +  (\bm{k} + \bm{q}_1)^2/(2M) + (\bm{k} - \bm{q}_2)^2/(2M) - i \epsilon.
\end{eqnarray}
\end{subequations}
In the top two rows of diagrams, both photons are emitted from the same charged-wino line by single-photon vertices. In the first row of diagrams, both photons are emitted from the same $w^+$ line. In the second row of diagrams, both photons are emitted from the same $w^-$ line. The matrix element for the first diagram is
\begin{equation}
\label{eq:MA}
\mathcal{M}_A =
-\frac{4 \pi \alpha}{M^2} \mathcal{Z}_1^{1/2}\mathcal{A}_{01}(E) 
\int_{\bm{k}} \frac{\bm{k} \!\cdot\! \bm{\varepsilon}_1^* (\bm{k}+\bm{q}_1) \!\cdot\!\bm{\varepsilon}_2^*}
{D_0 D_1 D_{12}}.
 \end{equation}
The matrix element for the second diagram is the same with $\bm{q}_1, \bm{\varepsilon}_1$ and $\bm{q}_2,  \bm{\varepsilon}_2$ interchanged. The matrix elements for the second row of diagram are the same as the matrix elements for the first row. In the third row of diagrams in figure~\ref{fig:OneLoop}, the two virtual photons are attached to oppositely charged-wino lines in the same loop. The matrix element for the first diagram in the row is
\begin{equation}
\label{eq:MB}
\mathcal{M}_B =
-\frac{4 \pi \alpha}{M^2} \mathcal{Z}_1^{1/2}\mathcal{A}_{01}(E) 
\int_{\bm{k}} 
\left(\frac{\bm{k} \!\cdot\! \bm{\varepsilon}_1^* \bm{k} \!\cdot\!\bm{\varepsilon}_2^*} {D_0 D_1 D_{12}'}
+ (\bm{q}_1 \leftrightarrow \bm{q}_2) \right).
\end{equation}
The matrix element for the second diagram in the row is the same. In the bottom row of diagrams in figure~\ref{fig:OneLoop}, the two virtual photons are attached to the same charged-wino line with a two-photon vertex. The matrix element for the first diagram is
\begin{equation}
\label{eq:MC}
\mathcal{M}_C =
\frac{4 \pi \alpha}{M} \mathcal{Z}_1^{1/2}\mathcal{A}_{01}(E) 
\bm{\varepsilon}_1^* \!\cdot\! \bm{\varepsilon}_2^*\int_{\bm{k}} 
\frac{1} {D_0 D_{12}}.
\end{equation}
The matrix element for the second diagram is the same.

\begin{figure}[t]
\centering
\includegraphics[width=0.45\linewidth]{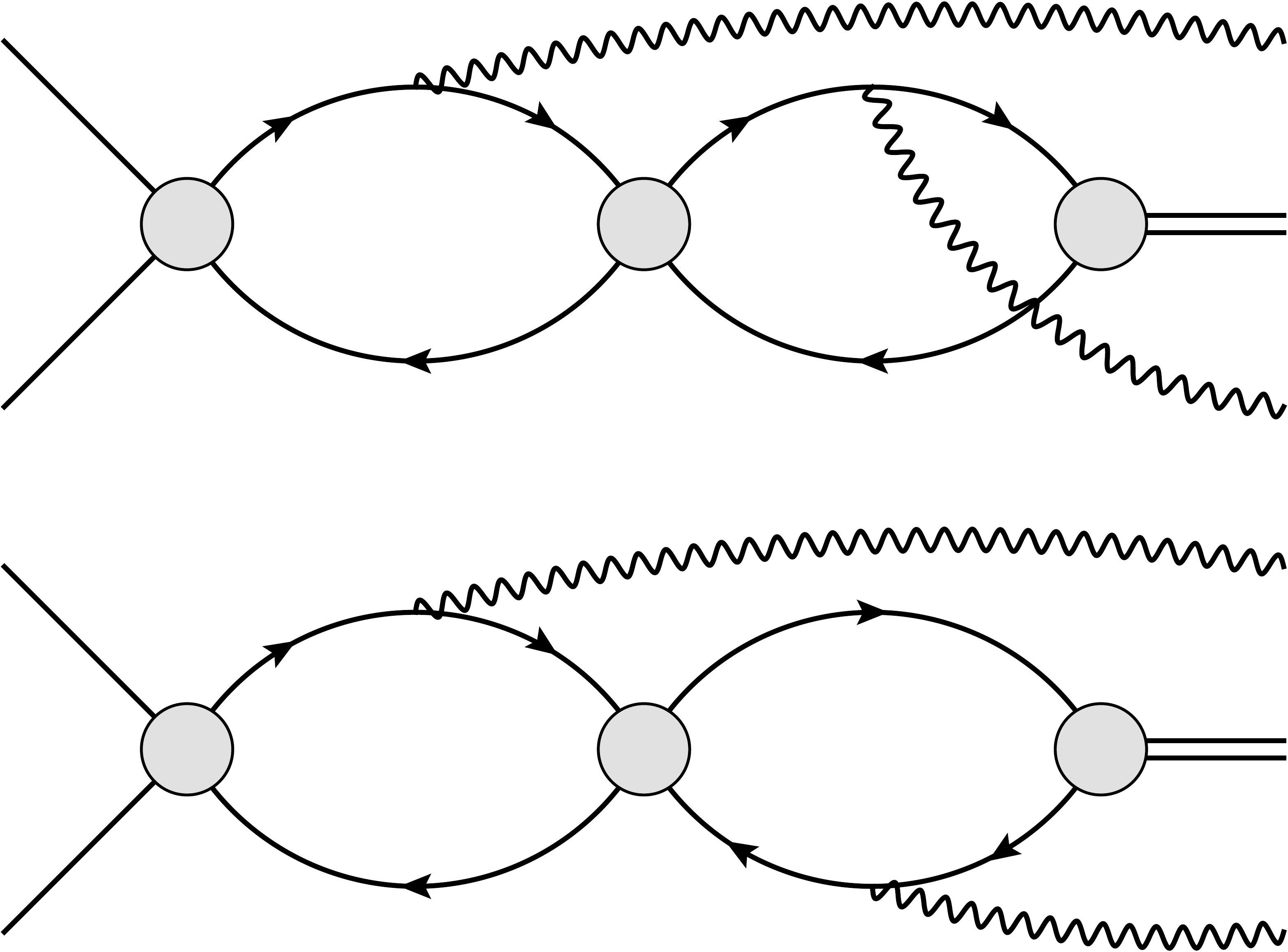}
\quad
\includegraphics[width=0.45\linewidth]{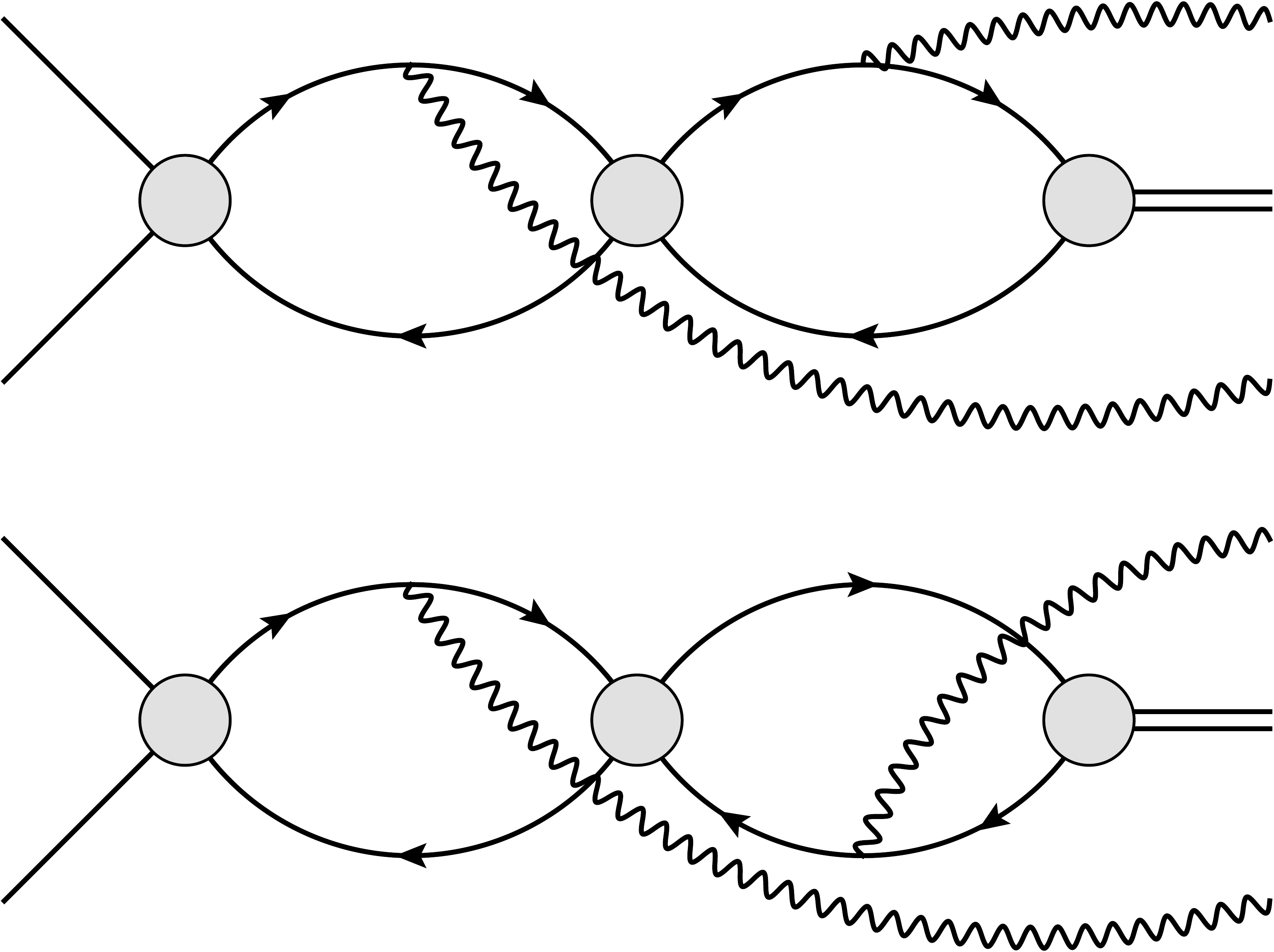}
\\[0.25cm]
\includegraphics[width=0.45\linewidth]{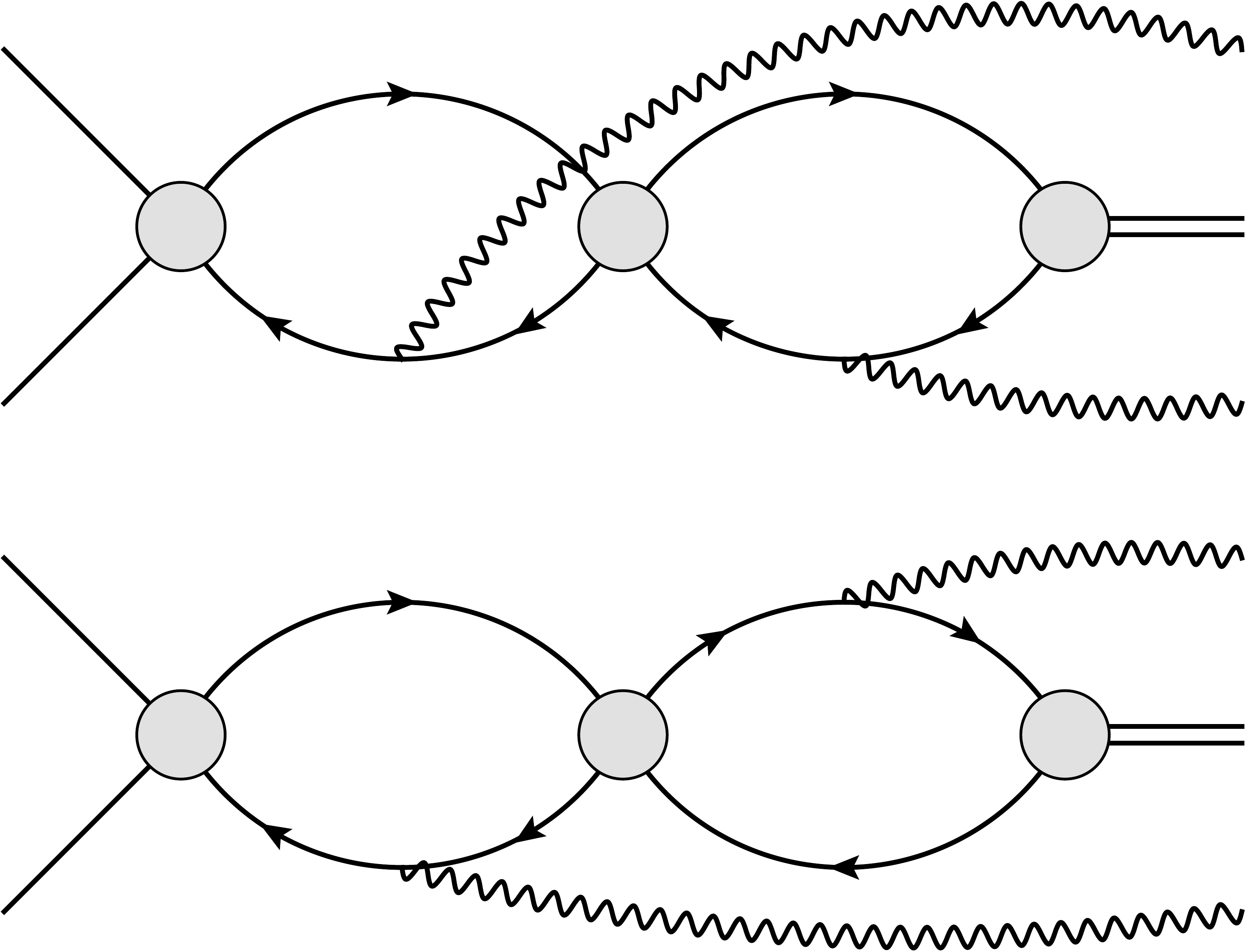}
\quad
\includegraphics[width=0.45\linewidth]{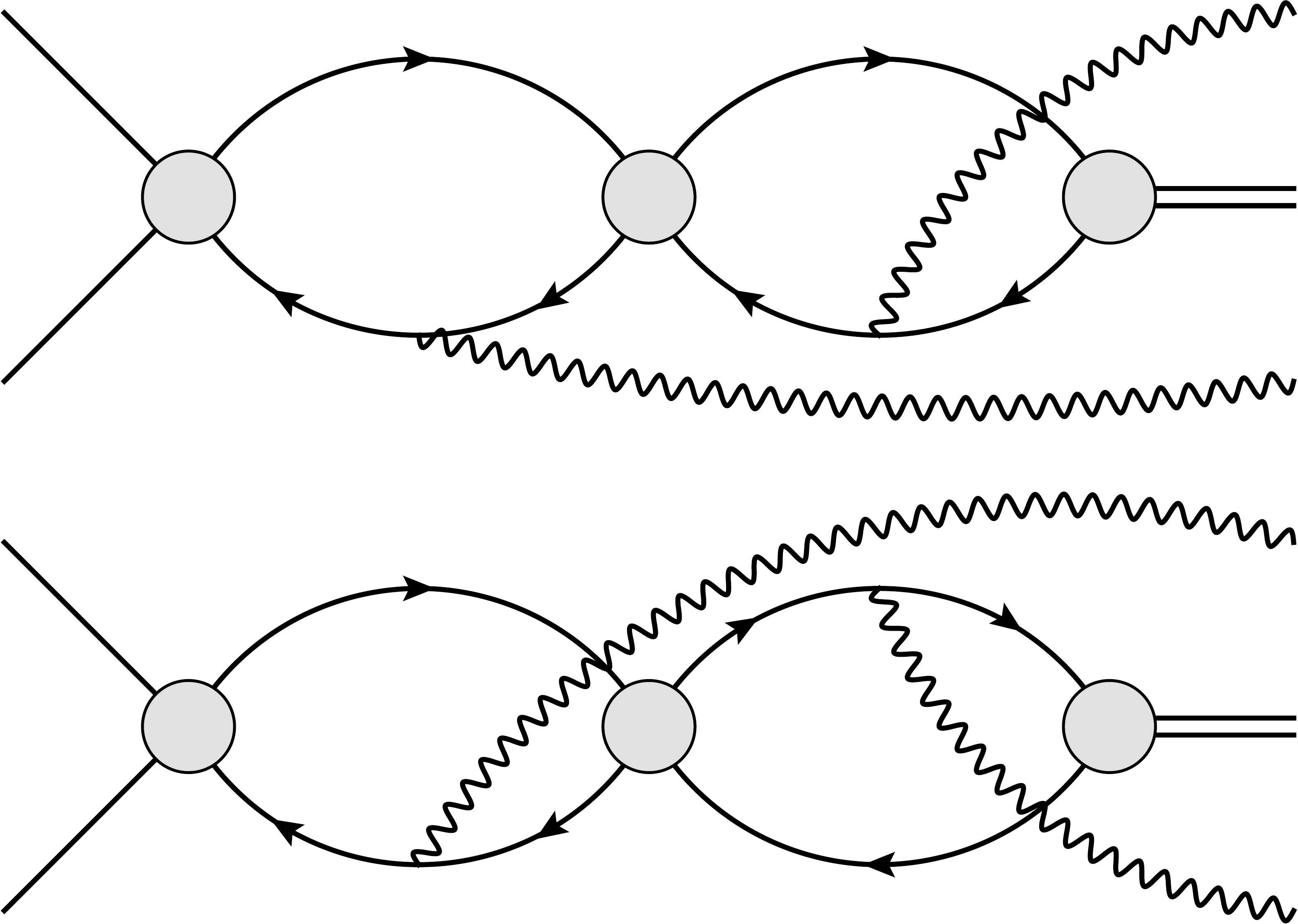}
\caption{Diagrams for $w^0 w^0 \to (ww) + \gamma\gamma$ in which the two photons are attached to charged-wino lines in separate loops. The sum of the two diagrams in which a photon is attached to a $w^+$ line and a $w^-$ from the same bubble is zero.
}
\label{fig:TwoLoop}
\end{figure}

There are eight diagrams in which the two photons are attached to charged-wino lines in separate loops. These diagrams are shown in figure~\ref{fig:TwoLoop}. The sum of a subdiagram in which a photon is attached to the $w^+$ line and the subdiagram in which the photon is attached to the $w^-$ line is 0. All the diagrams in figure~\ref{fig:TwoLoop} therefore cancel in pairs.

The complete matrix element in the center-of-momentum frame is the sum of the matrix elements in eqs.~\eqref{eq:MA}, \eqref{eq:MB}, and \eqref{eq:MC} and the three matrix elements obtained by interchanging $\bm{q}_1$ and  $\bm{q}_2$:
\begin{eqnarray}
\label{eq:Mtotal}
\mathcal{M} =
- \frac{8 \pi \alpha}{M^2} \mathcal{Z}_1^{1/2}\mathcal{A}_{01}(E) 
\int_{\bm{k}} \left(
\frac{\bm{k} \!\cdot\! \bm{\varepsilon}_1^* (\bm{k}+\bm{q}_1) \!\cdot\!\bm{\varepsilon}_2^*}
{D_0 D_1 D_{12}} 
+ \frac{\bm{k} \!\cdot\! \bm{\varepsilon}_1^* \bm{k} \!\cdot\!\bm{\varepsilon}_2^*} {D_0 D_1 D_{12}'} 
\hspace{0.5cm}\right.
\nonumber\\
\left.
- \frac{M \bm{\varepsilon}_1^* \!\cdot\! \bm{\varepsilon}_2^*}{2D_0 D_{12}} 
+ (\bm{q}_1, \bm{\varepsilon}_1 \leftrightarrow \bm{q}_2,  \bm{\varepsilon}_2) \right).
\end{eqnarray}
The momentum integrals can be evaluated by first combining the denominators using Feynman parameters, shifting the loop momentum $\bm{k}$ to eliminate terms in the denominator that are linear in $\bm{k}$, and then integrating analytically over $\bm{k}$. The integrands of the Feynman parameter integrals are functions of $2\delta-E$, $q_1$, $q_2$, and $q_3=|\bm{q}_1 + \bm{q}_2|$. Under the assumption that $E$ is at most of order $\delta$, the integrands can be expanded in powers of $q_i^2/M \delta$. The Feynman parameter integrals can then be evaluated analytically. Dropping terms suppressed by $q_i^2/M \delta$, the matrix element is
\begin{eqnarray}
\label{eq:Mlead}
\mathcal{M} \approx
-\frac{\alpha}{3} M^{1/2} \mathcal{Z}_1^{1/2}\mathcal{A}_{01}(E) 
\left( \frac{d_0 d_{12} + (d_0 + d_{12})d_1 - 3 d_1^2}{(d_0+d_1)(d_0+d_{12})(d_1+d_{12})} 
+ (d_1 \leftrightarrow d_2) \right) \bm{\varepsilon}_1^* \!\cdot\! \bm{\varepsilon}_2^*,
\end{eqnarray}
where the variables $d_i$ are defined by
\begin{subequations}
\begin{eqnarray}
d_0 &=& (2\delta - E - i \epsilon)^{1/2},
\\
d_1 &=& (2\delta - E +q_1- i \epsilon)^{1/2},
\\
d_2 &=& (2\delta - E +q_2- i \epsilon)^{1/2},
\\
d_{12} &=& (2\delta - E +q_1+q_2- i \epsilon)^{1/2}.
\end{eqnarray}
\end{subequations}

If we consider total energy $E$  small compared to $\delta$, the integrands of the Feynman parameter integrals can be expanded in powers of $E$, $q_1$, and $q_2$. Keeping terms through second order in $E$, including terms from expanding in $q_i^2/M \delta$, the matrix element is
\begin{equation}
\label{eq:Mexpand}
\mathcal{M} \approx
\frac{\alpha}{16M^2} \left( \frac{M}{2 \delta} \right)^{5/2} \mathcal{Z}_1^{1/2}\mathcal{A}_{01}(E) \,
q_1 q_2 \, \bm{\varepsilon}_1^* \!\cdot\! \bm{\varepsilon}_2^*.
\end{equation}
The structure of this term can be understood from an effective field theory perspective. If $E$ is very small, the matrix element can be reproduced by a low-energy effective field theory for the neutral winos, the wino-pair bound state, and photons. The leading term in the matrix element comes from an operator in the effective lagrangian that annihilates a pair of winos, creates the wino bound state, and creates two photons. Since the winos and the bound state are electrically neutral, the operators cannot have any covariant derivatives. Gauge invariance implies that the photons must be created by either $\bm{E}^2$ or $\bm{B}^2$, where $\bm{E}$ and $\bm{B}$ are the electric and magnetic fields. The term in eq.~\eqref{eq:Mexpand} comes from the $\bm{E}^2$ term. The term at next order in $q_i^2/M\delta$, which is proportional to $\bm{q}_2 \!\cdot\! \bm{\varepsilon}_1^*\, \bm{q}_1 \!\cdot\!\bm{\varepsilon}_2^* - \bm{q}_1 \!\cdot\! \bm{q}_2 \, \bm{\varepsilon}_1^* \!\cdot\! \bm{\varepsilon}_2^* $, comes from the $\bm{B}^2$ term.


\end{document}